\documentclass[twocolumn, usenatbib]{mnras}
\usepackage{savesym}
\usepackage{graphicx}
\usepackage{longtable}
\usepackage{changepage}

\expandafter\let\csname equation*\endcsname\relax
  \expandafter\let\csname endequation*\endcsname\relax 
\usepackage{subfig}
\usepackage{amsmath}
\usepackage{amssymb}
\usepackage{verbatim}
\usepackage[yyyymmdd,hhmmss]{datetime}
\usepackage{array}
\usepackage{times}
\usepackage{xcolor}

\DeclareRobustCommand{\DE}[3]{#2}
\let\DEthebibliography\thebibliography
\def\thebibliography{\DeclareRobustCommand{\DE}[3]{##3}\DEthebibliography}

\newcommand\mmb{{\cite{MummeryMori24}}\, }

\newcommand\fk{{{\tt fullkerr}\ }}

\title[Continuum emission from within the ISCO]{ Continuum emission from within the plunging region of black hole discs }
\author [Andrew Mummery, et al.]{Andrew Mummery$^1$\thanks{E-mail:
andrew.mummery@physics.ox.ac.uk}, Adam Ingram$^2$, Shane Davis$^3$, Andrew Fabian$^4$ 
\\
$^1$Oxford Theoretical Physics, Beecroft Building,  Clarendon Laboratory, Parks Road, Oxford, OX1 3PU, United Kingdom \\
$^2$School of Mathematics, Statistics and Physics, Newcastle University, Herschel Building, Newcastle upon Tyne, NE1 7RU, United Kingdom \\
$^3$Department of Astronomy, University of Virginia, Charlottesville, VA \\
$^4$Institute of Astronomy, Madingley Road, Cambridge CB3 0HA}

\date{}

\begin{document}

\pagerange{\pageref{firstpage}--\pageref{lastpage}} \pubyear{2023}

\maketitle

\label{firstpage}

\begin{abstract} 
 The thermal continuum emission observed from accreting black holes across X-ray bands has the potential to be leveraged as a powerful probe of the mass and spin of the central black hole. The vast majority of existing ``continuum fitting'' models neglect emission sourced at and within the innermost stable circular orbit (ISCO) of the black hole. { Numerical simulations, however, find non-zero} emission sourced {from} these regions. In this work we extend existing techniques by including the emission sourced from within the plunging region, utilising new analytical models which reproduce the properties of numerical accretion simulations. We show that in general the neglected intra-ISCO emission produces a hot-and-small quasi-blackbody component, but can also produce a weak power-law tail for more extreme parameter regions. A similar hot-and-small blackbody component has been added in by hand in an {\it ad-hoc} manner to previous analyses of X-ray binary spectra. We show that the X-ray spectrum of MAXI J1820+070 in a soft-state outburst is extremely well described by a full Kerr black hole disc, while conventional models which neglect intra-ISCO emission are unable to reproduce the data. We believe this represents the first robust detection of intra-ISCO emission in the literature, and allows additional constraints to be placed on the MAXI J1820+070 black hole spin which must be low $a_\bullet < 0.5$ to allow a detectable intra-ISCO region. Emission from within the ISCO is the dominant emission component in the MAXI J1820+070 spectrum between $6$ and $10$ keV, highlighting the necessity of including this region.  Our continuum fitting model is made publicly available. 
\end{abstract}

\noindent

\section{Introduction} 
Accretion onto astrophysical black holes liberates a substantial fraction of the rest mass energy of the accreting material, energy which is ultimately carried away as light, the majority of which is sourced from the very innermost disc regions. X-ray photons observed from galactic X-ray binaries therefore, in principle, carry characteristic observational signatures of the highly relativistic region of spacetime close to the event horizons of these black holes.  Even with the advent of multi-messenger probes of gravity \citep[e.g.,][]{Abbott16}, X-ray spectroscopy remains one of the principal techniques through which the physical properties (the mass and rate of rotation) of astrophysical black holes are probed. 

One of the two major observational techniques is named “continuum fitting”, and involves fitting the broad band spectral energy distribution (SED) observed from accreting systems across an X-ray bandpass.  As the details of the disc temperature profile, as well as various gravitational optics effects, are sensitive to both the mass and spin of the central black hole, these broad-band SEDs in principle contain sufficient information to constrain the black holes characteristics.  Indeed, this technique has been employed widely throughout the literature \citep[e.g.,][]{Shafee06, Steiner09, McClintock14, Zhao21}. Of course, inferences of black hole properties obtained using the continuum fitting procedure are model dependent, as some prescription for the locally liberated energy of the disc flow must be assumed. Systematic errors with classical theoretical descriptions of accretion flows could in principle lead to systematic errors in black hole parameter inference, and it is essential therefore to continually evaluate and look to extend theoretical models of black hole accretion.

One notable omission in classical {thin disc} models is the ignored  emission sourced at and within the ISCO (innermost stable circular orbit) of a black hole's spacetime. The ISCO is the outermost radial location at which circular test particle motion in the Kerr metric becomes unstable to inward perturbations. As inward perturbations rapidly grow in radial velocity, the ISCO is typically considered to be the ``inner edge'' of the disc.   Classical models of black hole accretion impose the so-called `vanishing ISCO stress' boundary condition \citep[e.g.,][]{SS73, PageThorne74}, which forces the temperature of the accretion flow to be zero at the ISCO, with no emission assumed to be produced between the ISCO and event horizon. In the original formulations of the relativistic thin disc equations \citep{NovikovThorne73, PageThorne74, Thorne1974} it was noted that this boundary condition may not be valid if there are substantial magnetic stresses on the flow in the innermost disc regions. At the time the physical origin of the turbulent stress which drives accretion was unknown. 

We now know that magnetic fields are essential to driving accretion in black hole discs, as accretion is mediated by angular momentum transport driven by the magneto-rotational instability \citep[MRI;][]{BalbusHawley91, BalbusHawley98}. It is therefore perhaps unsurprising that numerical simulations \citep[e.g.,][]{Noble10, Penna10, Zhu12, Schnittman16, Lancova19, Wielgus22} generically find non-zero temperatures at and within the ISCO, with non-zero associated thermal emission. This emission will inevitably modify the SED of an observed black hole disc, and may induce systematic errors in inferred black hole properties. 

The properties of the intra-ISCO thermal emission have previously only been examined in a limited number of general relativistic magnetohydrodynamic (GRMHD) simulations \citep[e.g.,][]{Noble11, Zhu12, Schnittman16}. This analysis is necessarily limited due to the great computational expense of running GRMHD simulations, which prevents global parameter studies.  Recently however,  \cite{MummeryBalbus2023} derived analytical models for the intra-ISCO disc thermodynamics, using solutions for the intra-ISCO kinematics of test particles \citep{MummeryBalbus22PRL}. It was demonstrated in \mmb that these analytical solutions reproduce the liberated energy of the full GRMHD simulations of \cite{Zhu12}, a result discussed in more detail below.  As such, it is now possible to derive continuum fitting models which include analytic expressions for the locally liberated energy {\it within} the ISCO, and examine the effects of this additional emission on modelling of astrophysical sources. Presenting and analysing such a continuum fitting model is the purpose of this paper.  

{While this work can in one sense be thought of as an extension of classical models based on a new theoretical development, it may in fact be more interesting to reverse this entire framework, and seek to use observations of Galactic X-ray binaries to probe a long-contentious question of theoretical physics.  The vanishing ISCO stress boundary condition has retained prominent advocates  \citep[e.g.,][]{Paczynski00}, who argue that a vanishing ISCO stress is a natural consequence of viscous angular momentum transport\footnote{It should be noted, of course, that angular momentum transport in an accretion flow is fundamentally not governed by viscosity, but by MHD turbulence which is a physically distinct process.}. Even if strictly true, in the notation of the remainder of this paper Paczy{\'n}ski argues that the ISCO stress should scale as $\delta_{\cal J} \sim \alpha (H/r)$, where $\alpha$ and $H/r$ are the disc alpha parameter and aspect ratio respectively,  we shall argue however that even a percent level ISCO stress can have important observational implications (e.g., Fig. \ref{fig:temp-comp}).  Furthermore,  differing simulations have found measured values for the ISCO stress which differ from each other by over an order of magnitude, and it has been suggested that the initial magnetic field topology employed in these simulations  may play an important role in the ultimately measured stress \citep[e.g.,][]{Penna10, Noble10}. It seems likely that high quality X-ray observations of black hole discs will contain signatures of the plunging region, much in the same way in which they can be leveraged to constrain black hole properties. It is the authors' hope that the models developed in this paper, coupled with future X-ray observations, may offer an observational handle through which  the complex physics of MHD turbulence in the inner regions of black hole discs may be probed.    }

{Indeed, observations of black hole systems are beginning to hint at cracks in pre-existing theories.} \cite{Fabian20} argued that observations of the X-ray binary system MAXI J1820+070 (hereafter MAXI J1820) in an extremely bright state could not be successfully modelled without the addition of {\it ad hoc} thermal emission components with small emitting areas and high emitting temperatures. The characteristics of this additional emission were argued to be similar to what might be expected from the intra-ISCO fluid. Similar additional components were seen in another MAXI source, MAXI J0637-430  \citep{Lazar21}, and it is possible that these components are wide spread but often ignored. Other observations that may hint at the need for additional physics beyond the standard approach include the extreme \citep[beyond the][limit]{Thorne1974} spins inferred from continuum fitting of the high mass X-ray binary Cygnus X-1 \citep{Zhao21}.  

We will demonstrate in this paper that the guess put forward by \cite{Fabian20} was correct, and MAXI J1820 is extremely well described by an accretion disc with a substantial emission component stemming from within the ISCO. The addition of an intra-ISCO component is required at a significance much stronger than $10\sigma$, with a change in $\chi^2$ statistic  of $\Delta \chi^2 = -630, -785, -820, -816$ respectively for each of the four epochs of MAXI J1820 NuSTAR data.  

The remainder of this paper is organised as follows. In sections 2, 3 and 4 we introduce the physics of the three components of the continuum fitting model. In section 2 we discuss the disc model, and how the locally liberated disc energy is computed. In section 3 we discuss the orbits and energy-shifts of photons emitted from disc fluid within the ISCO. The physics of the radiative transfer of the locally liberated energy are discussed in section 4, where we present the form of the rest frame energy spectrum of the intra-ISCO material. In section 5 we show example X-ray spectra, computed for a wide range of parameter space, highlighting the properties and importance of the intra-ISCO emission. In section 6 we show that previously poorly fit X-ray spectra of MAXI J1820 are very well described with the addition of intra-ISCO emission. {A discussion of the wider implications of these results are presented in section 7,} before we conclude in section 8. Some  results are presented in a set of Appendices.

\section{The disc model}

\begin{figure}
    \centering
    \includegraphics[width=\linewidth]{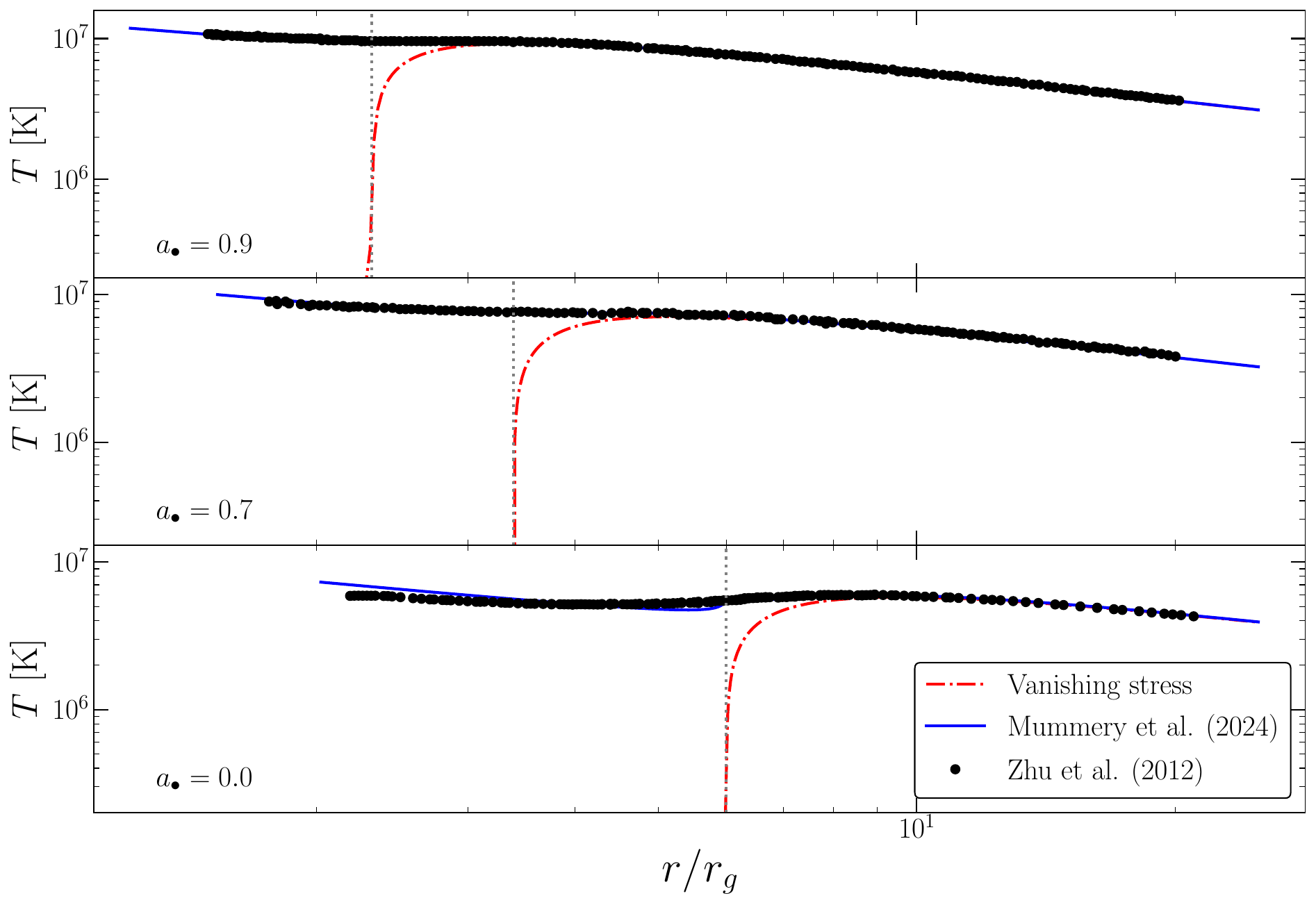}
    \caption{ The radiative temperature profiles of post-processed GRMHD simulations presented in \citealt{Zhu12} (black points; see text), compared to the analytical models developed in \citep[][\citealt{MummeryMori24}, blue curves]{MummeryBalbus2023} for three values of the black hole spin (displayed on plot). Also shown are vanishing ISCO stress models for comparison (red dashed curves). The simple analytical models reproduce the results of full GRMHD simulations. The ISCO stress parameters are $\delta_{\cal J} = 0.0055$ for $a_\bullet = 0.9$, $\delta_{\cal J} = 0.0085$ for $a_\bullet = 0.7$ and  $\delta_{\cal J} = 0.007$ for $a_\bullet = 0$. }
    \label{fig:temp-comp}
\end{figure}

The first key component of a continuum fitting model is a prescription for the locally liberated energy of the disc. Typically this locally liberated energy is a function of disc location and a few global physical free parameters of the system, such as the central black hole's mass and spin.  In this work we use the recent global thin disc solutions of \cite{MummeryBalbus2023}. This model was then extended in \mmb to better model the trans-ISCO properties of the flow. 

A detailed description of this new model is presented in these recent papers, and we only briefly summarise the key results and properties of them here (Appendix \ref{appA} contains a more detailed discussion).  On a fundamental level the model consists of the classical \cite{NovikovThorne73} description of the disc at radii larger than the ISCO $r > r_I$. The \cite{NovikovThorne73} model is the basis for many contemporary continuum fitting models \citep[e.g., the {\tt KERRBB} model in {\tt XSPEC}][]{Li05}. The vast majority of previous continuum fitting models are curtailed at the ISCO \citep[or attempt to self consistently compute the ISCO stress with an $\alpha$-model][which is likely to be a severe over-simplification]{Straub11}. In our extended model we solve for the fluid's evolution interior to the ISCO by assuming that the fluid elements undergo a geodesic plunge, with thermodynamic evolution driven by a combination of the radial stretching of the fluid (from the increasing intra-ISCO radial velocity), and the vertical compression of the flow (driven by the rapidly increasing vertical gravity of the flow as the event horizon is approached). The disc fluid is assumed to remain optically thick to scattering throughout the plunging region (an assumption which can be tested, and is satisfied provided the accretion rate is not too small $\dot m \equiv \dot M / \dot M_{\rm edd} \gtrsim 10^{-3}$; see equation \ref{scattmin}). The fact the flow remains optically thick to scattering allows the local radiative flux to be extracted from the disc's thermodynamic quantities. 

The joining of these two model solutions introduces an additional free parameter to the disc system: the value of the ISCO stress, denoted by $\delta_{\cal J}$. Physically $\delta_{\cal J}$ corresponds to the angular momentum the fluid passes back to the disc upon crossing the ISCO (normalised by the angular momentum of a circular orbit at the ISCO), where a  value of $\delta_{\cal J} = 0$ corresponds physically to no communication between the disc and plunging region and a vanishing ISCO stress. The  value $\delta_{\cal J} = 1$ represents maximal communication between the two regions, and fluid elements cross the event horizon with zero angular momentum\footnote{In principle magnetic field lines which cross a spinning black hole and the inner edge of the stable region of a disc can provide an {\it external} torque on the disc, with photon emission sourced directly from the black hole's spin-energy. In this case the parameter $\delta_{\cal J}$ represents the normalised angular momentum flux of this process, not an internal fluid torque. For black hole spins $a_\bullet > 0.3594$ all of the liberated energy of the disc can be sourced from this external torque with no accretion required, and $\delta_{\cal J}$ can take any positive value \citep{AgolKrolik00}.}. Numerical simulations \citep[e.g.][]{Shafee08, Noble10} place $\delta_{\cal J}$ somewhere in the range $\delta_{\cal J} \sim 0.02 - 0.2$.

The success of this model in reproducing the properties of full GRMHD simulations is displayed in Fig. \ref{fig:temp-comp}, where we compare the radiative temperature of these analytical models to those found in the GRMHD simulations of \cite{Zhu12}. \cite{Zhu12} extracted a radiative temperature from the local cooling rate computed in the GRMHD simulations run by \cite{Penna10}, and are displayed in Figure \ref{fig:temp-comp} by black dots, for three different black hole spins $a = 0$ (lower panel), $a = 0.7$ (middle panel) and $a=0.9$ (upper panel). The \cite{Zhu12} models all consist of an $M = 10 M_\odot$ black hole accreting at roughly $\dot M \sim 0.1 \dot M_{\rm edd}$. We take these parameters as input to the analytical model. Various ``effective’’ $\alpha$ parameters are reported by \cite{Zhu12} and we take their values ($\alpha = 0.1$ for $a = 0.7$ and $a = 0.9$, $\alpha = 0.01$ for $a=0$) for simplicity \citep[][their figure 6]{Zhu12}. The one free parameter of the model is then the ISCO parameters stress $\delta_{\cal J}$. We overplot in Figure \ref{fig:temp-comp} the vanishing ISCO stress radiative temperature curve (red dashed curves), which are forced to zero at the ISCO (contrary to the findings of the simulation), and in blue (solid curves) the analytical model.

\section{Photon energy shifts }
The specific flux density $F_\nu$ of the disc radiation, as observed by a distant observer at rest (subscript ${\rm ob}$), is given by
\begin{equation}
F_{\nu}(\nu_{\rm ob}) = \iint_{\cal S} I_\nu (\nu_{\rm ob}) \, \text{d}\Theta_{{{\rm ob}}} .
\end{equation} 
Here, $\nu_{\rm ob}$ is the photon frequency and $I_\nu(\nu_{\rm ob})$ the specific intensity,  both measured at the location of the distant observer.  We denote by ${\cal S}$ the limits of this integral, which is to be taken over the entire two dimensional disc surface.  The differential element of solid angle subtended on the observer's sky by the disk element is denoted $\text{d}\Theta_{{{\rm ob}}}$. 
Since $I_\nu / \nu ^3$ is a relativistic invariant \citep[e.g.,][]{MTW}, we may write
\begin{equation}
F_{\nu}(\nu_{\rm ob}) = \iint_{\cal S} f_\gamma^3 I_\nu (\nu_{\rm em}) \, \text{d}\Theta_{{{\rm ob}}},
\end{equation} 
where we define the photon energy shift factor $f_\gamma$ as the ratio of $\nu_{\rm ob}$ to the emitted local rest frame frequency $\nu_{\rm em}$:
\begin{equation}\label{redshift}
f_\gamma(r,\phi) \equiv \frac{\nu_{{\rm ob}}}{\nu_{\rm em}} = {p_\mu U^\mu\ ({\rm Ob})\over p_\lambda U^\lambda\ ({\rm Em})}=\left[ U^0 + \frac{p_{\phi}}{p_0} U^\phi + {p_r \over p_0} U^r \right]^{-1} ,
\end{equation}
where (Ob) and (Em) refer to observer and emitter, respectively.   The covariant quantities $p_r, p_\phi$ and $-p_0$ (on the far right) correspond to the radial momentum, angular momentum and energy of the {\em emitted photon} in the local (disc) rest frame.   Note that $p_\phi$ and $-p_0$ may be conveniently regarded as constants of the motion for a photon propagating through the Kerr metric.    

We will return to the properties of emitted spectrum  in the disc rest frame ($I_\nu(\nu_{\rm em})$) in a later sub-section, for now we focus our attention on the photon energy shifts $f_\gamma$. 

\begin{figure}
\includegraphics[width=0.45\textwidth]{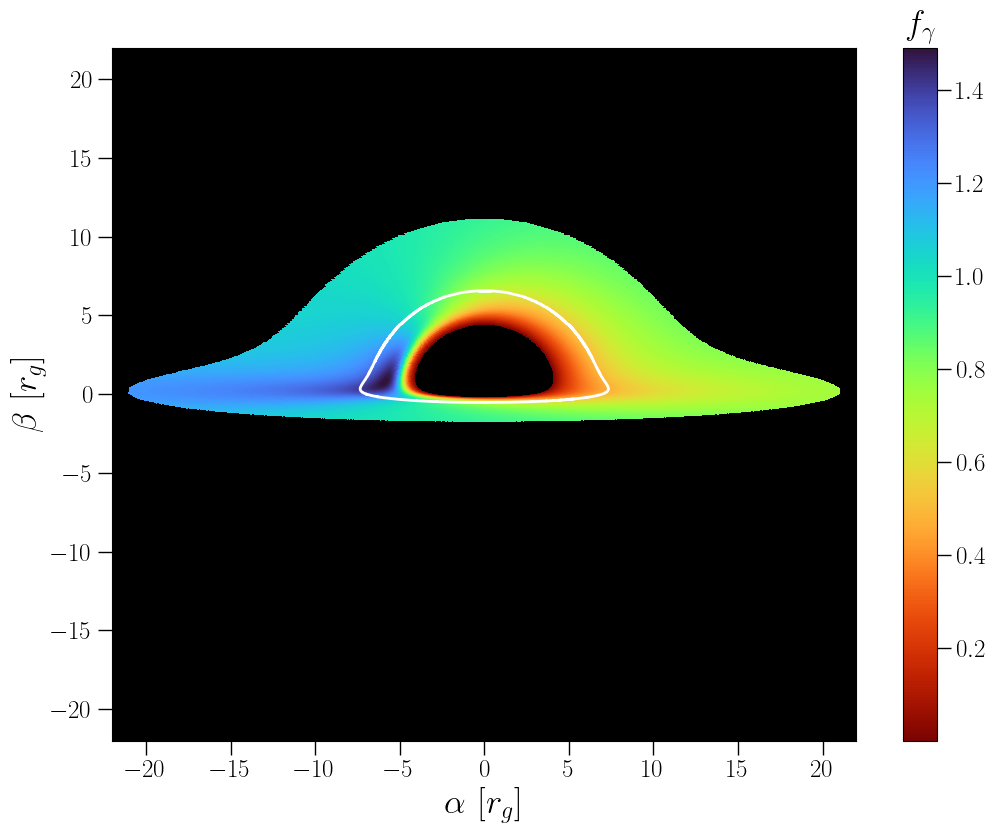}
\includegraphics[width=0.45\textwidth]{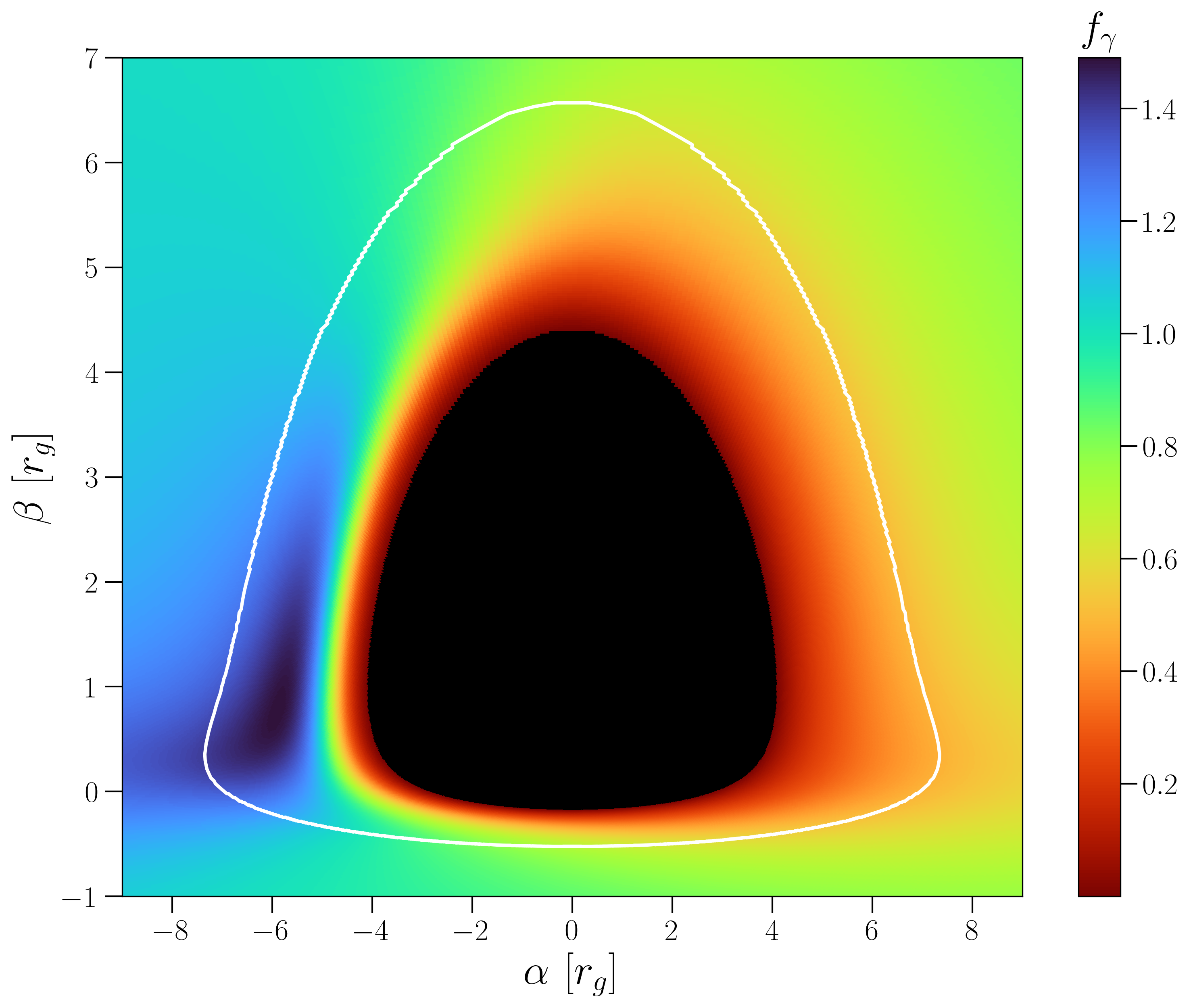}
\caption{Upper: An image plane view of the photon energy shift factor for a disc system about a Schwarzschild black hole $a = 0$, observed at $\theta_{\rm obs} = 85^\circ$. This figure also highlights the pronounced gravitational lensing of the near-ISCO region. The colour bar denotes the energy shift factor of the photons emitted from a given disc region and observed in the image plane at $(\alpha, \beta)$.  The white curve correspond to the image plane location of the ISCO. The observed photon energy-shift actually peaks within the ISCO, for these system parameters. Lower: A zoom-in of the region within the ISCO. }
\label{85}
\end{figure}

\begin{figure}
\includegraphics[width=0.5\textwidth]{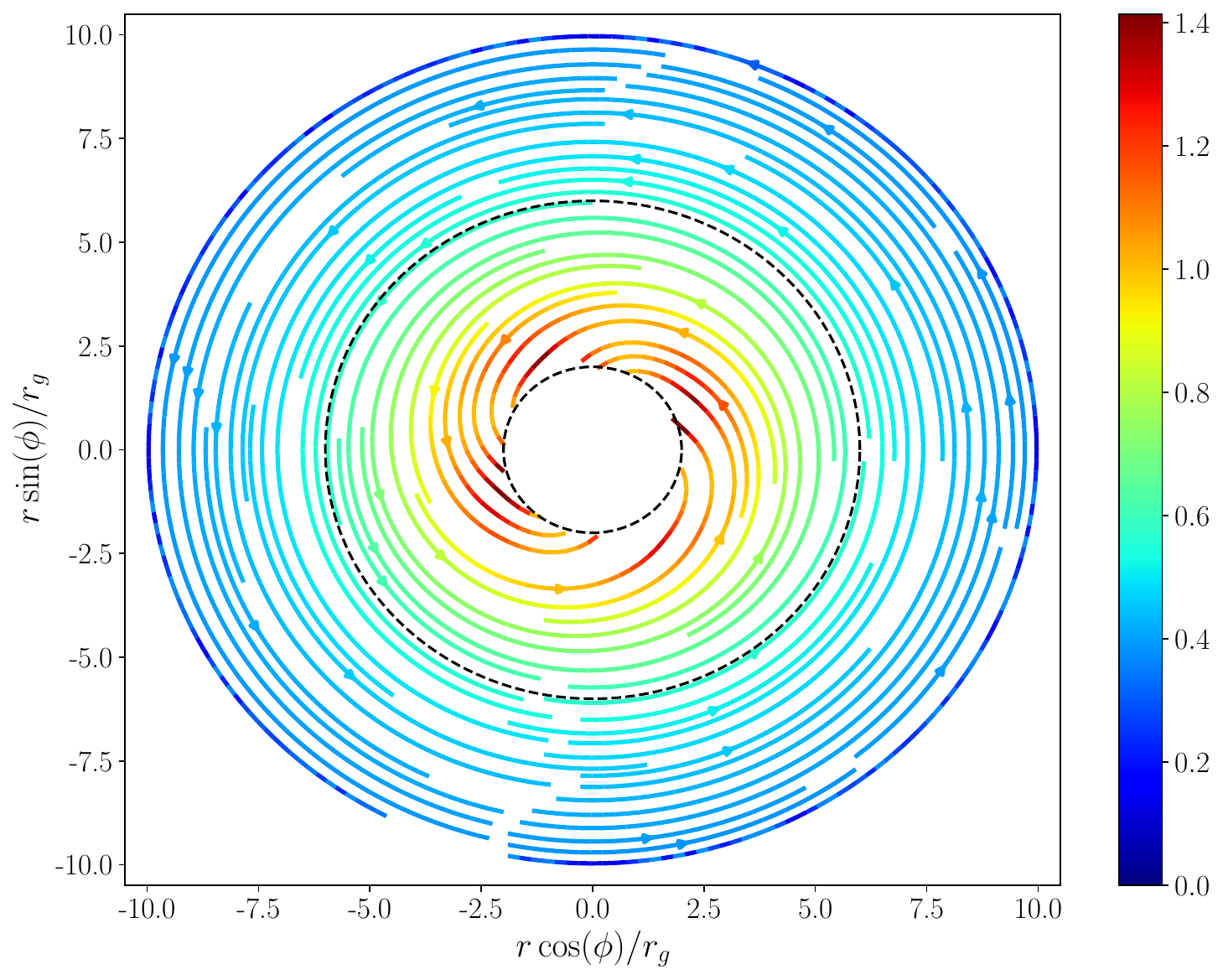}
\caption{Quasi-cartesian velocity streamlines of the disc fluid both within and outside of the ISCO. For this figure we take $a = 0$, and denote by a black dashed line the location of the ISCO. The colour bar displays the value of $\beta$, where $\beta^2 \equiv {\beta_X^2 + \beta_Y^2}$, where $\beta_{X}$ and $\beta_Y$ are defined in the text. Note that $\beta > 1$ for some radii, this is not unphysical, as neither $X$ nor $Y$ correspond  to physical proper distances. The velocity field of the disc fluid is still dominated by predominantly azimuthal motion, even within the ISCO.  }
\label{vel_strm}
\end{figure}

\begin{figure}
\includegraphics[width=0.5\textwidth]{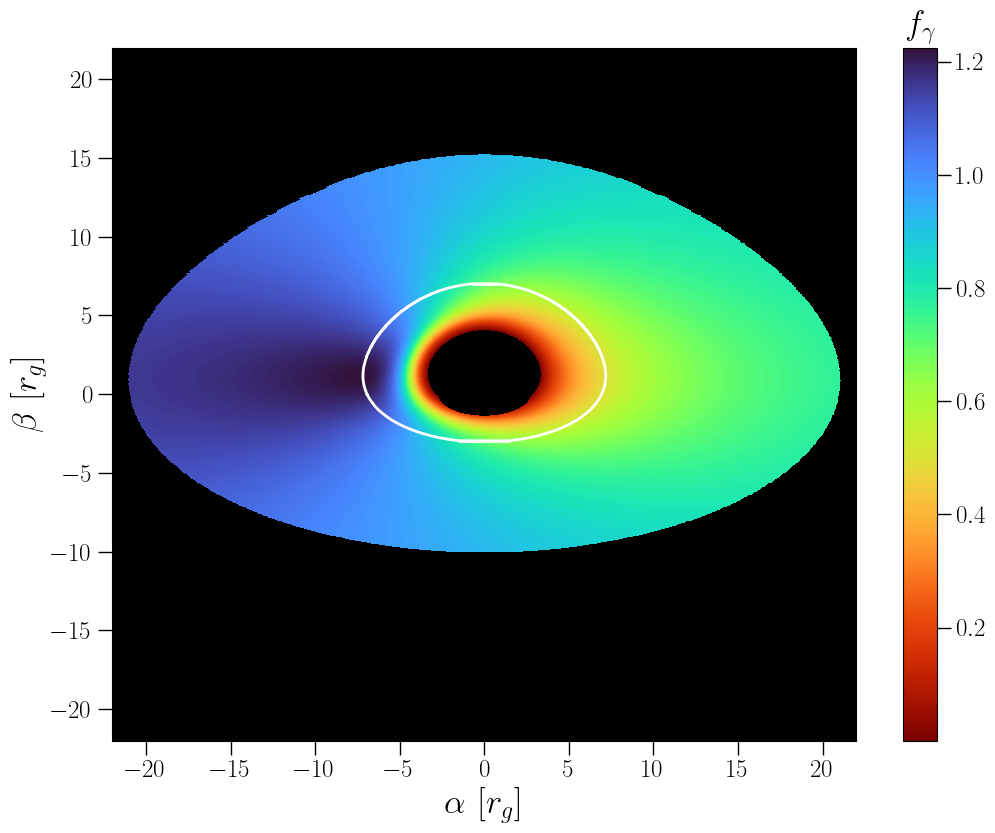}
\includegraphics[width=0.5\textwidth]{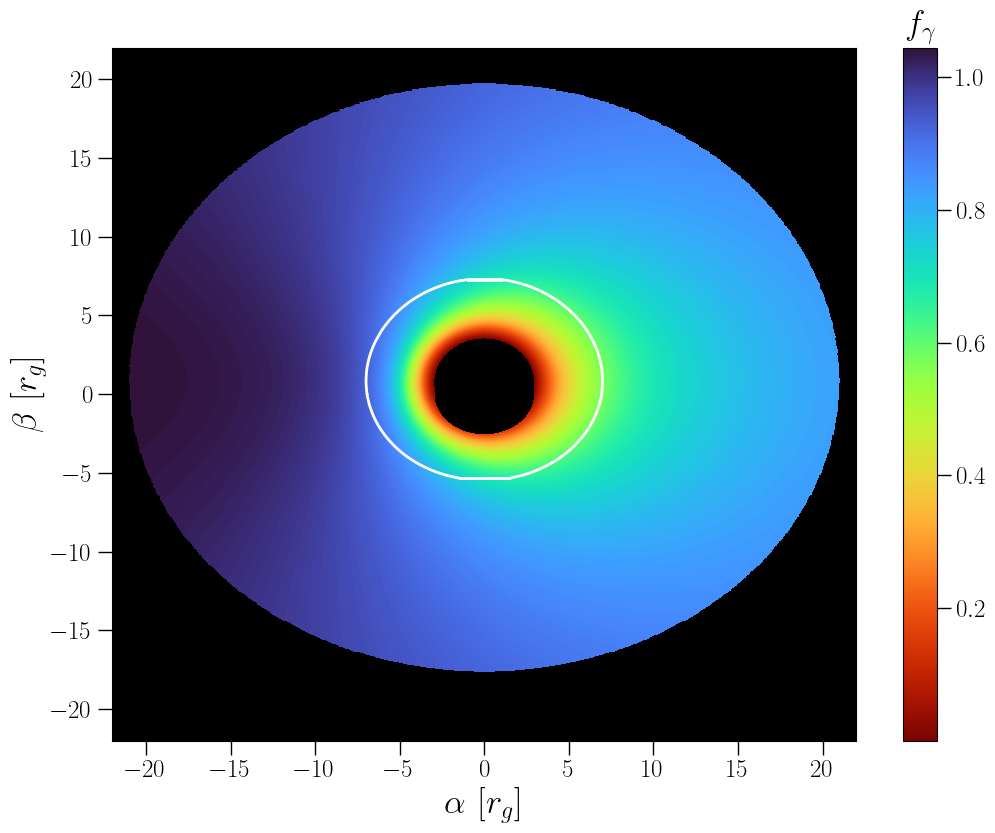}
\caption{An image plane view of the photon energy shift factor for a black hole disc system about a Schwarzschild black hole $a = 0$, observed at $\theta_{\rm obs} = 60^\circ$ (upper) and $\theta_{\rm obs} = 30^\circ$ (lower).  The colour bar denotes the energy shift factor of the photons emitted from a given disc region and observed in the image plane at $(\alpha, \beta)$.  The white curve correspond to the image plane location of the ISCO.  }
\label{60}
\end{figure}

The 4-velocity components of the disc fluid $U^0, U^\phi, U^r$ depend only on radius. Outside of the ISCO, the fluid moves on approximately circular orbits, with 4-velocity components  
\begin{align}
U^r &= 0,  \\
U^\phi &=  \frac{\sqrt{{GM}/{r^3}}}    {\left( 1 - {3r_g}/{r} + 2a\sqrt{{r_g}/{r^3} } \right)^{1/2}}, \\
U^0 &= \frac{1+a\sqrt{{r_g}/{r^3}}}{\left({1 - {3r_g}/{r} + 2a\sqrt{{r_g}/{r^3} } }\right)^{1/2}} ,
\end{align}
where $r_g \equiv GM/c^2$ is the black hole's gravitational radius, $M$ is the black hole mass and $a$ is the black hole spin parameter (with dimensions of length). 
Within the ISCO the fluid is assumed to spiral inwards while conserving its angular momentum and energy, which results in \citep{MummeryBalbus22PRL} 
\begin{align}
U^r &= - c \sqrt{2r_g \over 3 r_I} \left( {r_I \over r} - 1\right)^{3/2} , \label{ur} \\
U^\phi &= {2r_g\gamma a c + J(r -2r_g) \over r(r^2 - 2r_gr + a^2)} , \\
U^0 &=  {\gamma c (r^3 + ra^2 + 2r_ga^2) - 2J r_g a \over rc(r^2 - 2r_gr + a^2)} ,
\end{align}
where we denote by $r_I$ the ISCO radius. 
In these final two expressions the constants of motion $J$ and $ \gamma$ represent the specific angular momentum and energy of the ISCO circular orbit respectively, and  are equal to \citep{MummeryBalbus22PRL}
\begin{align}
J &= 2\sqrt{3} r_g c \left( 1 - {2a\over 3 \sqrt{r_g r_I}}\right), \\
\gamma &= \sqrt{1 - {2r_g\over 3 r_I}}.
\end{align}

Starting from a finely spaced grid of points described by cartesian impact parameters $(\alpha, \beta)$ in the image plane, we trace the geodesics of each photon back to the disc plane, recording $(r_f, p_r, p_\phi, p_0)$ for each photon. We use the code \verb|YNOGK| \citep{YangWang13}, which is based on \verb|GEOKERR| \citep{DexterAgol09}, to numerically solve for the photon geodesics \citep[see][for a more detailed description of the grid used]{Ingram19}.   The parameter $r_f$ is the radius at which the photon crosses the $\theta = \pi/2$ disc equatorial plane. Together these parameters uniquely determine the energy-shift factor $f_\gamma$. The differential element of solid angle subtended by the disc surface is simply ${\rm d}\Theta_{\rm ob} = {\rm d}\alpha {\rm d}\beta /D^2$, for an observer at a distance $D$ from the disc.

As an interesting first example, we plot in Fig. \ref{85} the energy shift factor $f_\gamma$ as observed in the image plane, $f_\gamma(\alpha, \beta)$. This figure was produced for a Schwarzschild black hole $a = 0$, observed at $\theta_{\rm obs} = 85^\circ$. Note also the pronounced gravitational lensing of the near-ISCO region. The colour bar denotes the energy shift while the white curve correspond to photons emitted from precisely the ISCO radius. An extremely interesting, and potentially surprising,  result is that the observed photon energy-shift actually peaks within the ISCO, for these system parameters (see the lower panel of Fig. \ref{85} for a zoom-in of the intra-ISCO region).  This result is simple to understand qualitatively, as the fluid flow within the ISCO retains a strong  azimuthal character, and is not moving on a purely ``radial'' plunge. 

To understand this further, in Figure \ref{vel_strm} we plot quasi-cartesian  velocity streamlines of the disc fluid both within and outside of the ISCO. We define quasi-cartesian coordinates 
\begin{equation}
X = r\cos(\phi), \quad Y = r \sin(\phi), 
\end{equation}
with corresponding dimensionless velocities 
\begin{align}
\beta_X &= {1\over c} {{\rm d} X \over {\rm d} \tau } = {1\over c} \left( U^r \cos(\phi) - r U^\phi \sin(\phi)\right) , \\
\beta_Y &= {1\over c} {{\rm d} Y \over {\rm d} \tau } = {1\over c} \left( U^r \sin(\phi) + r U^\phi \cos(\phi)\right) .
\end{align}
These velocities define streamlines of the flow.  It is clear to see in Fig. \ref{vel_strm} that the fluid flow retains a strong azimuthal character, even within the ISCO (denoted by a black dashed curve). The magnitude of the velocity itself increases within the ISCO, such that large Doppler boosting  can overcome the increasing gravitational redshift of the near horizon spacetime. This leads to an increase in the photon energy shifts within the ISCO for inclined disc systems (e.g., those shown in Figs. \ref{85}). 

The relative importance of the different effects of Doppler and gravitational energy shifting depend predominantly on the inclination of the disc system. In Fig. \ref{60} we plot Schwarzschild photon energy shift maps for inclinations of $60^\circ$ (upper panel) and $30^\circ$ (lower panel). Photons emitted from within the ISCO may still be blue-shifted for more moderately inclined systems (upper panel, Fig. \ref{60}), but more face-on discs are dominated by the gravitational red-shift, and photons emitted from within the ISCO reach the observer at lower energies (lower panel, Fig. \ref{60}).

Photon energy shift maps for the Kerr metric with general spin parameter $a$ all show qualitatively similar behaviour to the Schwarzschild metric displayed explicitly here, and are not shown explicitly. 

{Finally, we note that we have set up the calculation here in such a way that only those photons which escape to infinity contribute to the observed flux. This is important as within the ISCO  fluid elements pick up a large velocity component in the direction of the black hole's event horizon. A sizeable fraction of the photons emitted by  a fluid element which is emitting isotropically in it's own rest frame will be beamed into the black hole in the ``lab'' frame. While this beaming is taken into account in our calculations, we do not explicitly calculate the ratio of captured-to-escaped photons.  We direct the reader interested in this effect to \cite{AgolKrolik00}, who calculated the captured-photon fraction for discs with a finite ISCO stress (but no intra-ISCO emission).   }

\section{Photon starvation: radiative transfer inside the ISCO}
{The disc fluid within the ISCO remains optically thick to scattering right down to the event horizon of the black hole, a result which holds provided that the Eddington-normalised accretion rate of the disc is moderate. To see this, combine the definition of the scattering optical depth $\tau_{\rm scat} = \kappa_{\rm es} \Sigma$, with the constraints of mass conservation $\dot M = 2\pi r U^r \Sigma$, and the rapid radial acceleration of the intra-ISCO region, which leads to a radial velocity $U^r = c \sqrt{2r_g  (r_I - r_+)^3 / 3 r_I r_+^3}$ as the fluid approaches the event horizon $r\to r_+$ (see equation \ref{ur}). The minimum scattering optical depth is therefore }
\begin{equation}
    \tau_{\rm scat, min} = {\kappa_{\rm es} \dot M c \over 2\pi  G M} \sqrt{ 3 r_I r_+ r_g \over 2 (r_I - r_+)^3 } , 
\end{equation}
{and upon substituting for $\dot M = \dot m \dot M_{\rm edd} = \dot m L_{\rm edd}/(\eta c^2)$, with Eddington luminosity $L_{\rm edd} = 4\pi G M c / \kappa_{\rm es}$, leaves the simplified result ($\dot m$ is the Eddington-normalised accretion rate, $\eta$ is the mass-to-light accretion efficiency) }
\begin{equation}\label{scattmin}
    \tau_{\rm scat, min} = {\dot m \over \eta }  \sqrt{ 6 r_I r_+ r_g \over (r_I - r_+)^3 } . 
\end{equation}
{The remaining factors are all order unity for low spins, and we see that we have optically thick (to scattering) fluid down to the event horizon provided $\dot m \gtrsim \eta$. This result highlights the fact that while there is a substantial drop in the density across the plunging region, the density required to maintain a substantial Eddington ratio is sufficiently high that this drop does not lead to optically thin (to scattering) material in the plunging region.  }

{Photons travelling through a material which is optically thick to scattering will in general thermalise their energy. However, } the energy liberated from the disc fluid can only be carried out as purely thermal photon emission as long as there are a sufficient number of photons to carry away the total energy budget with the energy distribution prescribed by the blackbody spectral profile.  If there are an insufficient total number of photons available, each emitted photon will have to carry a (potentially significantly) larger energy, so that the total energy budget can be radiated away. The number density of photons required to carry away the energy budget of a blackbody at temperature $T$ is $n_\gamma \propto T^3$ and therefore, for the nearly isothermal effective temperature profiles we find for the intra-ISCO region (Fig. \ref{fig:temp-comp}), this corresponds to a roughly constant requirement of photon number density throughout the plunging region. 

However, while intra-ISCO flows require a near constant number density of photons, the availability of photons rapidly decreases across the ISCO. This is because photons are produced by free-free processes, the emissivity (energy produced per unit volume per unit time) of which ($\eta_{\rm ff}$) depends on the {\it square} of the disc density 
\begin{equation}
    \eta_{\rm ff} = \eta_0 \, m_p^{-2}\,  T_c^{1/2} \rho^2 , 
\end{equation}
where $\eta_0 \simeq 1.4 \times 10^{-27}$ erg/s/cm$^3$ caries the dimensions of $\eta_{\rm ff}$, $m_p$ is the protons mass and $T_c$ and $\rho$ are the disc fluids central temperature and density respectively (and should take dimensionless values appropriate for the cgs units system). The density of the disc fluid $\rho \simeq \Sigma/H$ drops rapidly across the intra-ISCO region, owing to the constraint of a constant radial mass flux $\dot M = 2\pi r \Sigma U^r$, coupled with a rapidly growing radial velocity $U^r$. Therefore as the disc fluid is accelerated over the ISCO it quickly loses the ability to produce the required number of photons to maintain a thermal emission profile. A fluid which is unable to produce sufficient photons to radiate away all of its  energy budget as a thermal spectrum is said to be ``photon starved''.   Intra-ISCO emission is photon-starved, and will therefore be characterised by extreme ``spectral hardening''.

\begin{figure}
    \centering
    \includegraphics[width=\linewidth]{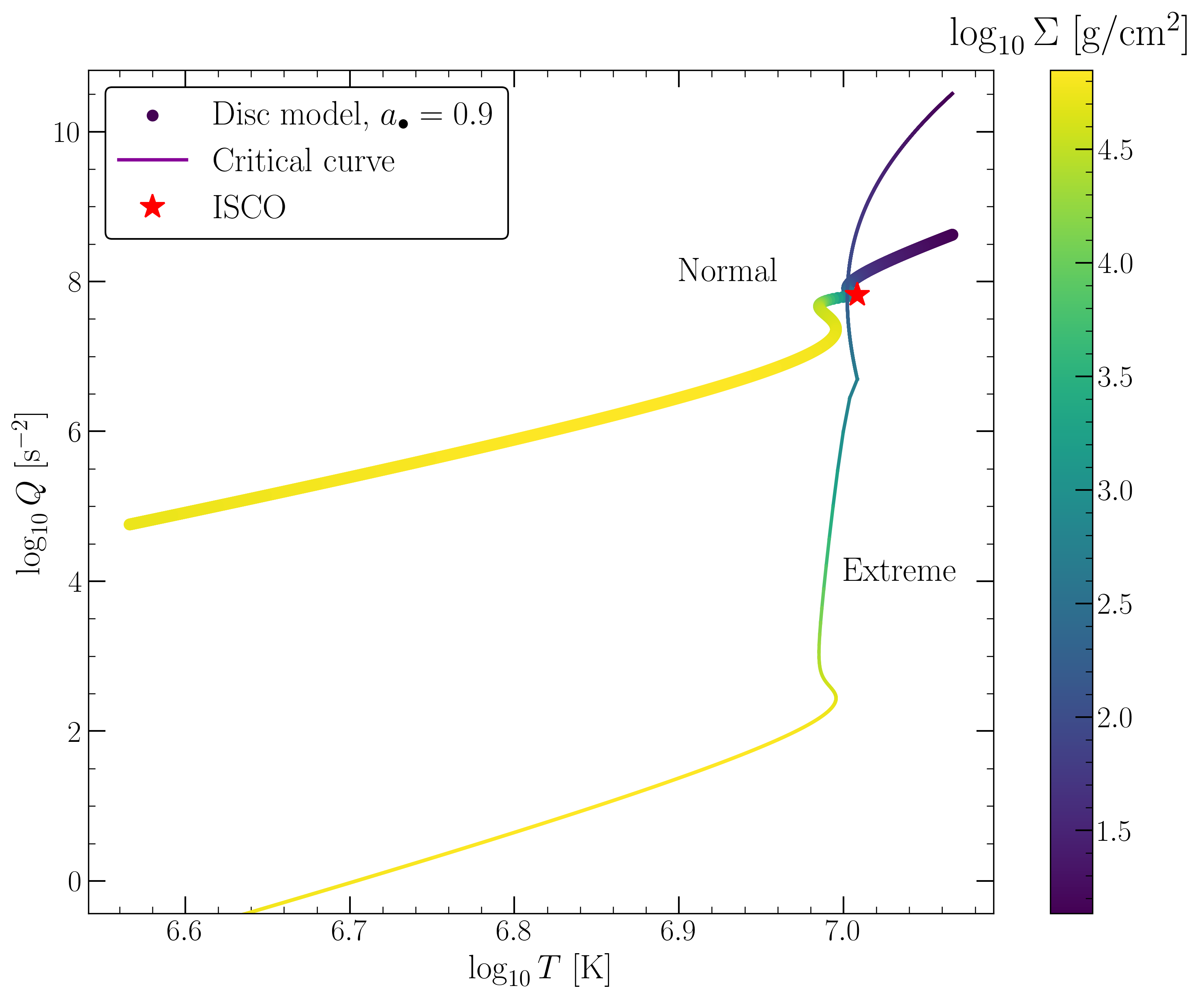}
    \includegraphics[width=\linewidth]{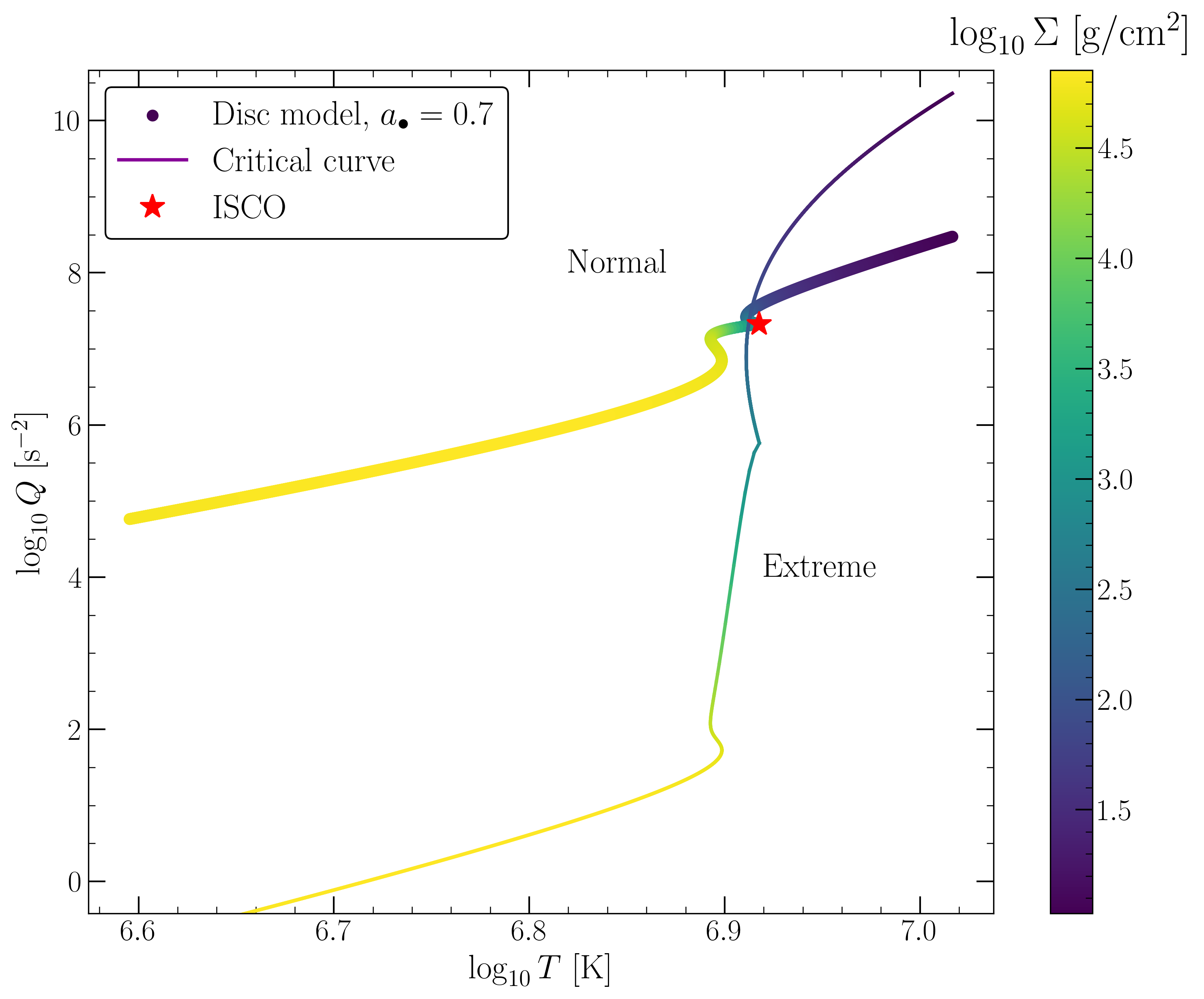}
    \caption{Phase space ($Q$, $T$, $\Sigma$) plots of the thin disc solutions displayed in Fig. \ref{fig:temp-comp} (thick curve), and the ``critical curve'' of extreme spectral hardening described in {\citealt{Davis19}} (thin curve). Points above the critical curve (at the same surface density, displayed by the colourbar) will display normal spectral hardening, while points below the curve will show extreme hardening. The transition point of both solutions is very close to the ISCO (red star). Intra-ISCO fluid can be identified by smaller surface densities, and larger vertical gravity.  }
    \label{fig:spec-hard}
\end{figure}

It is possible to derive a critical curve in disc parameter space which demarcates the approximate transition from ``normal'' to ``extreme'' spectral hardening. The flux (energy per unit area per unit time) of photons produced by free-free processes across the disc atmosphere is approximately $\eta_{\rm ff} H$, and if this becomes lower than the flux required to be carried away for thermal emission, or explicitly when 
\begin{equation}
    \eta_{\rm ff} H \simeq \sigma T^4,
\end{equation}
 there will be an insufficient photon number density in the flow and each photon must carry a  larger energy to radiate away the disc energy budget\footnote{We assume here that during the time taken for photons to traverse the disc atmosphere $H$ a negligible amount of energy is advected into the black hole. This is a reasonable assumption for thin discs as the vertical diffusion time $\Delta t \sim H \tau / c \sim H \kappa \dot M /r U^r$ shrinks by the same factor with which the fluid's radial velocity grows, and $H$ is small compared to $r$.}. Using standard disc scaling laws \cite{Davis19} showed that extreme spectral hardening occurs for disc systems which cross the critical curve 
\begin{equation}\label{critcurve}
    \log_{10} Q_{\rm crit} = \log_{10} Q_0 + 7.5 \log_{10} T - 2.125 \log_{10} \Sigma, 
\end{equation}
where all discs with $Q \lesssim Q{\rm crit}$ will show extreme hardening. The parameter $Q_0$ depends only on fundamental constants 
\begin{equation}
    Q_0 \equiv {m_p^2 \, \kappa_{\rm es}^{7/8} \, \sigma^2 \over \eta_0 \, c} ,
\end{equation}
and $Q$ is the  vertical gravity of the black hole  experienced by the disc fluid which is defined through the equation of vertical hydrostatic equilibrium\footnote{This is somewhat of a misnomer, as $Q$ depends on the four-momentum  of the orbiting disc fluid as well as the properties of the central black hole.} 
\begin{equation}
    -{{\rm d}P \over {\rm d}z} = \rho Q z, 
\end{equation}
where $P$ and $\rho$ are the total pressure and density of the fluid respectively, and $z$ is the vertical coordinate. In the Kerr metric $Q$ is given by \citep{Abramowicz97} 
\begin{equation}
    Q = {GM_\bullet \over r^3 } {\cal R}_z(r, a_\bullet) ,
\end{equation}
where ${\cal R}_z$ is a (dimensionless) relativistic correction term, listed in Appendix \ref{appA}.

For typical disc parameters the transition to extreme spectral hardening occurs almost precisely at the ISCO radius. This can be seen in Fig. \ref{fig:spec-hard}, where we plot the vertical gravity $Q$ against the effective temperature of the disc, for the analytical disc solutions which reproduce the \cite{Zhu12} GRMHD simulation results (i.e., those profiles displayed in Fig. \ref{fig:temp-comp}). We also display the \cite{Davis19} critical curve (i.e., equation \ref{critcurve}), with $T$ and $\Sigma$ computed at each radius in the disc. Regions of the disc (at fixed surface density, denoted by the colour of the plot) which are above the critical curve display ``normal'' spectral hardening, while regions of the disc below the critical curve will show extreme spectral hardening. As is clear to see in Fig. \ref{fig:spec-hard}, the interception point between the disc profile and the critical curve is almost exactly the ISCO radius (denoted by a red star in Fig. \ref{fig:spec-hard}). This means that the emission sourced from within the ISCO will be observed to be significantly hardened compared to the emission sourced from the main body of the disc. 

\begin{figure}
    \centering
    \includegraphics[width=\linewidth]{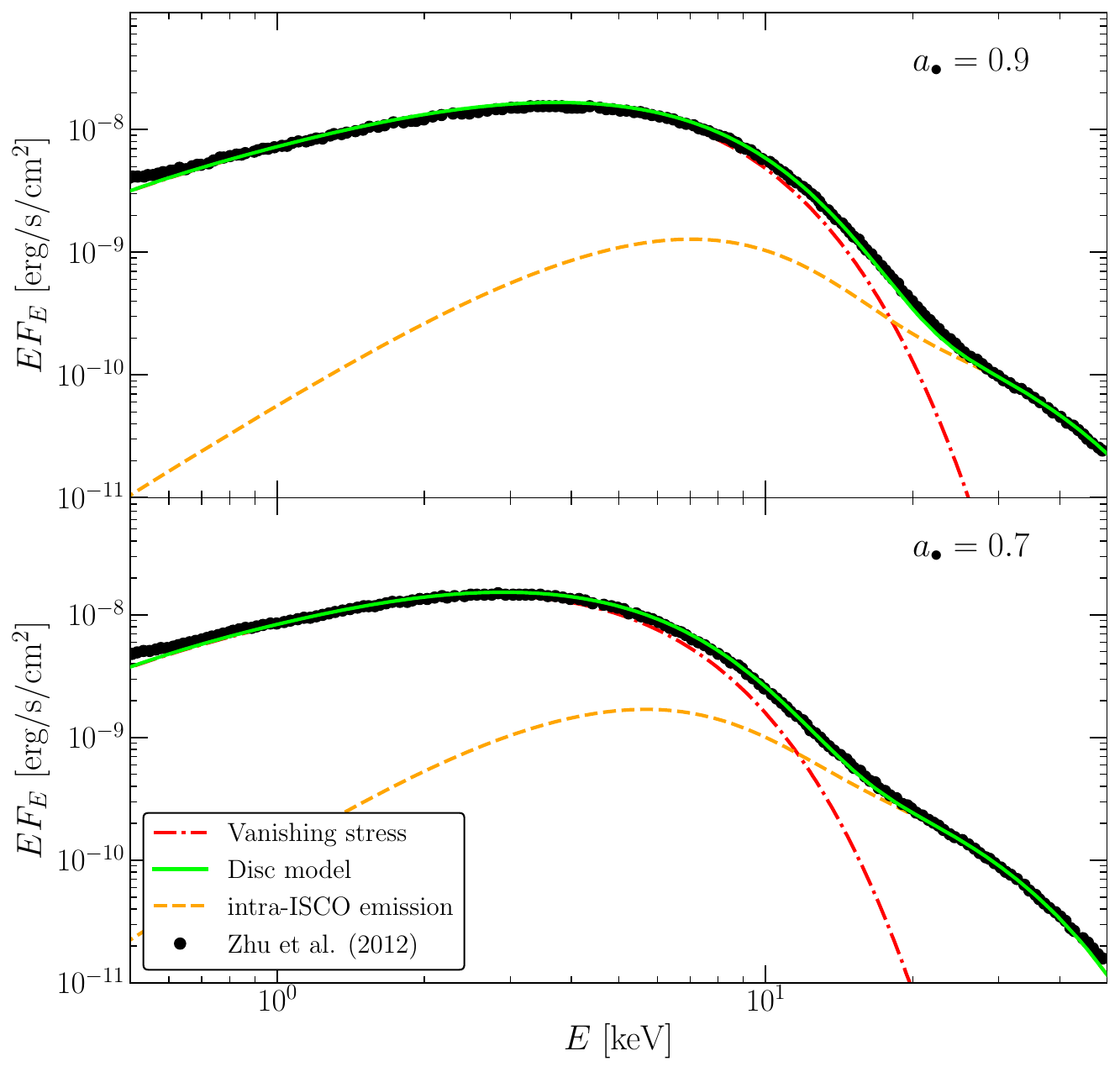}
    \caption{The X-ray spectra computed from {\tt TLUSTY} post-processing of GRMHD simulations \citep[black points]{Zhu12}, compared to the colour-corrected analytical models developed here (green solid curve). The effective temperature profiles are those displayed in Fig. \ref{fig:temp-comp}, and the colour-correction parameters are $f_I = 4, \xi = 6$ ($a_\bullet = 0.9$), and $f_I = 3, \xi = 4$ ($a_\bullet = 0.7$).  }
    \label{fig:spec-comp}
\end{figure}

This behaviour was observed in \cite{Zhu12}, where the local {\tt TLUSTY} spectra \citep[e.g.,][]{Hubeny95, Davis06, DavisHubeny06} sourced from the intra-ISCO region (i.e., spectra post processed from the \citealt{Penna10} GRMHD simulations) peaked (in the source rest frame) at energies up to $E_\gamma \sim 100$ keV, despite the effective temperature of the flow being at significantly lower energies $kT \sim 1-10$ keV (see Fig. \ref{fig:spec-comp}).

Spectral hardening in the continuum fitting community is normally modelled with a single spectral hardening factor $f_{\rm col} \geq 1$, which modifies the locally emitted spectrum from thermal in the following simple manner 
\begin{equation}
    I_\nu(\nu_{\rm em}) = \left({1 \over f_{\rm col}}\right)^4 B_\nu(\nu_{\rm em}, f_{\rm col} T ) ,
\end{equation}
where $B_\nu$ is the usual Planck function. 
Note that the normalisation factor of $f_{\rm col}^{-4}$ means that the total luminosity of the emitted photons is unchanged by spectral hardening (and therefore total energy is conserved), but that typical photons are emitted with a larger energy $h\nu \sim f_{\rm col} k T \geq kT$. 

It is not immediately clear that the complex physics of photon starvation (captured by the full {\tt TLUSTY} calculations) will be in any way well described by a simple colour-corrected blackbody model.  We do however have a well defined test case to probe this physics, as the \cite{Zhu12} spectra are generated from effective temperature profiles which we know are well described by the analytical disc models of \cite{MummeryBalbus2023}. Therefore, if we are able to reproduce in detail the observed spectra of \cite{Zhu12}, then the simple colour-corrected emission model must be capturing the full physics of the {\tt TLUSTY} radiative transfer calculations. 

We experimented numerically with a number of colour-correction prescriptions for the photon starved intra-ISCO region. A single constant $f_{\rm col} = f_I \gg 1$ did a reasonable job at reproducing the \cite{Zhu12} disc spectra, but the following functional form 
\begin{equation}
    f_{\rm col} = f_I \left({r\over r_I}\right)^{-\xi} , \quad r \leq r_I,
\end{equation}
where $f_I \geq 1$ and $\xi\geq 0$ are constant fitting parameters, did an excellent job. {We will henceforth refer to $\xi$ as the ``photon starvation parameter''.} We display in Fig. \ref{fig:spec-comp} X-ray spectra generated using the above prescription, and the disc effective temperature profiles of Fig. \ref{fig:temp-comp}, compared to the {\tt TLUSTY} spectra generated from temperature profiles extracted from GRMHD simulations, presented in \cite{Zhu12}. The simple models presented in this paper do a remarkable job at reproducing the  complex physics of post-processed GRMHD simulations. 

In Fig. \ref{fig:spec-comp} we split the individual contributions to the total X-ray spectrum (green solid curve) into disc regions exterior to the ISCO (red dot-dashed) and interior to the ISCO (orange dashed). While the bulk of the total luminosity originates from extra-ISCO disc regions, above $E_\gamma \sim 10$ keV the emission is entirely dominated by emission sourced from within the ISCO.  

Naturally, a tuneable colour-correction still represents a significant simplification of the full radiative transfer of the intra-ISCO region. In a future analysis we wish to explore the physical properties of the radiative transfer within the ISCO in significantly more detail, and it seems likely that this prescription can be improved upon.  However, existing tools such as {\tt TLUSTY} are liable to crash in the particular region of parameter space relevant for intra-ISCO accretion \citep[see e.g.,][for a discussion of this]{Zhu12, Davis19}, and developing new radiative transfer tools lies beyond the scope of this current paper. While this treatment may need to be revisited in future studies, it is clear that this modelling does capture the physics of the plunging region (Fig. \ref{fig:spec-comp}). In its current form this model offers a route to analysing intra-ISCO emission in X-ray binaries, which is a valuable addition. 

Finally, it is worth noting that {\tt TLUSTY} itself has a number of simplifying assumptions which may not be formally valid within the ISCO. Firstly, {\tt TLUSTY} assumes that the disc electrons obey a thermal distribution (with a  single temperature identical to the ion temperature). \cite{Hankla22} have recently argued that this assumption might break down, and the intra-ISCO electron population may in fact be better described by a non-thermal distribution.  In addition, {\tt TLUSTY} neglects irradiation of the disc atmosphere from above, and within the ISCO self-irradiation is likely to be important at small radii. Models developed for post processing GRMHD simulations \citep[e.g.,][]{Narayan16, Kinch21}  include these effects, and in principle a future extension of these models should include these effects. 

\section{Example X-ray spectra}
Figure \ref{fig:spec-comp} highlights the main observational effect of including intra-ISCO emission, namely the appearance of an additional hot and small (compared to the main body of the disc) quasi-blackbody component and a weak and soft power-law tail extending to much higher energies. In this section we examine these properties in more detail. 

There are of course two different modifications to the disc emission which occur when a disc has a non-zero ISCO stress. The first results simply from additional flux sourced from the stable disc regions ($r>r_I$) which escapes to the observer \citep[e.g.,][]{AgolKrolik00}, while in this paper we will focus on the extra emission sourced from {\it within} the ISCO. 

\begin{figure}
    \centering
    \includegraphics[width=\linewidth]{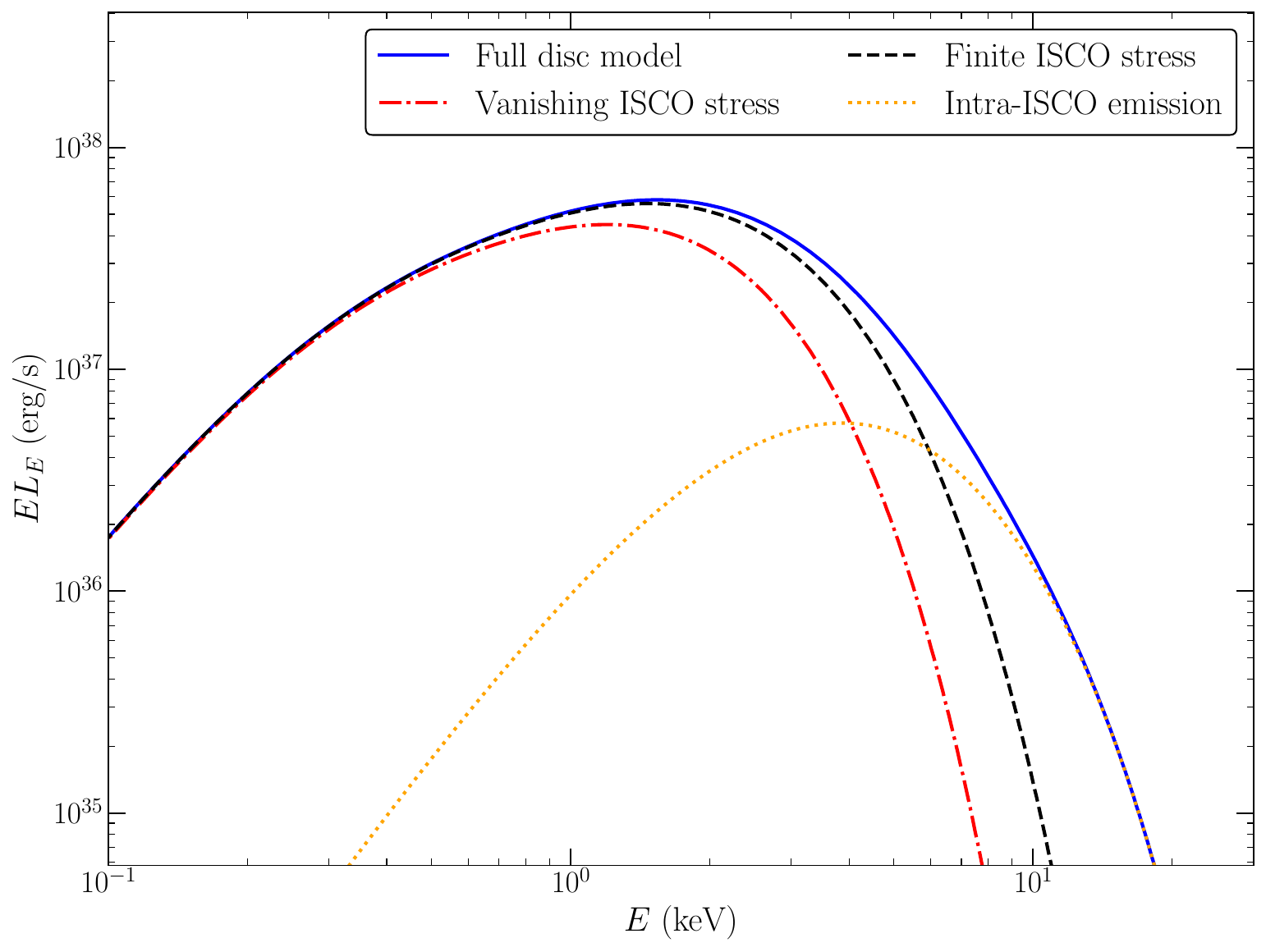}
    \caption{The different spectral effects of modifying the vanishing ISCO stress boundary condition. The extra emission from a disc with a finite ISCO stress comes in two main components. Additional emission from the main body of the disc (outside of the ISCO) simply hardens the main disc spectrum (contrast the red dot-dashed and black dashed spectra). In contrast, the intra-ISCO emission typically presents itself as an additional hot and small quasi-blackbody component (orange dashed curve). For physical parameter values see the main body of the text.  }
    \label{fig:canon}
\end{figure}

\begin{figure}
    \centering
    \includegraphics[width=\linewidth]{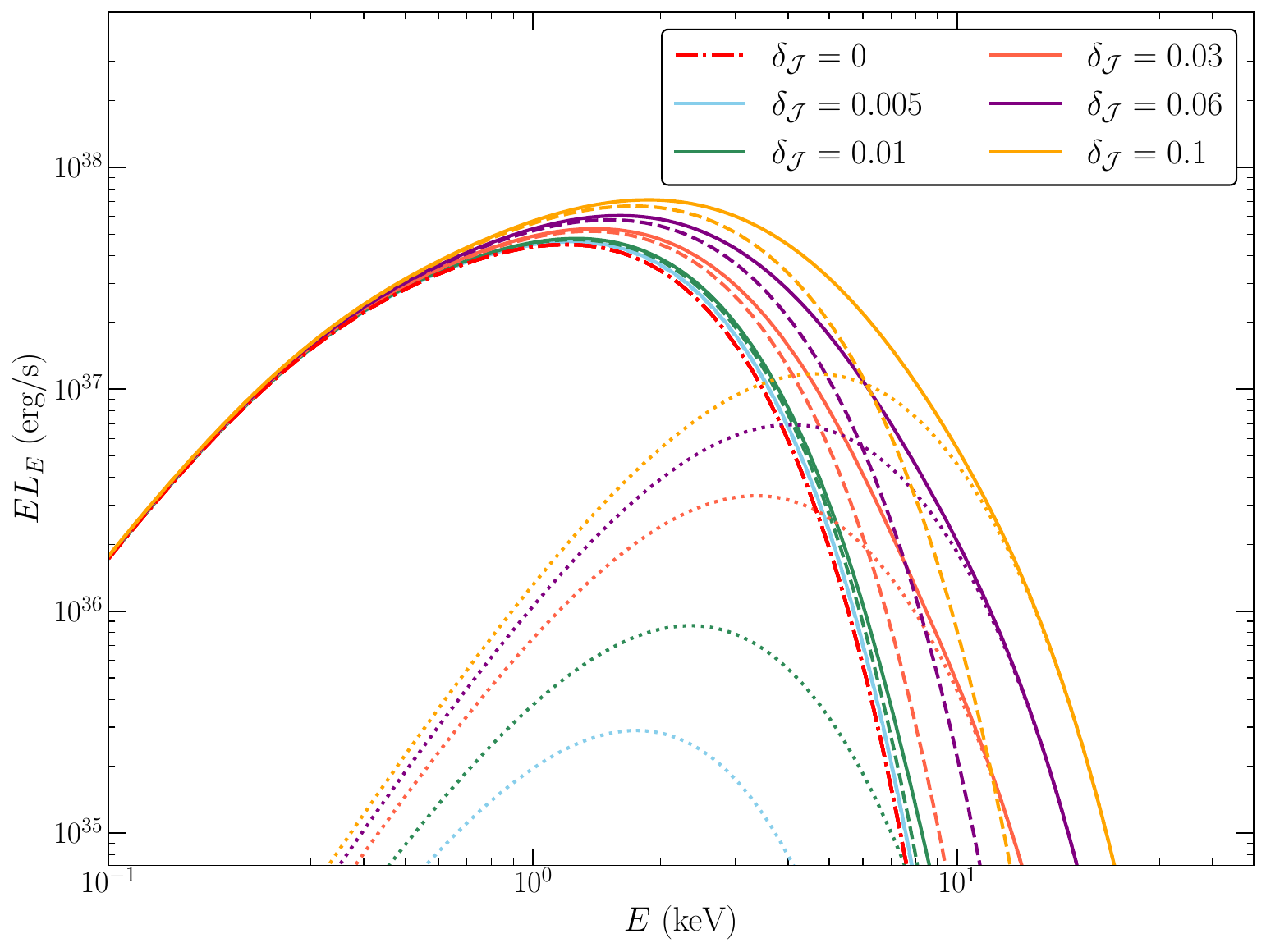}
    \caption{The effect of varying the disc's ISCO stress parameter on the emergent X-ray spectrum. As the ISCO stress is increased there is a small increase in emission from outside of the ISCO (contrast the dashed curves with the red dot-dashed vanishing ISCO stress curve), but the principle modification is the increasing flux from the intra-ISCO region. Increasing the ISCO stress dramatically increases both the ``temperature'' and ``emitting area'' of the emission within the ISCO (dotted curves). All non-displayed parameters are identical to those in Fig. \ref{fig:canon}.   }
    \label{fig:vary_dj}
\end{figure}

In Figure \ref{fig:canon} we highlight the different properties of these two additional emission components. Each X-ray spectrum is produced assuming a $M = 10M_\odot$ black hole, which is not rotating ($a_\bullet = 0$), and is fed by a constant mass accretion rate $\dot M = L_{\rm edd}/c^2$, where $L_{\rm edd}= 1.26 \times 10^{38} (M/M_\odot)$ erg/s (this corresponds to a bolometric disc luminosity of $L \sim 0.1 L_{\rm edd}$, typical for soft-state X-ray binaries). The disc is observed at an inclination of $i = 45^\circ$. For simplicity we take a colour-correction factor in the main body of the disc of $f_d = 1.7$, a value we also take for $f_I$. We take $\xi = 2$, similar to the value found from fits to MAXI J1820 discussed later. 

\begin{figure}
    \centering
    \includegraphics[width=\linewidth]{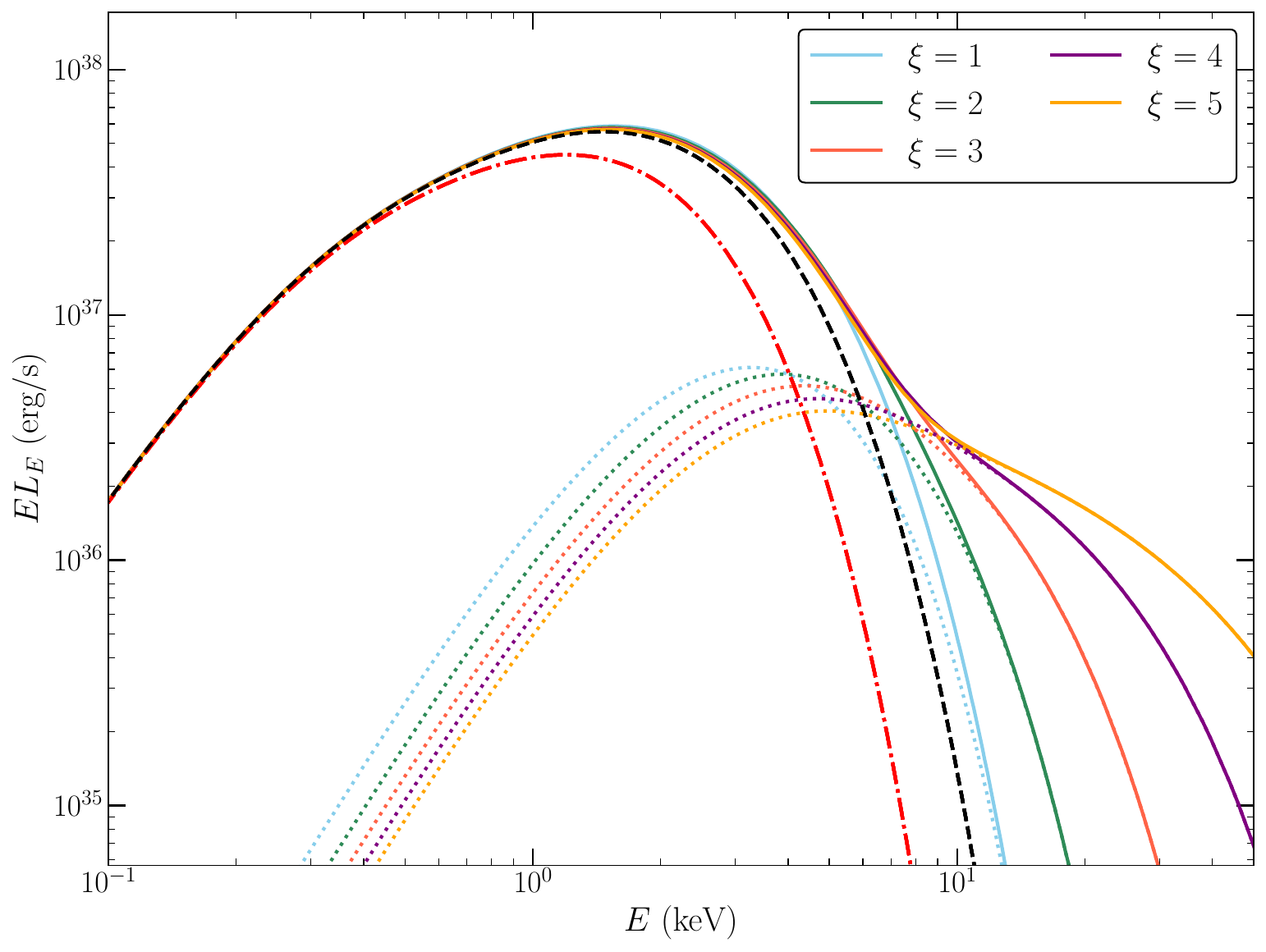}
    \caption{ The effect of varying the intra-ISCO photon starvation parameter $\xi$ on the emergent X-ray spectrum. When $\xi \lesssim 3$ the intra-ISCO emission is well described by additional quasi-blackbody emission, while for even harder emission the additional spectrum becomes more power-law like.   }
    \label{fig:vary_xi}
\end{figure}

In red (dot-dashed) we show the X-ray spectrum which would be observed if the disc had a vanishing ISCO stress ($\delta_{\cal J}= 0$). In black (dashed) we show the spectrum observed from the disc assuming a non-zero ISCO stress $\delta_{\cal J} = 0.05$, but {\it still with no emission sourced from within the ISCO}. The inclusion of this additional extra-ISCO emission simply hardens the disc spectrum slightly, and is in general degenerate with a change in $\dot M$ or source-observer distance \citep{Li05}. However, when emission within the ISCO is included (blue solid curve) we see that intra-ISCO emission has a different character to extra-ISCO emission, and in general dominates over all other emission regions for moderately high photon energies $E \sim 5$ keV. The emission sourced from within the ISCO is displayed by an orange dotted curve, and is reasonably well described by a single temperature blackbody. {The physical origin of this quasi-single-temperature appearance is chiefly related to the interplay between the rising colour-temperature of the flow (e.g., Fig. \ref{fig:temp-comp}) being counteracted by the rising gravitational redshifting of the emitted photons (e.g., Fig. \ref{60}) of this region. This leads to only a small region from within the ISCO contributing the majority of the additional flux.   }

{It is clear to see from Figure \ref{fig:canon} that a non-zero ISCO stress increases the radiative efficiency of a black hole  accretion flow, typically defined as $\eta \equiv L/\dot M c^2$. This result is well known, and in fact the ISCO stress can itself be parameterised entirely as an additional accretion efficiency $\Delta \eta$ \citep[see][who do not include emission from within the ISCO in their calculation]{AgolKrolik00}. As can be seen in Fig. \ref{fig:canon}, the extra emission from {\it within} the ISCO represents a small fraction of the total emitted luminosity  \citep[as has been argued previously, e.g.,][]{Zhu12}, but it represents a huge increase in the flux {\it at certain observed frequencies}. We will demonstrate in the following section that this  is of crucial observational significance.  }

Thermal emission sourced from within the ISCO is particularly sensitive to four free parameters, the ISCO stress $\delta_{\cal J}$, the photon starvation parameter $\xi$, the disc-observer inclination $i$ and the black hole spin $a_\bullet$. We now examine the effects of each of these parameters in turn, keeping to the same figure conventions: vanishing ISCO stress discs are shown by dot-dashed curves, finite ISCO stress discs with no intra-ISCO emission by dashed curves, the emission from within the ISCO is shown by dotted curves and the total disc flux is shown by a solid curve. All parameters (except when explicitly stated) are the same as those used in Fig. \ref{fig:canon}.

\begin{figure}
    \centering
    \includegraphics[width=\linewidth]{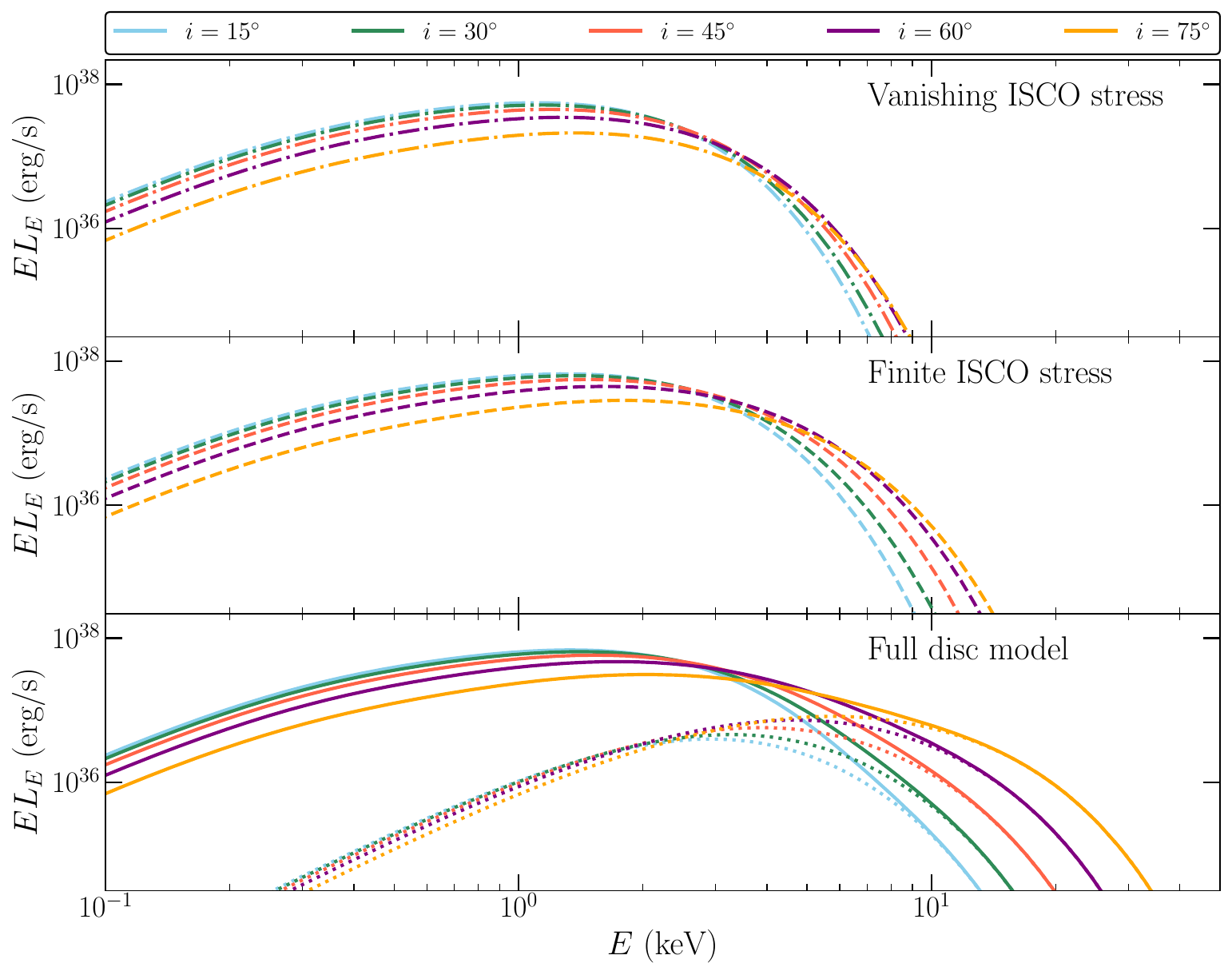}
    \caption{ The effect of varying the disc-observer inclination on the properties of the emission sourced from within the ISCO. For the Schwarzschild black holes considered in this figure, the intra-ISCO emission is always important at energies $E\gtrsim 10$ keV. The emission from within the ISCO grows with importance with increasing inclination, a result of Doppler beaming of the inner disc (see e.g., Fig. \ref{85}). The intra-ISCO disc region can produce the majority of the flux above $E \sim 5$ keV for moderate inclinations $i \gtrsim 60^\circ$. Upper panel: vanishing ISCO stress discs; middle panel: discs with a finite ISCO stress and no intra-ISCO emission; lower panel: the full disc model, with intra-ISCO emission displayed as a dotted curve.   }
    \label{fig:vary_i}
\end{figure}

\begin{figure}
    \includegraphics[width=\linewidth]{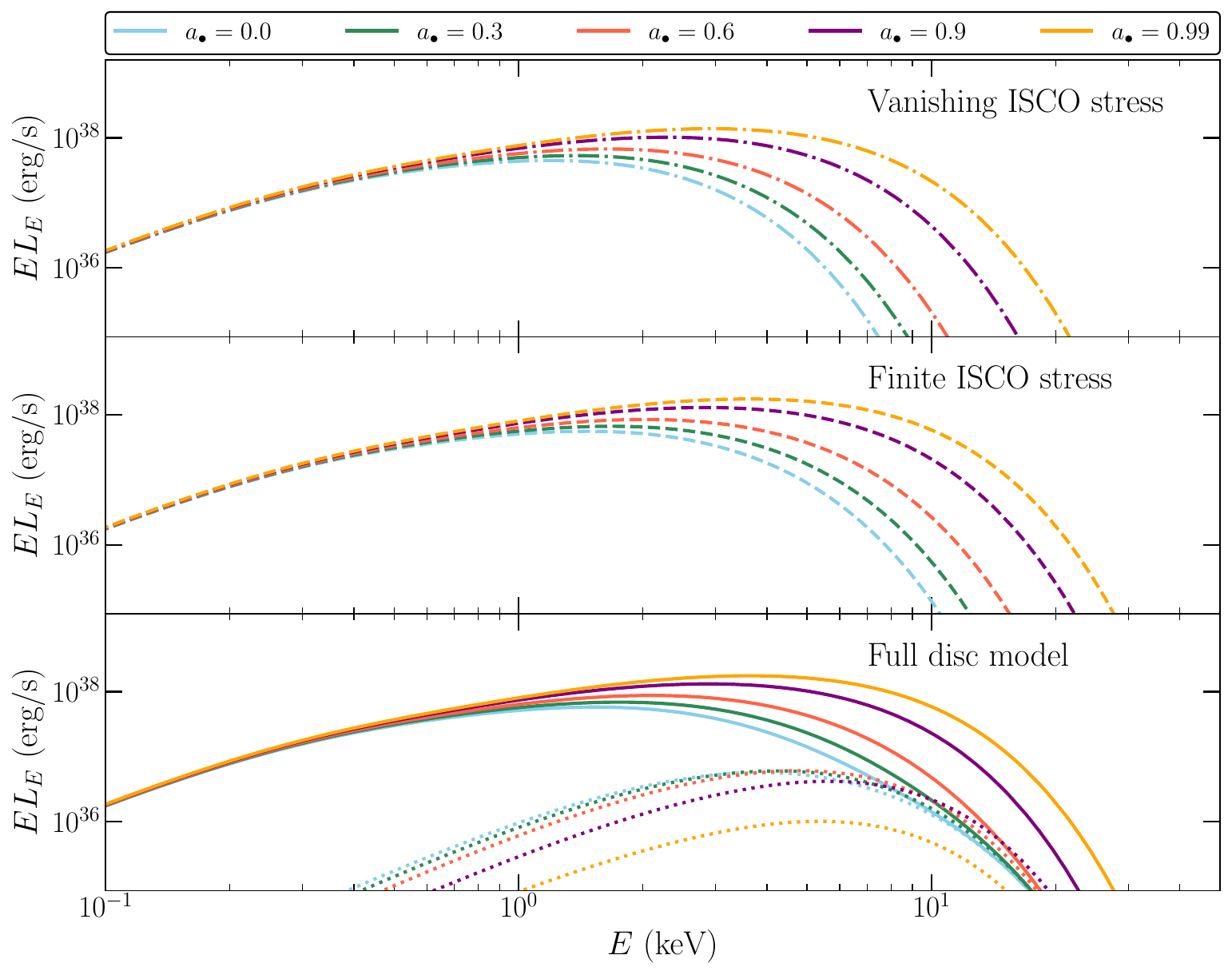}
    \caption{ The effect of varying the black hole spin parameter $a_\bullet$ on the X-ray spectrum, and the properties of the emission sourced from within the ISCO. Intra-ISCO emission is most important at lower black hole spins, a result simply of the larger event horizon-to-ISCO radial range at low spins. With intra-ISCO emission included, the differences between different spin parameters are less pronounced.
    Upper panel: vanishing ISCO stress discs; middle panel: discs with a finite ISCO stress and no intra-ISCO emission; lower panel: the full disc model, with intra-ISCO emission displayed as a dotted curve.  }
    \label{fig:vary_a}
\end{figure}

The most obvious parameter on which the intra-ISCO emission depends is the ISCO stress parameter $
\delta_{\cal J}$. In Fig. \ref{fig:vary_dj} we display the effect of varying this parameter on the X-ray spectrum observed at $i = 45^\circ$. For low ISCO stress values \citep[e.g., values much lower than is seen in GRMHD simulations][]{Noble10, Schnittman16} the X-ray spectrum looks near indistinguishable from a vanishing ISCO stress disc, and the intra-ISCO emission is that of a cool and very small blackbody (light blue curve). However, as the ISCO stress is increased the intra-ISCO emission gets both hotter and larger in amplitude (it has a larger ``emitting area'' in the language of a blackbody profile). This can be clearly seen from observing the dotted profiles (Fig. \ref{fig:vary_dj}) growing in importance and ultimately dominating the high energy emission for $\delta_{\cal J} \gtrsim 0.03$. 

It is extremely important to stress that the intra-ISCO emission is much more sensitive to $\delta_{\cal J}$ than the additional emission sourced from finite ISCO stress discs from $r \geq r_I$. This can be seen most clearly in Fig. \ref{fig:vary_dj}, as the finite ISCO stress discs (dashed curves) show a moderate range of behaviours for increasing $\delta_{\cal J}$, while the intra-ISCO emission (dotted curves) shows a much stronger dependence. This means that accounting for the impact of deviations from a no-torque boundary condition is even more important for obtaining a robust spin estimate than previously models concluded \citep[{e.g., a simple change in $\dot M$ cannot mimic intra-ISCO emission, although it can mimic the extra emission sourced from exterior to the ISCO which results from a non-zero stress}][]{Li05} . These degeneracies will be examined in more detail in a follow up work.

The precise observed form of the intra-ISCO emission depends on how extreme the photon starvation becomes within the ISCO. This can be seen in Fig. \ref{fig:vary_xi}. In Fig. \ref{fig:vary_xi} we show X-ray spectra (produced with otherwise identical parameters to the models discussed above), but with different values of the starvation parameter $\xi$. For values $\xi \lesssim 3$ the intra-ISCO emission is observed to have a quasi-blackbody form, however, if $\xi$ reaches values larger than this then the observed emission becomes increasingly well approximated by a power-law. As we discussed in section 4, the precise physics of the radiative transfer within the ISCO is uncertain, and will need to be addressed with dedicated calculations. It appears that intra-ISCO emission is best approximated by a hot-and-small blackbody, as strong power-law tails require extremely large colour-corrections (of order $f \gtrsim 100$), and even large colour-corrections (like the $\xi = 3$ case) show quasi-thermal emission. We do note however that there is sufficient energy liberated within the ISCO to power pronounced power-law components. We stress that this ``power-law'' appears from a sum of blackbodies, each with a progressively higher colour-correction, and not any intrinsic change in the modelling of the emission (a similar effect produces a power-law spectrum from thermal Comptonization).

Finally, in Figures \ref{fig:vary_i} and \ref{fig:vary_a} we examine the effect of varying the disc-observer inclination and black hole spin on the emergent flux sourced from within the ISCO. In each Figure the upper panel shows a vanishing ISCO stress disc spectrum, the middle panel the spectrum resulting from a disc with a finite ISCO stress and no intra-ISCO emission, while the lowest panel shows the full disc model, with intra-ISCO emission displayed as a dotted curve.  

In Fig. \ref{fig:vary_i} we display X-ray spectra observed at different inclinations (displayed in Figure legend). For these Schwarzschild discs the intra-ISCO emission is always important at energies $E\gtrsim 10$ keV. The emission from within the ISCO grows with importance with increasing inclination, a result of Doppler beaming of the inner disc (see e.g., Fig. \ref{85} and the discussion of section 3). The intra-ISCO disc region can produce the majority of the flux above $E \sim 5$ keV for moderate inclinations $i \gtrsim 60^\circ$. While the differences between the extra-ISCO emission of finite and vanishing ISCO stress discs show a weak dependence on inclination, the dependence of the {\it intra}-ISCO emission on inclination is much stronger. This suggests that disc models which include intra-ISCO emission may provide tighter constraints on source-observer inclinations.

Finally, in Fig. \ref{fig:vary_a} we display the effects of varying the black hole spin parameter $a_\bullet$ on the intra-ISCO emission. Perhaps unsurprisingly, intra-ISCO emission is most important at lower black hole spins, a result simply of the larger event horizon-to-ISCO radial range of low spin spacetimes. Interestingly, the intra-ISCO emission  (for those physical parameters given by those of Fig. \ref{fig:canon}) is relatively insensitive to black hole spin (see the dotted curves in Fig. \ref{fig:vary_a}), with each quasi-blackbody component having roughly similar temperature and area. This stops being true at the very highest spins (orange curves), where the effective area is strongly suppressed. 

It is interesting, and important, to note that with the inclusion of intra-ISCO emission, the differences between X-ray spectra produced by black holes with different spin parameters becomes less pronounced. This property will have relevance for the ability of continuum fitting to place constraints on black hole spins. 

To summarise, intra-ISCO emission is always important, but is particularly important for black holes with low spin parameters (Fig. \ref{fig:vary_a}).  The emission from within the ISCO is characterised by a hot-and-small quasi-blackbody profile, which grows in prominence for larger ISCO stress parameters (Fig. \ref{fig:vary_dj}) and inclinations (Fig. \ref{fig:vary_i}). For large values of the photon starvation parameter $\xi$ the intra-ISCO emission instead appears as a power-law component (sourced from a sum of blackbodies with rapidly growing colour-correction), but further study is required to determine whether this level of photon starvation is applicable to real systems (Fig. \ref{fig:vary_xi}). If this level of spectral hardening is relevant then  there is sufficient energy liberated within the ISCO to power a possibly substantial power-law fraction. 

{Finally, we discuss some apparent discrepancies between the results of this section and the results of \cite{Zhu12}. \cite{Zhu12} argue that their Schwarzschild simulations showed the smallest difference to standard vanishing ISCO stress models -- while we have argued that a large intra-ISCO area should maximise the discrepancy.  It is important to stress that \cite{Zhu12} were principally concerned with potential degeneracies in black hole spin measurements in models with and without intra-ISCO emission, while we are concerned with the more general features of the spectrum. Most important however is that we have explored how the spectrum varies as black hole spin is changed whilst $\delta_{\cal J}$ and $\xi$ are held constant. In the particular case of the \cite{Zhu12} analysis however, the Schwarzschild simulation showed substantially lower spectral hardening (see e.g., their Figure 9), suggesting a decrease in $\xi$ with lower spin. We caution against reading too much into this scaling however, other simulations \citep[e.g.,][which are somewhat better resolved]{Noble11} find much larger discrepancies between simulations of Schwarzschild discs and standard vanishing ISCO stress models. In addition, both of the \cite{Zhu12} and \cite{Noble10} simulations use cooling functions, while modern numerical methods solve the full equations of radiation hydrodynamics. Future comparisons of the models presented here and modern simulations is of great interest. 

Finally, we again stress the general philosophy of the approach put forward in this paper -- which is that observations may have a lot to teach us about intra-ISCO accretion, particularly in the most theoretically uncertain regions of parameter space, such as what the expected values of $\delta_{\cal J}$ and $\xi$ are as a function of other system parameters. While it is an important bench-marking test that the models developed in this paper can reproduce the thermodynamic (Fig. \ref{fig:temp-comp}) and observed (Fig. \ref{fig:spec-comp}) properties of individual GRMHD simulations, modelling of observations should be allowed to stand alone and at some level inform theoretical understanding of this region.    }

\section{Emission from within the ISCO of MAXI J1820 }\label{sec:maxij1820}

\begin{figure*}
    \centering
    \includegraphics[width=.49\linewidth]{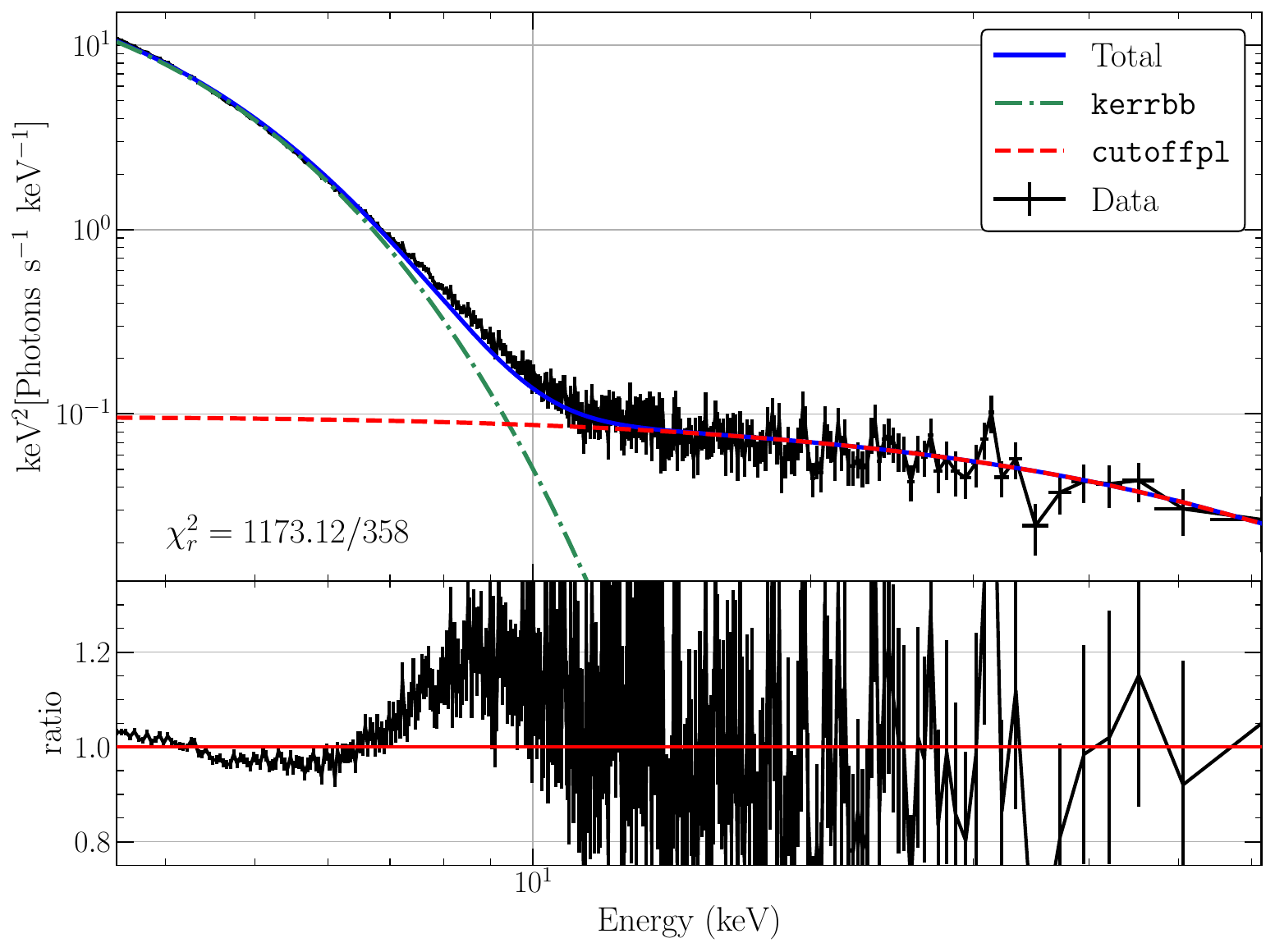}
    \includegraphics[width=.49\linewidth]{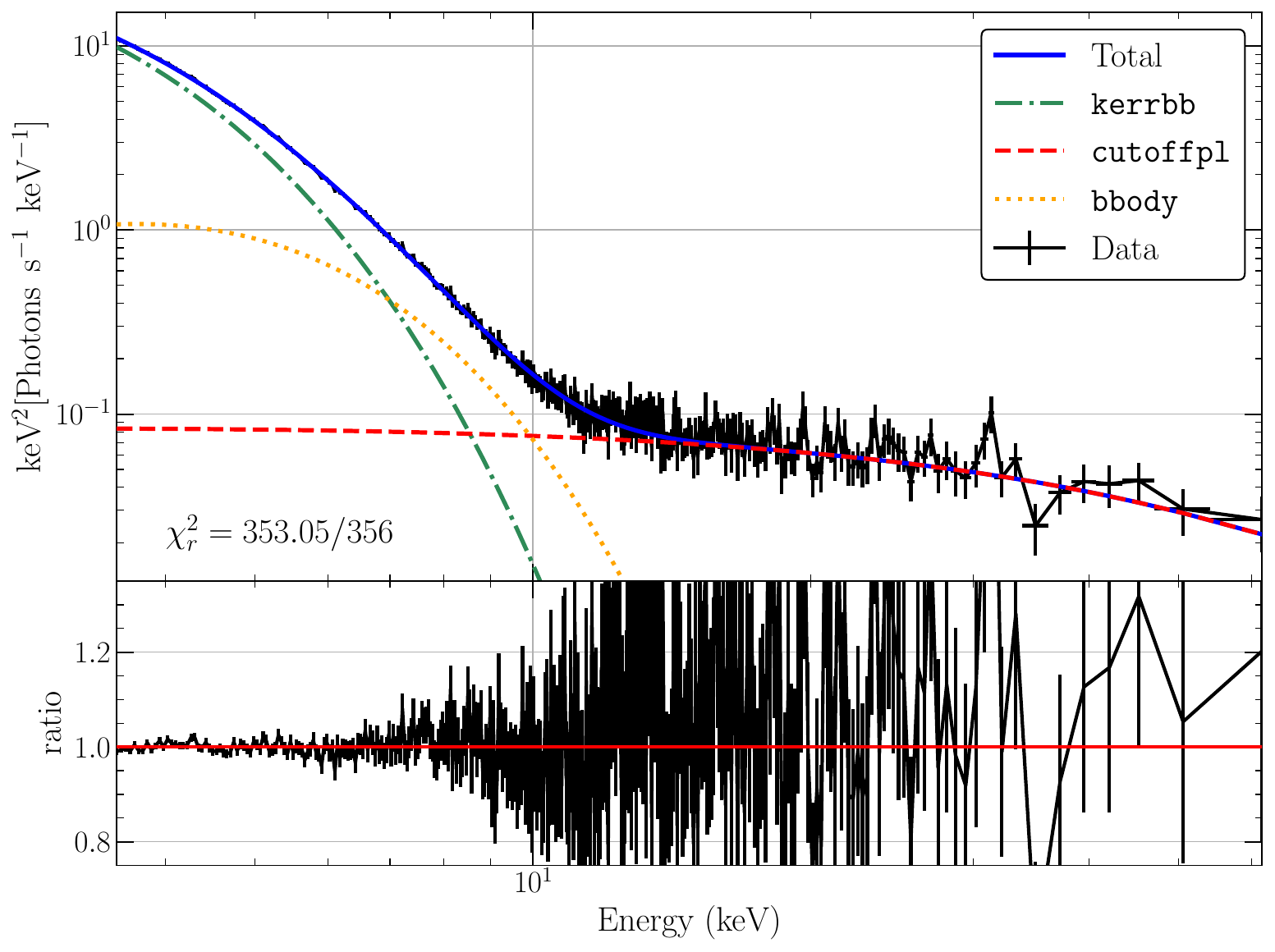}
    \includegraphics[width=.67\linewidth]{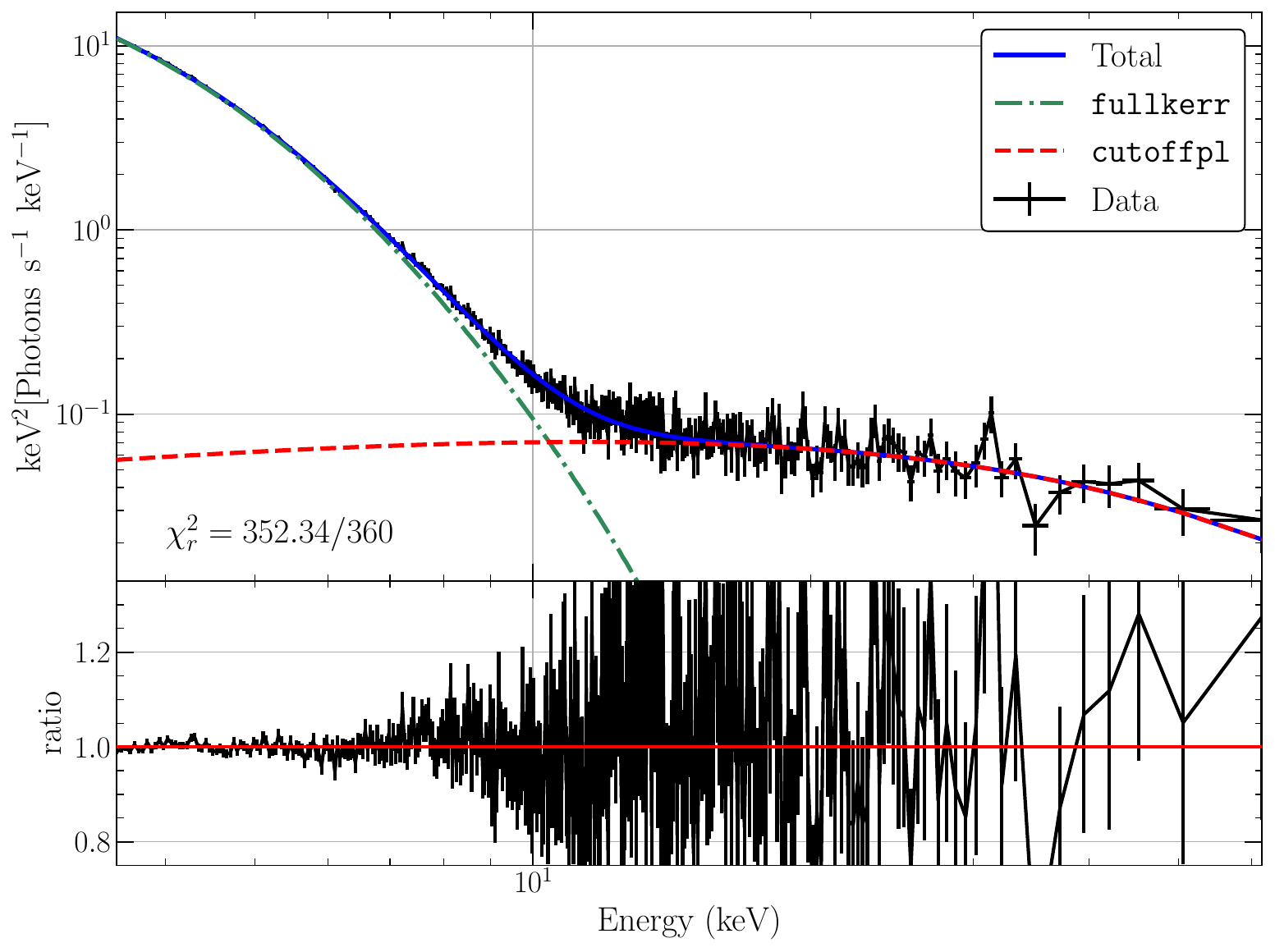}
    \caption{The Nu29 X-ray spectrum of MAXI J1820 (black points), compared to various accretion models for the emission. For each plot we use the model {\tt cutoffpl} to model the high energy power-law tail. In the upper left plot we display the best fitting {\tt kerrbb} model (which employs a vanishing ISCO stress boundary condition), which provides an extremely poor fit to the data. Adding an {\it ad-hoc} single temperature blackbody component to {\tt kerrbb} produces a formally excellent fit (upper right panel). However, using a full model for the disc (one which includes intra-ISCO emission) removes the requirement for the addition of {\it ad-hoc} components. This is displayed in the central panel, where we show the excellent fit of {\tt fullkerr} to the MAXI J1820 data.   }
    \label{fig:maxij1820}
\end{figure*}

\begin{figure*}
    \centering
    \includegraphics[width=0.67\linewidth]{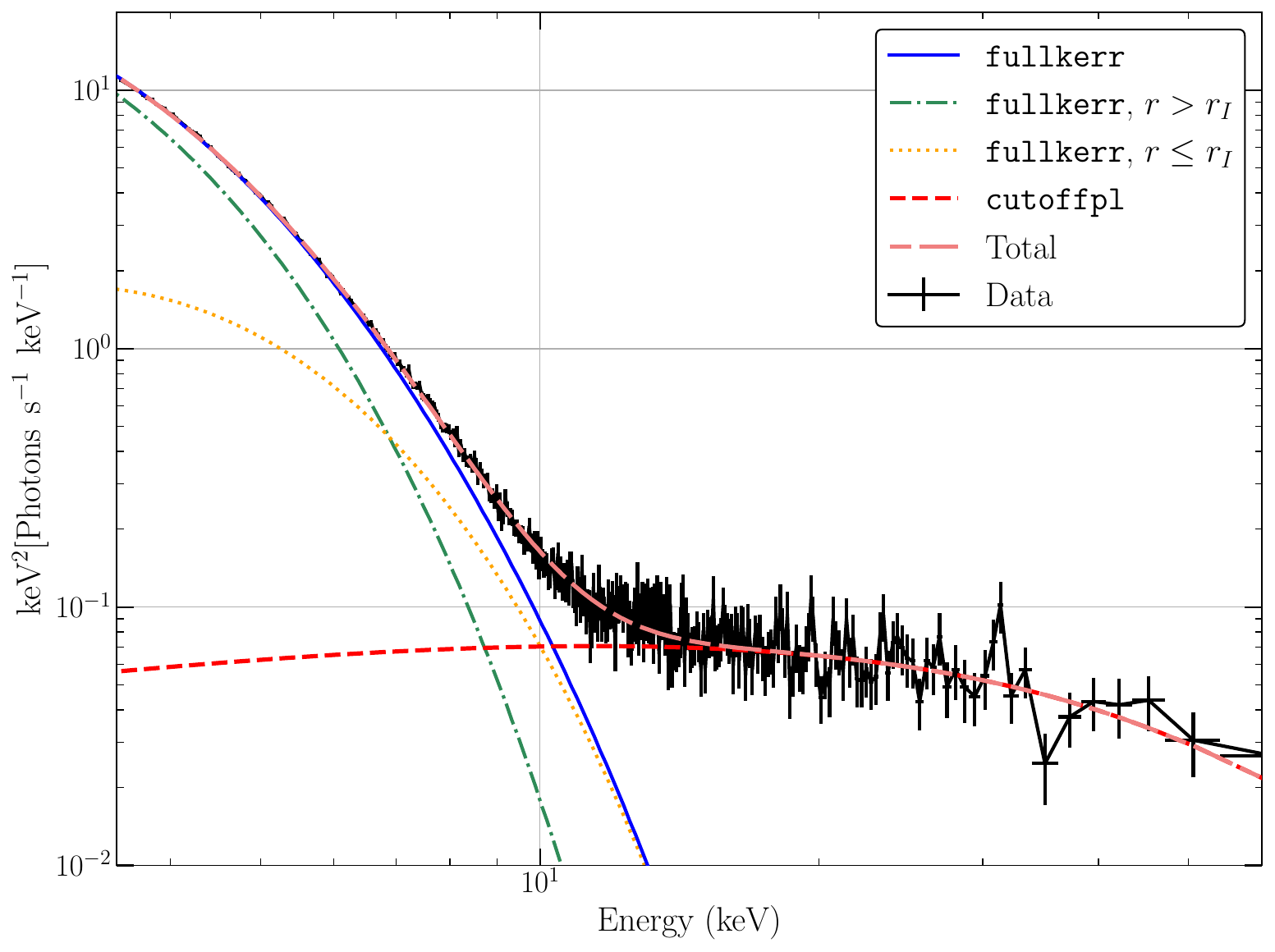}
    \caption{The physical origin of the different emission fit to MAXI J1820. We split the disc emission (blue solid curve) into components sourced outside (green dot-dashed) and inside (orange dashed) the ISCO. The intra-ISCO emission provides the hot-and-small thermal component previously added {\it ad-hoc} to vanishing ISCO stress accretion models (e.g., the upper right panel of Fig. \ref{fig:maxij1820}).   }
    \label{fig:components}
\end{figure*}

\begin{figure}
    \centering
    \includegraphics[width=\linewidth]{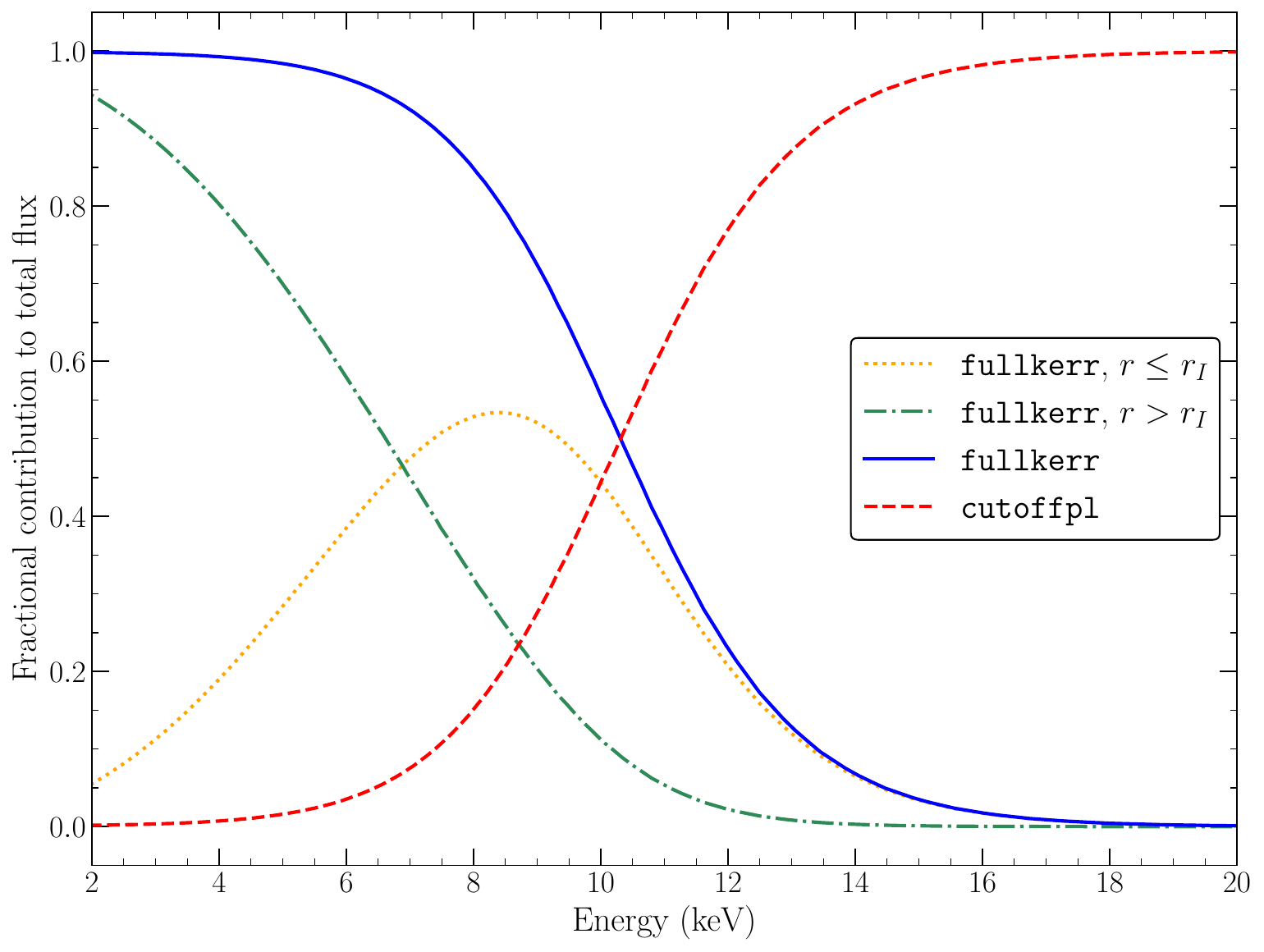}
    \caption{The relative contributions to the total flux of the disc (blue solid), power-law (red dashed) and extra-ISCO (green dot-dashed) and intra-ISCO (orange dashed) emission components. The intra-ISCO emission is the dominant source of flux for the photon energy range $E \sim 6-10$ keV. }
    \label{fig:frac_cont}
\end{figure}

With the properties of the intra-ISCO continuum emission now discussed, we proceed by analysing the X-ray spectrum of the low mass black hole X-ray binary MAXI J1820 \citep{Tucker18, Kawamuro18} taken during an outburst in 2018 \citep{Shidatsu19}. This thermal-dominated X-ray spectrum has previously been modelled by \cite{Fabian20}, who found that they were unable to describe the soft-state spectrum of MAXI J1820 with conventional disc models \citep[e.g.,][]{Li05}, owing to the presence of additional hot-and-small thermal emission. \cite{Fabian20} speculated that this additional emission may be sourced from within the ISCO, and as such this is an ideal observation with which to test the new accretion model developed in this paper. 

A representative X-ray spectrum is shown in Fig. \ref{fig:maxij1820}, in black we display the data, which is taken from the NuSTAR observation Nu29 \citep[different NuSTAR observations are distinguished by the last two digits in their OBSIDs; we refer the reader to][for details on the individual observations]{Fabian20}. We observe a clear thermal continuum and a high energy power-law tail (presumably produced by the Compton up-scattering of seed disc photons by a hot electron corona). In the upper left panel we show a best fit of {\tt kerrbb + cutoffpl}\footnote{None of the models considered in this section required the addition of a component to describe absorption of photons by neutral hydrogen (e.g., {\tt tbabs}). This is due to the relatively high starting point of the NuSTAR bandpass ($\sim 3$ keV) and the low galactic hydrogen column density in the direction of MAXI J1820.} to the data. As we are not interested in the best fitting parameters of the {\tt kerrbb} fit, merely in its overall performance, we allow all parameters (except the source-observer distance which we fix to $3$ kpc, and the ISCO stress which we fix to 0) to freely vary during the fit. We do not report the best fitting {\tt kerrbb} parameters, as the fit is so poor \citep[see also][]{Fabian20}.  In the canonical picture {\tt kerrbb} should well describe the thermal continuum, with {\tt cutoffpl} modelling the power-law tail.  The total fitted model is shown as a  blue solid curve, with the {\tt kerrbb} contribution shown by the green dot-dashed curve, and the {\tt cutoffpl} model component shown by a red dashed curve. The lower panel of the upper left plot shows the ratio between the best fitting model and the data, and highlights the clear structured residuals between the model and the data. The fit is formally extremely poor, with a reduced chi-squared of $\chi^2_r = 1173.12/358$. This is precisely in keeping with the finding of \cite{Fabian20}.

In the upper right panel we show that with the addition of a hot and small blackbody (modelled in {\tt XSPEC} with the {\tt bbody} model), the structured residuals are removed from the fit, and a formally excellent fit is found $\chi^2_r = 353.1/356$. Of course, {\tt bbody} is simply a phenomenological addition, characterised by a temperature $T_{\rm bb}$ and radius $R_{\rm bb}$, and has no physical content. 

In contrast, in the central panel of Fig. \ref{fig:maxij1820} we display a joint fit of {\tt fullkerr + cutoffpl}, where {\tt fullkerr} is our new continuum fitting model which includes the intra-ISCO emission (described in Appendices \ref{appA}, \ref{appB}). As we are now interested in the physical parameters of the fit (in contrast to the {\tt kerrbb} fit), we fix various parameters to those with prior constraints. The source-observer distance is fixed to $D= 3$ kpc \citep{Fabian20, Tetarenko23}, the black hole mass is fixed to that inferred from optical spectroscopy $M_\bullet = 8.5 M_\odot$ \citep{Torres19, Tetarenko23}, the black hole spin and source-observer inclination are fixed to values inferred from an analysis of the hard-state X-ray spectrum, namely $a_\bullet = 0.2, i = 35^\circ$ \citep[][see also the recent work of \citealt{Dias24}]{Buisson19}. We also fix the colour-correction factor in the main body of the disc and at the ISCO to $f_d = f_I = 1.7$, a canonical value for X-ray binary studies \citep[e.g.,][]{Li05, Davis06}. This leaves 3 fitting parameters, the mass accretion rate $\dot M$, the ISCO stress $\delta_{\cal J}$ and the photon starvation parameter $\xi$. The best fitting values of these parameters are $\dot M = \left[0.86 \pm 0.02\right] L_{\rm edd}/c^2$, $\delta_{\cal J} = \left[3.85 \pm 0.14\right] \times 10^{-2}$, and $\xi = 1.7 \pm 0.06$, where the uncertainties correspond to 90$\%$ confidence intervals. This accretion rate corresponds to an Eddington luminosity ratio of order $l \equiv L_{\rm disc}/ L_{\rm edd} \sim 0.1$, as expected for a soft state spectrum. The {\tt cutoffpl} parameters were fit to the high energy tail,  with resulting parameter values $\Gamma = 1.54$, $E_{\rm cut} = 24.7$ keV, $N = 0.036$ ($N$ is a flux normalisation at 1 keV). These parameters were not allowed to vary after an initial fit to the high energy excess. The ISCO stress parameter $\delta_{\cal J} \sim 0.04$ lies in the range typically seen in GRMHD simulations \citep[e.g.,][]{Noble10, Penna10, Schnittman16}, {and is broadly consistent with the \cite{Paczynski00} estimate $\delta_{\cal J} \sim \alpha (H/r)$.} 

We see that the addition of intra-ISCO emission removes the requirement for additional {\it ad-hoc}  thermal components, and the fit is formally excellent $\chi^2_r = 352.34/360$. When contrasted with a fit to {\tt kerrbb} alone, the addition of intra-ISCO emission reduces the $\chi^2$ parameter by over 800, and is required at extremely high statistical significance. We believe that this represents the first robust detection of intra-ISCO emission in the literature.

The fact that the thermal excess at around $E = 6 - 10$ keV is sourced from within the ISCO is shown in Fig. \ref{fig:components}, where we explicitly split out the contributions to the total {\tt fullkerr} flux (blue solid curve) from radii outside of the ISCO (green dot dashed curve), and within the ISCO (orange dotted curve). Contrasting Fig. \ref{fig:components} with the upper right panel of Fig. \ref{fig:maxij1820} we see that the intra-ISCO emission is providing precisely the additional emission modelled out by {\tt bbody} in the work of \cite{Fabian20}.

In Fig. \ref{fig:frac_cont} we display the fractional contributions of each of the different emission regions to the total model flux fit to MAXI J1820. We see that the intra-ISCO emission dominates the total emitted flux for a relatively broad ($E \sim 6-10$ keV) range of observed energies.  {While the total fraction of the $0.05-50$ keV disc luminosity sourced from within the ISCO is only $\simeq 4.1\%$ for this epoch, its observational effect is pronounced in the ``right'' energy bands. } {What constitutes the right energy band is somewhat dependent on the free parameters of the model (e.g., Fig. \ref{fig:vary_dj}), but is typically several times the peak temperature of the main body of the disc and therefore roughly $E\sim 6-10$ keV for typical X-ray binary parameters.  }

\begin{figure}
    \centering
    \includegraphics[width=\linewidth]{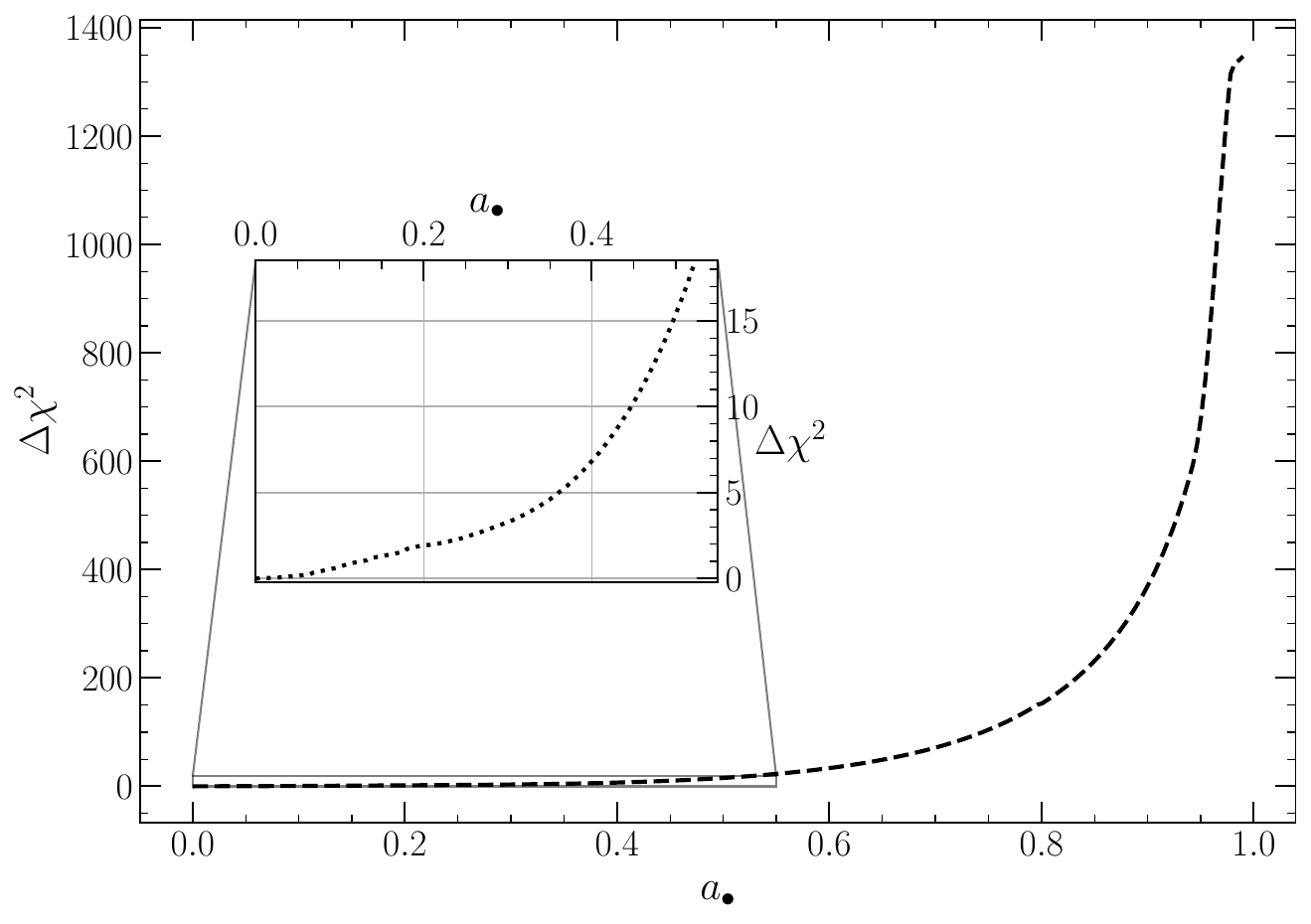}
    \includegraphics[width=\linewidth]{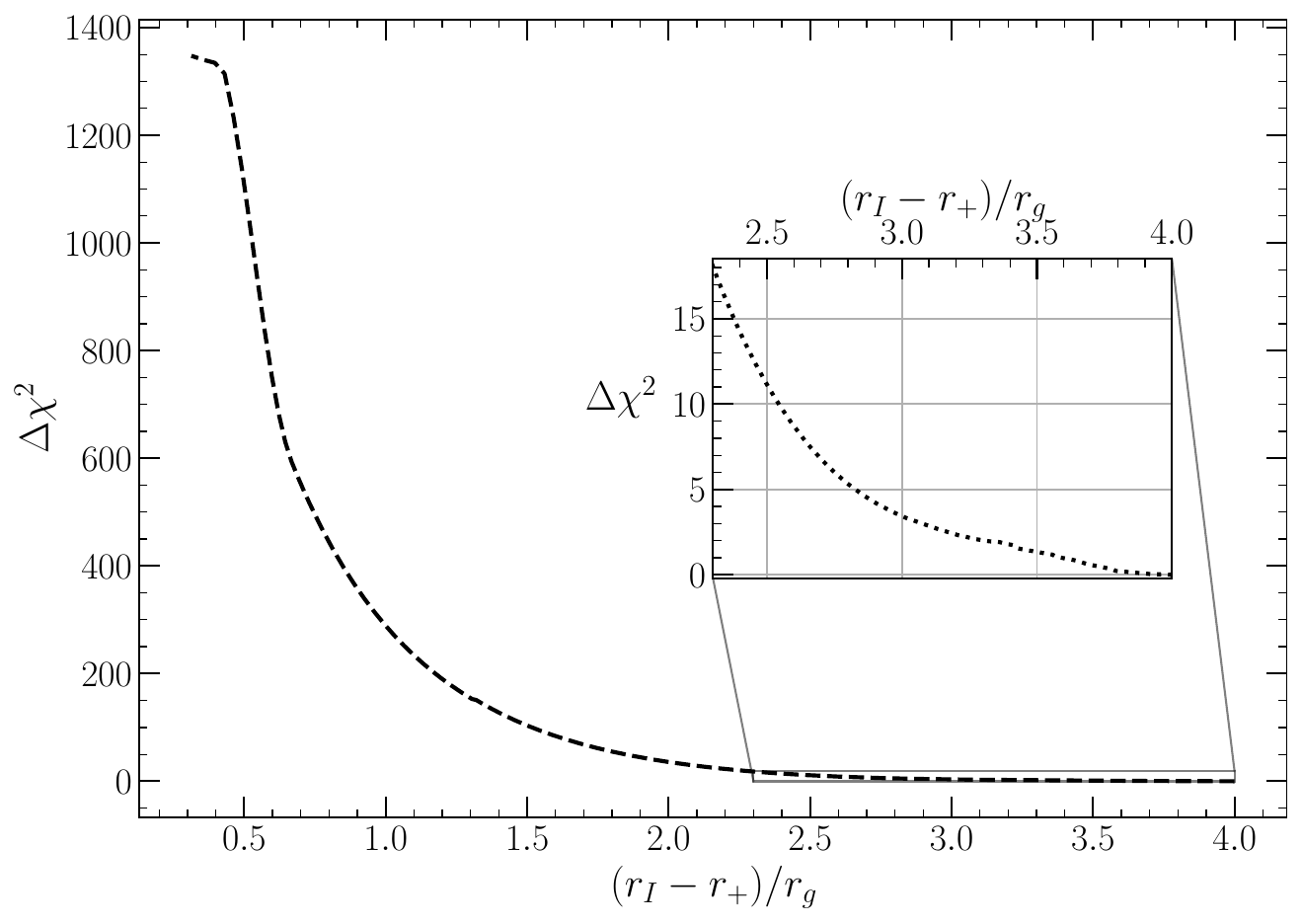}
    \caption{The change in $\chi^2$ fit statistic as a function of assumed black hole spin (upper), or ISCO-to-event horizon distance (lower). If the spin of MAXI J1820 is too high, there is insufficient disc area to power the required hot-and-small thermal component, and the fit rapidly becomes very poor. The two zoomed-in insets display the region of parameter space acceptable at a $99.9\%$ confidence level. We can constrain the black hole spin in MAXI J1820 to be low $a_\bullet < 0.5$, in accordance with reflection modelling in the hard state.  }
    \label{fig:spin_const}
\end{figure}

Finally, we note that the fact that observational requirement of an observable intra-ISCO region can be leveraged to place strong constraints on the {\it uppermost} value the black hole spin in MAXI J1820 can take. For too large a spin, the size of the ISCO-to-event horizon region shrinks, and the hot-and-small thermal component becomes unobservable (e.g., Fig. \ref{fig:vary_a}).  In Fig. \ref{fig:spin_const} we refit the MAXI J1820 X-ray spectrum for spin parameters covering the prograde range $0 < a_\bullet < 0.99$, and display the change in $\chi^2$ statistic as a function of spin (upper) and ISCO-to-event horizon radial distance (lower). In these fits the inclination $i$, accretion rate $\dot M$, ISCO stress $\delta_{\cal J}$ and photon starvation parameter $\xi$ were allowed to vary freely. The quality of fit is relatively insensitive to spin for low spins $a_\bullet \lesssim 0.4$, but rapidly becomes extremely poor for spins $a_\bullet \gtrsim 0.5$. Using these results we are able to place constraints on the MAXI J1820 black hole spin, which must be below $a_\bullet \leq 0.52$ at 99.9$\%$ confidence level (Fig. \ref{fig:spin_const}). This result is in good agreement with the hard state reflection modelling of \cite{Buisson19} {and \cite{Dias24}, although disagrees with the results of \cite{Draghis2023} (showing of course that these hard state models are inconsistent with each other). It is unclear why \cite{Draghis2023} find such a high spin for MAXI J1820 ($a_\bullet = 0.988^{+0.006}_{-0.028})$, in tension with other analyses which use the same models and data, but it may be related to the accretion disc density they assumed, which was lower than that expected in X-ray binary discs $n_e \sim 10^{20}$ cm$^{-3}$. These higher values of the density were used in \cite{Dias24}, who favour lower (and even negative) spins. }

MAXI J1820 was observed by \cite{Fabian20} to require an additional {\it ad-hoc} blackbody emission component in additional epochs to just Nu29 considered in this section. In Appendix \ref{other_epochs} we show that, with the same fixed black hole, source-observer, and colour-correction parameters as in this section, the epochs Nu 25, Nu27 and Nu31 can also be described by the {\tt fullkerr + cutoffpl} model. Each epoch had a similar ISCO stress parameter $\delta_{\cal J} \simeq 0.04$, and the different epochs are principally distinguished by different accretion rates $\dot M$. Each {\tt fullkerr + cutoffpl} fit represents a similar improvement over {\tt kerrbb + cutoffpl} as the Nu29 epoch discussed in this section, with fit statistic improvements of order $|\Delta \chi^2| \sim 600-800$. 

\section{Discussion} 
{In the previous section we focused on NuSTAR observations of the black hole X-ray binary MAXI J1820 during a bright soft-state outburst. In this section we discuss MAXI J1820 in the context of the broader X-ray binary population, and include other data sets (in particular a NICER observation of MAXI J1820) in our analysis.  }

{During it's 2018 outburst, MAXI J1820 was also observed by the NICER telescope, roughly simultaneously with the observation presented in section \ref{sec:maxij1820}. NICER probes a lower energy range to NuSTAR (NICER observes at photon energies $1-10$ keV), and so contains more information on the peak of the disc spectrum than the NuSTAR observations.  While the key signature of intra-ISCO emission is located at photon energies $E \sim 6-10$ keV,  data at energies substantially above $E = 10$ keV is crucial to revealing it's presence, as a proper handle on the properties of the comptonised power-law tail is required so that the intra-ISCO feature is not simply modelled out with a power-law. For this reason we have focused on NuSTAR data, which gives us this proper handle, but it is an interesting exercise to see how the {\tt fullkerr} model copes with the lower energy interval probed by NICER.   }

\begin{figure}
    \centering
    \includegraphics[width=\linewidth]{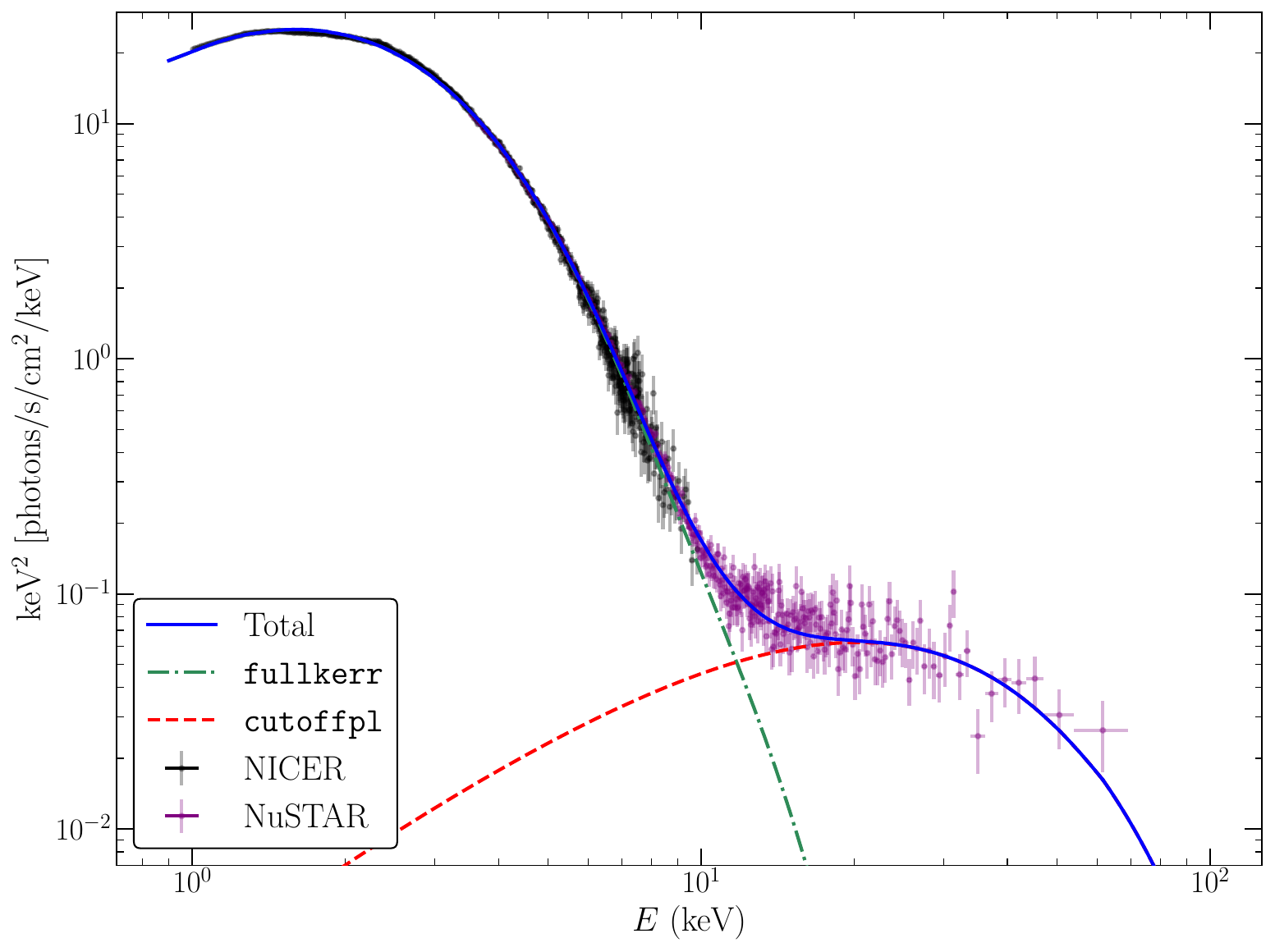}
    \includegraphics[width=\linewidth]{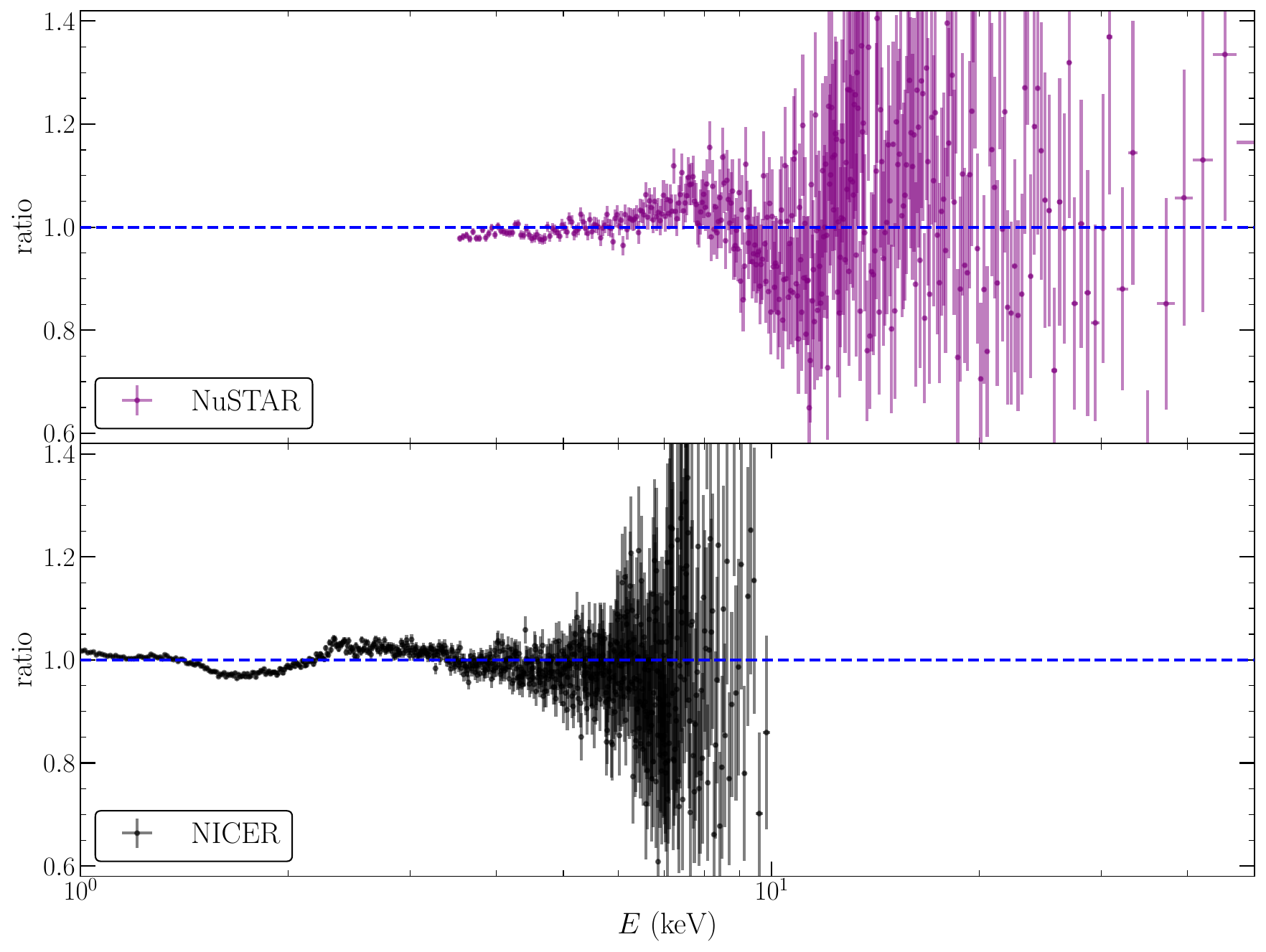}
    \caption{ Upper: a joint fit to simultaneous NICER (black) and NuSTAR (purple) data of MAXI J1820, with the model components denoted by different line styles (see legend). In the lower panel we show the ratio of the data to the total model for each observatory. The joint fit is statistically acceptable. The NICER data has been corrected by a factor $1/{\tt constant}$ with ${\tt constant} \simeq 0.43$ (see text) to account for the fact that over half of the mirror modules of NICER were switched off during this observation. }
    \label{fig:nustar+nicer}
\end{figure}

{In Figure \ref{fig:nustar+nicer} we plot a joint fit to the NICER (black points) and NuSTAR (purple points) observations of MAXI J1820. Including this NICER data adds a couple of additional complications which we now discuss. Firstly, as MAXI J1820 was so bright during it's outburst, over half of the mirror modules of NICER were
switched off during this observation. (This was done simply to deal with the unusually high photon count rate of the source.) We therefore include a multiplicative amplitude in our modelling (the model {\tt constant} in {\tt XSPEC} notation) to our fitting function, which is fixed to unity for the NuSTAR data but is best fit by ${\tt constant} \simeq 0.43$ for the NICER data. In addition, the lower bandpass of NICER probes an energy range where the effects of the absorption of X-ray photons by neutral hydrogen between the source and observer become more significant. To model this more properly, we include the {\tt tbabs} model  in {\tt XSPEC} \citep{Wilms00}.  In full therefore our model is, in the notation of {\tt XSPEC}, given by }
\begin{equation}
{\tt model} = {\tt constant} \times {\tt tbabs} \left({\tt fullkerr} + {\tt cutoffpl} \right) .
\end{equation}
{Finally, at the lower energies probed by NICER, systematic errors in the calibration of X-ray telescopes become important. Unlike Optical telescopes, which can use  nearby featureless white dwarfs, no standard objects exist in X-ray astronomy which can be  observed and used for calibration at the percent level. As such, it is standard practice to add percent level systematic errors to the X-ray data taken from a source, to prevent  bumps in the X-ray spectrum at around $1-3$ keV from dominating the statistics of the fit (this may be particularly relevant when over half of the mirror modules are switched off).   }

{Using the same set of free parameters as before (in other words, freezing the parameters  $M_\bullet = 8.5 M_\odot$, $a_\bullet = 0.2$, $i = 35^\circ$ and $D = 3$ kpc), and including a systematic error of $2\%$, we find a best fit to the joint NICER+NuSTAR data which is consistent (in terms of the values of the physical parameters $\dot M, \delta_{\cal J}$ and $\xi$) with our fit to the NuSTAR data alone, and which is   statistically acceptable $\chi_r^2 = 1142.19 / 1050$. Letting any of the free parameters $a_\bullet$, $i$, or the disc colour-correction factor $f_d$ freely vary always provides a statistically excellent fit (for example, a freely varying colour-correction factor in the main body of the disc has best fitting reduced chi-squared parameter $\chi_r^2= 986.28/1049$; this fit is shown in Fig. \ref{fig:nustar+nicer}). If we allow all three of these parameters to vary freely, we find an excellent fit even if the systematic errors in the data collection are lower (at the $1.5\%$ level). 

A key result is that the improvement of the {\tt fullkerr} fit over the model }
\begin{equation}
    {\tt model} = {\tt constant} \times {\tt tbabs} \left({\tt kerrbb} + {\tt cutoffpl} \right) , 
\end{equation}
{grows with the addition of the NICER data. Allowing all parameters in the {\tt kerrbb} model (other than the source-observer distance and the vanishing ISCO stress boundary condition) to freely vary results in a best-fitting reduced chi-squared value of $\chi_r^2 = 2657.24/1050$, which is $\Delta \chi^2 = 1515.05$ higher than our canonical model fit. }

{It is also worth noting that the source MAXI J1820 potentially has, as has been suggested in a number of works \citep[e.g.,][]{Buisson19, Fabian20, Thomas22, Poutanen22}, a large misalignment between the binary orbital inclination and the disc's inclination with respect to the black hole spin axis. We do not believe that this complication has any effect on the interpretation put forward in this paper for two reasons. Firstly, we have verified that by fixing the inclination of the disc to that inferred from the binary orbit inclination ($i \simeq 70^\circ$) we are still able to find excellent fits to the NuSTAR data by allowing the black hole spin $a_\bullet$, accretion rate $\dot M$ and intra-ISCO parameters $\delta_{\cal J}$ and $\xi$ to vary.  Secondly, while the accretion flow may well show some large scale warping owing to this misalignment, it is likely that the inner disc will be aligned with the black hole's equatorial plane at small radii (an assumption when solving the disc equations in the {\tt fullkerr} model), owing to the Bardeen-Peterson effect \citep{Bardeen75}, a theoretical model of disc-black hole alignment with numerical support \citep{Liska21}.   }

{Finally, it is worth stressing the findings of \cite{Fabian20}, namely that the excess flux observed in the spectrum of MAXI J1820 at $E \sim 6-10$ keV cannot plausibly be modelled out with alternative explanations. Adding additional power-law components would require power-law indices which are implausibly steep $\Gamma \sim 7-8$, additional iron line components can only reproduce the data if the geometry of the irradiating source changes by significant factors on short timescales, and an additional Gaussian emission component would require an equivalent width which is too large to be plausible \citep[we direct the interested reader to][for further discussion]{Fabian20}.     }

{In an ideal world, a final test of this model would be the continued detection of this excess component as the disc cools into an even lower luminosity state.  This would rule out any other effects related to the moderate $L \sim 0.1 L_{\rm edd}$ luminosity of the source in the bright soft state. Unfortunately, as MAXI J1820 {continued to dim at  epochs later than the final epoch studied in this paper  (Nu31)} the power-law component {thereafter} grew {in amplitude (and relative importance)}, and {subsequently} dominated the spectrum rendering such a test impossible {(see e.g., Figure 1 of \citealt{Fabian20} for the full MAXI J1820 light curve).} It is important to note however that the properties of the power-law tail and {\it ad-hoc} blackbody component added by \cite{Fabian20} showed no correlation in properties, despite the power-law component varying by over a factor of 40. This suggests they are not linked physically.   }

{We do believe however that the components discussed in this paper should be detectable in a wider population of sources, which may allow further tests of this model at a range of accretion rates. Of particular interest are sources suspected to have low spins and large inclinations, where the effects of plunging-region emission are expected to be maximal (Figs. \ref{fig:vary_i}, \ref{fig:vary_a}). Sources with these characteristics are known, such as LMC X-3 and H1743-322 \citep[see e.g.,][for a review of previous spin estimates]{Bambi18}, and will be interesting avenues for future research.   }

\subsection{ Iron K-$\alpha$ flux from within the plunging region }
\begin{figure}
    \centering
    \includegraphics[width=\linewidth]{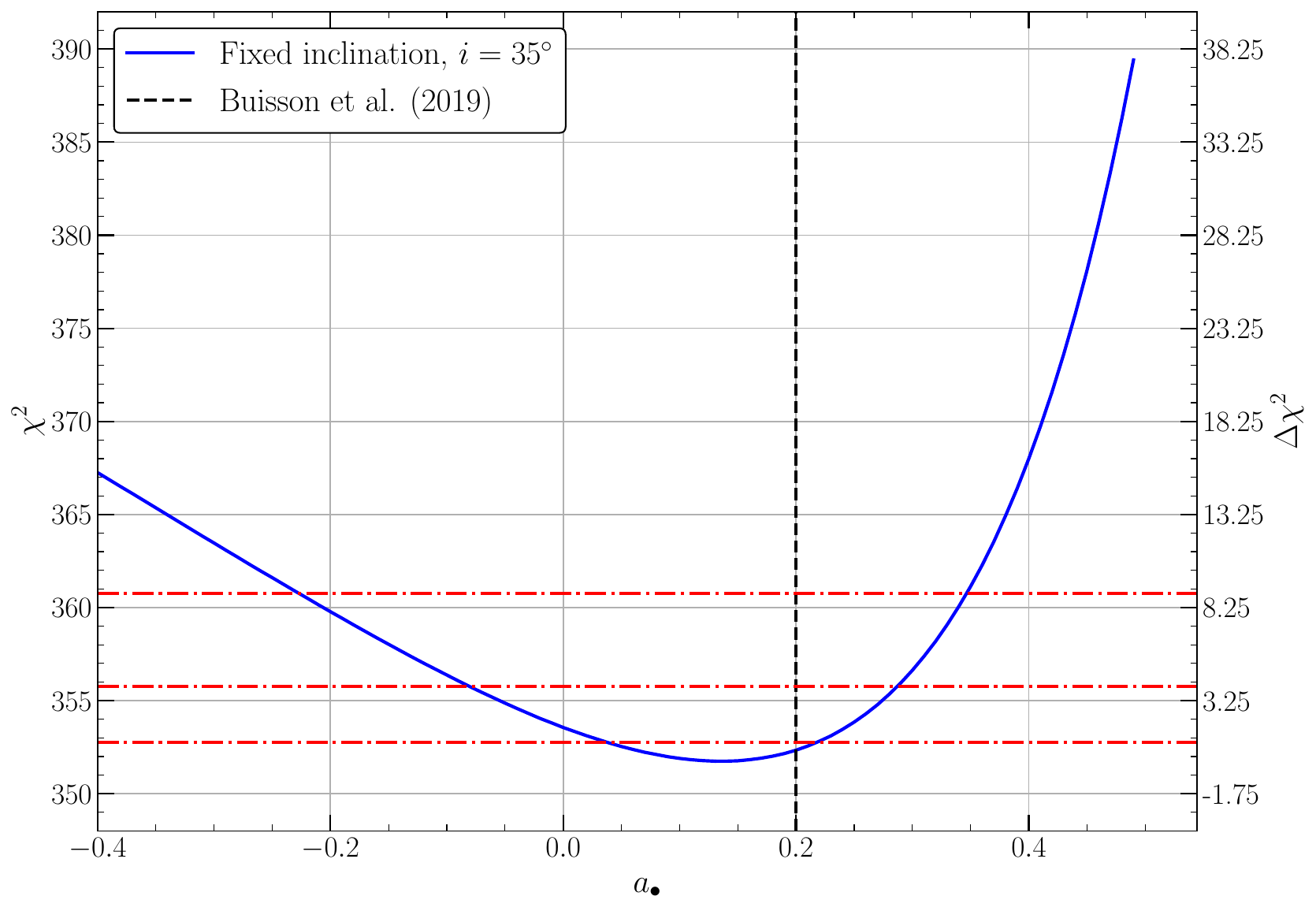}
    \caption{The chi-squared fit statistic of the {\tt fullkerr + cutoffpl} model fit to the NuSTAR data shown in Figure \ref{fig:maxij1820}, as a function of black hole spin parameter $a_\bullet$, with the inclination fixed to $i = 35^\circ$. This inclination was the best fitting value found by \citealt{Buisson19} when modelling the hard state emission of MAXI J1820. We see that the spin value inferred from our continuum fitting analysis is consistent with that found from the hard state analysis of \citealt{Buisson19}, $a_\bullet = 0.2$.  Red dot-dashed lines show the 1, 2 and 3$\sigma$ intervals respectively.  }
    \label{fig:spin_val_comp}
\end{figure}

{While the focus of this work has been on continuum emission sourced from within the plunging region, here we briefly discuss the implications of a non-zero density and emission for iron K-$\alpha$ studies of black hole accretion discs. One major difference between continuum fitting and reflection spectroscopy studies is that, while the radiative temperature of the disc remains high within the plunging region (e.g., Fig. \ref{fig:temp-comp}), the density of the flow does drop much more rapidly \citep[a result of the rapid increase in the radial velocity and the fixed radial mass flux $\dot M = 2\pi r U^r \rho H$ e.g.,][]{MummeryBalbus2023}. A key implication of this result is that the disc ionization fraction $\xi_{\rm Fe} \propto F_X/\rho$ grows rapidly (the parameter $F_X$ is the irradiating X-ray flux, which typically grows with decreasing radius as a power-law{, although this behaviour may change at the smallest length scales comparable to the size of the corona}), and for typical parameters the flow quickly becomes over ionized within the ISCO. Indeed, it has been argued that  this region therefore  typically contributes minimally to the time averaged X-ray reflection spectrum \citep[e.g.][]{Wilkins20}.  }

{That being said, X-ray binaries represent a particularly interesting probe of intra-ISCO reflection spectroscopy, owing to their large electron number densities in the main body of the disc $n_e \sim 10^{20}$ cm$^{-3}$, which makes them less susceptible to over ionization (when compared to AGN discs, for example).  As can be clearly seen in Figs. \ref{85} and \ref{60}, the red and blue shifting of photons emitted from the plunging region can be extreme, and may well offer interesting probes of the intra-ISCO region \citep{Reynolds97}. These calculations are being actively pursued \citep[e.g.,][]{Dong23}.  }

{Finally, one particularly interesting avenue for future studies will be in examining the concordance between reflection spectroscopy and continuum fitting black hole spin measurements, when plunging region emission is included. As a first demonstration of these model tests we plot in Figure \ref{fig:spin_val_comp} the chi-squared fit statistic of the {\tt fullkerr + cutoffpl} model fit to the NuSTAR data shown in Figure \ref{fig:maxij1820}, as a function of black hole spin parameter $a_\bullet$, with the inclination fixed to $i = 35^\circ$. This inclination was the best fitting value found by \cite{Buisson19} when modelling the hard state emission of MAXI J1820. We see that the spin value inferred from our continuum fitting analysis is consistent with that found from the hard state analysis of \cite{Buisson19}, $a_\bullet = 0.2$.  Red dot-dashed lines show the $1, 2$ and $3\sigma$ confidence intervals respectively (with larger confidence intervals at higher $\chi^2$ values).  See Appendix \ref{other_epochs} for this same calculation performed at different epochs.  }

\section{Conclusions }
In this paper we have extended existing X-ray continuum fitting models to include emission sourced from within the ISCO of Kerr black hole discs. GRMHD simulations of black hole accretion  find such emission to be a generic feature of these systems \citep[e.g.,][]{Noble10, Zhu12, Schnittman16, Wielgus22}, driven by magnetic stresses which couple the stable and plunging disc regions. 

The extra luminosity sourced from within the ISCO is generally observed to take the form of a hot-and-small quasi-blackbody component (e.g., Figs. \ref{fig:canon}, \ref{fig:vary_dj}, \ref{fig:vary_i}), but can for more extreme regions of parameter space power a power-law tail (e.g., Figs. \ref{fig:spec-comp}, \ref{fig:vary_xi}). This additional emission broadens and hardens the total disc spectrum, and should in principle be observable for a wide range of parameter space. This is true provided that black holes are not all near-maximally rotating, as in the maximal spin limit the event horizon-to-ISCO region shrinks and the intra-ISCO emission becomes less important (Fig. \ref{fig:vary_a}).  

We have then shown that the addition of intra-ISCO emission resolves some problems for conventional disc models. In particular, we are able to describe the emission from MAXI J1820 during a 2018 soft-state outburst, something which is not possible with conventional models truncated at the ISCO (Fig. \ref{fig:maxij1820}). The intra-ISCO emission reproduces the requirements of previously added {\it ad-hoc} blackbody profiles (Fig. \ref{fig:components}). The dominant component of the emission between 6 and 10 keV in the MAXI J1820 spectrum results from emission sourced from within the ISCO, highlighting the importance of this region (Fig. \ref{fig:frac_cont}). As far as the authors are aware  this work represents the first robust detection of intra-ISCO emission in the literature. 

The detection of intra-ISCO emission allows constraints to be placed on the MAXI J1820 black hole spin, which we have demonstrated must be low $a_\bullet \leq 0.52$ (99.9$\%$ confidence level) to allow a detectable intra-ISCO region. This result is in good accordance with the hard state reflection modelling of \cite{Buisson19} and \cite{Dias24}. Modelling of intra-ISCO emission has the potential to be a powerful diagnostic of black hole parameters in the future. 

As other sources \citep[for example MAXI J0637-430][]{Lazar21} have also been fit with {\it ad-hoc} blackbody components, it seems to us likely that observations of intra-ISCO emission may be widespread, but previously modelled out or neglected.

\section*{Acknowledgments}
 The authors would like to thank the reviewer for a detailed report which improved the manuscript. 
 AM gratefully acknowledges support and hospitality from the Institute for Advanced Study, where some of this work was completed.  This work was supported by a Leverhulme Trust International Professorship grant [number LIP-202-014]. For the purpose of Open Access, AM has applied a CC BY public copyright licence to any Author Accepted Manuscript version arising from this submission. AI acknowledges support from the Royal Society. This research has made use of data from the \textit{NuSTAR} mission, a project led by the California Institute of Technology, managed by the Jet Propulsion Laboratory, and funded by the National Aeronautics and Space Administration.  We acknowledge the use of public data from the \textit{NICER} data archive.

\section*{Data availability}
The {\tt XSPEC} model \fk is available at the following GitHub repository:  \url{https://github.com/andymummeryastro/fullkerr}. At this link the X-ray spectrum of MAXI-J1820 fit in the manuscript can also be found. 

\bibliographystyle{mnras}
\bibliography{andy}

\appendix

\section{The disc model}\label{appA}
In this Appendix we list every analytical result used in the {\tt fullkerr} model, so that this paper remains self contained. More detailed treatments of both the intra- and extra-ISCO disc regions are presented elsewhere in the literature. 
\subsection{Orbital motion and ISCO location }
The key orbital component relevant for the study of thin accretion flows is the angular momentum of a fluid element undergoing circular motion, as this is an excellent approximation to the dynamical behaviour of the fluid in the stable disc regions. For the Kerr metric standard methods \citep[e.g.,][]{Hobson06} lead to the following circular angular momentum profile   
\begin{equation}\label{orbang}
U_\phi = (GMr)^{1/2}{(1+a^2/r^2-2ar_g^{1/2}/r^{3/2}) \over \sqrt{ 1 -3r_g/r +2ar_g^{1/2}/r^{3/2}}} .
\end{equation}
The (dimensionless) energy of a circular orbit is similarly given by
\begin{equation}\label{orben}
U_t/c^2 = - { (1-2r_g/r +ar_g^{1/2}/r^{3/2}) \over \sqrt{ 1 -3r_g/r +2ar_g^{1/2}/r^{3/2}} }.
\end{equation}
{In this paper we shall define the $\phi$ coordinate to increase in the direction of the orbital motion of the fluid.  This means that $U_\phi(r, a) \geq 0$ for all orbits, and retrograde motion is identified with negative Kerr spin parameters $a < 0$. }

A thin disc will persist in its classical form wherever these orbits are stable. The stability of circular orbits  requires $\partial_r U_\phi > 0$, the relativistic analogue of the Rayleigh criterion.   Evaluating this gradient, we find 
\begin{equation}\label{grad_j}
{\partial U_\phi \over \partial r}  =
{ (GM)^{1/2}\over 2r^{1/2}{\cal D}}\left(1 +{ar_g^{1/2}\over r^{3/2}}\right)  \left( 1 - {6r_g\over r} -{3a^2\over r^2} +{8ar_g^{1/2}\over r^{3/2}}\right) ,
\end{equation}
where 
\begin{equation}
{\cal D} = \left( 1 -3r_g/r +2ar_g^{1/2}/r^{3/2}\right)^{3/2}.
\end{equation} 
Note that ${\cal D}$ is finite and positive for all radii $r > r_I$ of interest {(where we denote the ISCO location by $r_I$)}, so equation (\ref{grad_j}) passes through zero at $r=r_I$,  where the final factor vanishes, i.e.:
\begin{equation}\label{a}
r_I^2  -    6r_gr_I + 8a\sqrt{r_gr_I} - 3a^2  = 0.
\end{equation}
The location of the ISCO as a function of Kerr spin $r_I(a)$ can be written in closed form, and is given by \citep{Bardeen72} 
\begin{equation}
{r_I \over r_g}  = 3 + Z_2  - {\rm sgn}(a_\bullet) \sqrt{(3 - Z_1)(3 + Z_1 + 2 Z_2)},
\end{equation}
where 
\begin{equation}
a_\bullet \equiv a / r_g,
\end{equation}
is the dimensionless spin, and we have defined 
\begin{equation}
Z_1 = 1 + (1-a_\bullet^2)^{1\over3}  \left((1+a_\bullet)^{1\over3} + (1-a_\bullet)^{1\over3}\right) 
\end{equation}
and
\begin{equation}   
Z_2 = \sqrt{3a_\bullet^2 + Z_1^2} .
\end{equation}
Another orbital motion component of use is 
\begin{equation}
    U^t(r, a) = {1 + a\sqrt{r_g/r^3} \over \left(1 - 3r_g/r + 2a\sqrt{r_g/r^3}\right)^{1/2}}.
\end{equation}

\subsection{Extra-ISCO temperature profile}
In the stable regions of the disc, the temperature profile of a thin accretion flow fed by a constant mass flux $\dot M$ in the general Kerr metric $g^{\mu \nu}$ is given by \citep{NovikovThorne73, PageThorne74, Balbus17}
\begin{equation}\label{temp_def}
\sigma T^4 = - {1 \over 2\sqrt{|g|}} \left( U^t \right)^2 {{\rm d} \Omega  \over  {\rm d} r} \left[ {\dot M \over 2\pi} \int^r {U_\phi'  \over U^t } \, {\rm d}r + {{\cal F}_{\cal J} \over 2\pi} \right] ,
\end{equation}
where 
\begin{equation}    
\Omega \equiv {U^{\phi} \over U^t}  = {\sqrt{GM/r^3} \over 1 + a\sqrt{r_g/r^3}} ,
\end{equation}
$g$ is the metric determinant, and $\dot M$ is the constant mass flux through the disc, while ${\cal F}_{\cal J}$ is an integration constant relating to the angular momentum flux through the disc.
The temperature of the main body of the accretion disc is therefore found by solving the integral 
\begin{multline}
I(r, a) = {\dot M \over 2\pi} \int^r {U_\phi'  \over U^t } \, {\rm d}r + {{\cal F}_{\cal J} \over 2\pi}  \\ = {\dot M  c \over 4\pi} \int^r \sqrt{r_g \over r} {r^2 - 6r_gr - 3a^2 + 8a\sqrt{r_g r} \over r^2 - 3r_gr + 2a\sqrt{r_gr}}\, {\rm d}r + {{\cal F}_{\cal J} \over 2\pi}
\end{multline}
with $x \equiv \sqrt{r/r_g}$ and $a_\bullet = a/r_g$, we have
\begin{equation}
I(r, a) = {\dot M r_g c \over 2 \pi } \int^r  {x^4 - 6x^2 - 3a_\bullet^2 + 8a_\bullet x \over x^4 - 3x^2 + 2a_\bullet x }\, {\rm d}x + {{\cal F}_{\cal J} \over 2\pi} ,
\end{equation}
which can be solved with standard partial fraction techniques, as was first shown in \cite{PageThorne74}. The roots of the lower cubic 
\begin{equation}
x^3 - 3x + 2a_\bullet = 0,
\end{equation}
denoted $\{x_\lambda\}$ are 
\begin{equation}
x_\lambda = 2 \cos\left[ {1\over 3} \cos^{-1}(-a_\bullet) - {2\pi\lambda\over3}\right] 
\end{equation}
with $\lambda = 0, 1, 2$. The solution is then 
\begin{equation}\label{zeta_sol}
I(r, a) = {\dot M r_g c \over 2 \pi } \left[ x - {3a_\bullet \over 2} \ln(x) + \sum_{\lambda = 0}^{2} k_\lambda \ln\left|x - x_\lambda\right|\right] + {{\cal F}_{\cal J} \over 2\pi}, 
\end{equation}
where 
\begin{equation}
 k_\lambda \equiv {2 x_\lambda - a_\bullet(1 + x_\lambda^2)  \over 2(1 - x_\lambda^2)} .
 \end{equation}
We now choose the physical values of our integration constants. The mass flux ${\dot M}$ simply scales the amplitude of each parameter, while ${\cal F}_J$ determines the ISCO boundary condition. The vanishing ISCO stress condition enforces 
\begin{multline}
- {{\cal F}_{{\cal J}, {\cal V}} \over \dot M  } \equiv {\cal J}_{\cal V}= \sqrt{GMr_I} \Bigg[ 1- {3a_\bullet \over 2 x_I} \ln(x_I) \\ + {1\over x_I} \sum_{\lambda = 0}^{2} k_\lambda \ln\left|x_I - x_\lambda\right|\Bigg] ,
\end{multline} 
and the general boundary condition we shall write as 
\begin{equation}
- {{\cal F}_{\cal J} \over \dot M  } \equiv {\cal J}_{\cal V}(1 - \delta_{\cal J}).
\end{equation}
{The parameter ${\cal J}_{\cal V}$ corresponds to the specific angular momentum that would have to be carried by each fluid element from the ISCO to the horizon so that there is no communication between the two disc regions. } This expression defines $\delta_{\cal J}$, which is our free inner disc stress parameter. We define our inner disc parameter in this manner as $\delta_{\cal J}$ is a number that can be readily determined from GRMHD simulations of accretion discs. Physically $\delta_{\cal J}$ can be thought of as the fraction of the specific angular momentum `passed back' into the disc from the $r \leq r_I$ region by whatever process is generating the stress at the ISCO. 

{While $\delta_{\cal J}$ has a clear physical interpretation, it is not immediately clear what value it should take, or upon which other physical parameters it should depend. As we discuss in the main body of the text, the physical origin of $\delta_{\cal J}$ is ultimately expected to be angular momentum transport driven by magnetic turbulence, and no simple scaling arguments for its parameter dependence is currently known. Simulations provide a rough scale of $\delta_{\cal J}$, placing it in the range $\sim 0.01 - 0.2$ \citep{Shafee08, Penna10, Noble10}, but also suggest that $\delta_{\cal J}$ may depend on factors including the magnetic field topology of the simulation \citep[for example, whether the initial field geometry comprised of 1 or 4 field loops][]{Penna10}. Intuitively, it seems likely that $\delta_{\cal J}$ will depend on factors such as the overall disc magnetisation (a ``MAD'' or ``magnetically arrested disc'' can in some ways be thought of as an extreme ``stress'' sourced from magnetic field, for example). \cite{Gammie99} derives a parameterisation for the ISCO stress entirely in terms of a magnetic field strength at the ISCO, but makes the assumption of a cold (zero temperature) gas, which is not ideal for our continuum fitting purposes. 

Given the anticipated scaling of $\delta_{\cal J}$ with properties of the disc magnetic field, we anticipate that otherwise similar black holes (in terms of mass $M$ and spin $a_\bullet$) may have quite different values of $\delta_{\cal J}$. We reiterate that observational studies, and in particular parameter inference studies of $\delta_{\cal J}$, may have much to teach us about this region.  }

Folding this result through to the key radiative temperature profile, we find (eq. \ref{temp_def})
\begin{multline}\label{TRaaa}
\sigma T_R^4 = {3 GM\dot M \over 8 \pi r^3} \Bigg[ 1- {3a_\bullet \over 2 x} \ln(x) + {1\over x} \sum_{\lambda = 0}^{2} k_\lambda \ln\left|x - x_\lambda\right| \\ - {{\cal J}_{\cal V} (1 - \delta_{\cal J})  \over \sqrt{GMr} }  \Bigg] \left[ 1 -{ 3 \over  x^{2} } + { 2 a_\bullet \over x^{3}} \right]^{-1} .
\end{multline}
{This is the extra-ISCO temperature profile used in the \fk model, and also serves to determine the thermodynamic boundary condition for the post-ISCO plunge. }
\subsection{Intra-ISCO disc}
Unlike the extra-ISCO disc regions, inside the ISCO the evolution of the disc temperature depends on all thermodynamic quantities, not just the local mass flux $\dot M$ and angular momentum flux ${\cal F}_{\cal J}$. Assuming an ideal fluid stress energy tensor 
\begin{equation}
    T^{\mu\nu} = \left(\rho + {P+ e\over c^2}\right)U^\mu U^\nu + P g^{\mu\nu} + T^{\mu\nu}_{\rm EM} ,
\end{equation}
where $P, e$ and $\rho$ correspond to the total pressure, energy density and (rest) mass density of the fluid respectively, and $T^{\mu\nu}_{\rm EM}$ is the electromagnetic stress-energy tensor, which does not enter the energy equation in the {ideal} MHD limit \citep{MummeryBalbus2023}. The energy equation of the fluid can be expressed 
\begin{equation}
    U_\nu \nabla_\mu T^{\mu\nu} = 0 , 
\end{equation}
or explicitly 
\begin{equation}\label{energy_eqn}
    {{\rm d} e \over {\rm d} r} - \left({P+ e \over \rho}\right) {{\rm d} \rho \over {\rm d}r} = 0. 
\end{equation}
This is the fundamental thermodynamic equation within the ISCO which we solve. We must supplement this equation with the following expressions valid inside of the ISCO 
\begin{align}
c_{s}^2 &\equiv {P \over \rho} , \label{D1}\\
H &= c_{s} \sqrt{r^4 \over 2 G M_\bullet r_I} 
 ,\label{D2} \\
\rho &\equiv {\Sigma \over H}, \label{D3}\\
P &= P_g + P_r \equiv {\rho k T_{c} \over \mu m_p} + {4 \sigma T_{c}^4 \over 3 c} ,\label{D4} \\
T_{c}^4 &= \left({3\kappa \Sigma\over 8} \right) T_{R}^4 , \label{D5}\\
\Sigma &= {\dot M \over 2\pi r |U^r|} ,\label{D6} \\
e &= {3\over 2} P_g + 3 P_r \label{D7} , \\
U^r &= - c \sqrt{2r_g \over 3 r_I} \left({r_I \over r} - 1\right)^{3/2} - \alpha^{1/2} c_{s, I} \label{D8} . 
\end{align}
These are, in order, the definition of the speed of sound, the solution of vertical hydrostatic equilibrium, the definition of the disc density and pressure, the approximate solution of radiative transfer in the disc atmosphere,  the conservation of mass in the disc, the definition of the energy density  and the geodesic plunge solution with trans-ISCO velocity given by the typical velocity fluctuation scale. 

Equations \ref{D1} -- \ref{D8} are sufficient to close equation \ref{energy_eqn}, which becomes a first order ordinary differential equation for $T_c(r)$. We solve this equation numerically in {\tt fullkerr}. Equation \ref{D5} then gives the radiative temperature used in the disc model (cf. Fig. \ref{fig:temp-comp}). 

In certain limits the radiative temperature can be written as an analytical function of radius{, an analysis of which was performed in \cite{MummeryBalbus2023}}. If radiation pressure dominates over gas pressure then \citep{MummeryBalbus2023}
\begin{equation}
T_R = T_{R, I}   \left({r_I \over r}\right)^{17/28}  \left[{\varepsilon}^{-1}\left({r_I \over r} - 1 \right)^{3/2} + 1 \right]^{-1/28} ,
\end{equation}
while if gas pressure dominates then \citep{MummeryBalbus2023}
\begin{equation}
    T_R = T_{R, I}   \left({r_I \over r}\right)^{5/4}  \left[{\varepsilon}^{-1}\left({r_I \over r} - 1 \right)^{3/2} + 1 \right]^{-1/4} .
\end{equation}
In these expressions anything with a subscript $I$ is evaluated at the ISCO, and 
\begin{equation}
    \varepsilon \equiv {\alpha^{1/2} c_{s, I} \over c} \sqrt{3r_I \over 2r_g} \ll 1. 
\end{equation}
In both of these expressions we have assumed that electron scattering opacity dominates over the free-free opacity. 

{In addition to the radiative temperature, all thermodynamic quantities of the disc can be solved for within the ISCO by suitable manipulation of equations \ref{energy_eqn}--\ref{D8}. Of relevance for the physics of photon starvation (section 4) is the vertical structure of the disc, as the scale height sets both the photon diffusion timescale and  ultimately the number of photons which can be produced by free-free processes across the disc thickness. For an arbitrary adiabatic equation of state $e = P/(\Gamma - 1)$ the disc scale height can be solved for exactly \citep{MummeryBalbus2023}, and is given by }
\begin{equation}
    \left({H \over H_I}\right)^{1 + \Gamma} = \left({r_I \over r} \right)^{\Gamma - 5} \left[ \varepsilon^{-1} \left({r_I \over r} - 1\right)^{3/2} + 1\right]^{1 - \Gamma} .
\end{equation}
{An important, and potentially counter-intuitive, result of the above theory is that the scale height of the disc drops over the plunging region (i.e., the disc gets thinner over the plunge). }

\subsection{Vertical gravity}
The (dimensionless) relativistic vertical gravity correction factor \citep{Abramowicz97} is given by 
\begin{equation}
    {\cal R}_z(r, a) = \left[ {U_\phi^2/(r_g^2 c^2) - a_\bullet^2 \left(U_t^2/c^2 - 1\right) \over r/r_g}  \right] . 
\end{equation}
A result valid at all radii. Outside of the ISCO the fluid's orbital 4-momentum elements are given by the results listed in equations (\ref{orbang}) and (\ref{orben}). Inside the ISCO, when $U_\phi$ and $U_t$ are conserved and given by their ISCO values, this result simplifies dramatically to \citep{MummeryBalbus2023}
\begin{equation}
    {\cal R}_z(r, a) = {2 r_I \over r} . 
\end{equation}

\section{Parameters of the {\tt fullkerr} continuum fitting {\tt XSPEC} model }\label{appB}
\begin{table}
    \renewcommand{\arraystretch}{2}
    \centering 
    \begin{tabular}{|p{2.0cm} p{2cm} p{2cm}|}
    \hline
       Parameter  & Units & Allowed range  \\
    \hline 
       $M$ & $M_\odot$ &  $M > 0$ \\ 
       $a_\bullet$ & & $-1 < a_\bullet < 1$ \\
       $i$ & Degrees & $0^\circ < i < 90^\circ$ \\ 
       $\dot M$ & $L_{\rm edd}/c^2$ & $\dot M > 0$ \\ 
       $\delta_{\cal J}$ & & $\delta_{\cal J} \geq 0$ \\
       $f_d$ & & $f_d \geq 1$ \\
       $f_I$ & & $f_I \geq 1$ \\
       $\xi$ & & $\xi > 0$ \\
       {\tt norm} & $1/$kpc$^2$ &  {\tt norm} $ > 0 $ \\ 
       \hline 
    \end{tabular}
    \caption{The parameters of the \fk model,  their units and allowed ranges. The accretion rate parameter is defined in terms of the Eddington luminosity, which equals $L_{\rm edd} = 1.26 \times 10^{38} (M/M_\odot)$ erg/s. {One sided ranges denote physically allowed values, e.g., positive black hole masses. }   }
    \label{fkt}
\end{table}

The {\tt fullkerr} continuum fitting model takes as an input the free parameters listed in Table \ref{fkt}. We treat the colour-correction in the main extra-ISCO body of the disc by a single number $f_d$, which is in principle a free parameter in the fits (although it was fixed to $f_d = 1.7$ for the MAXI J1820 fits in this paper). In the {\tt XSPEC} model we also allow for the use of the \cite{Done12} temperature dependent model (specified in {\tt XSPEC} by $f_d = -1$), but as the NuSTAR spectra probed high photon energies $E > 3$ keV, we found no sensitivity to the exact extra-ISCO $f_{\rm col}$ parameterisation for MAXI J1820. 

The {\tt fullkerr} model assumes a \cite{SS73} $\alpha$-parameter of $\alpha = 0.1$. Again, the spectra are only very weakly dependent on $\alpha$, and this is hard coded for convenience. In this model we restrict the model to purely black hole spacetimes (i.e., $|a_\bullet| \leq 1$), and do not consider naked singularity spacetimes \citep[cf.][]{MummeryBalbusIngram24}. The reason for this is the intra-ISCO behaviour of the disc fluid near $r = 0$ is observable for a naked singularity disc, and the models developed in \cite{MummeryBalbus2023} break down in this region.

\section{Fits to other epochs of MAXI J1820 }\label{other_epochs}
In Table \ref{epochs} we list the best fitting {\tt fullkerr} fitting parameters to the other MAXI J1820 epochs. We also list the  fit statistic of this model and the improvement in the fit statistic compared to a {\tt kerrbb + cutoffpl} model. For these fits we fix the black hole parameters to $M = 8.5 M_\odot$, $a_\bullet = 0.2$, the source-observer parameters to $i = 35^\circ$, $D = 3$ kpc, and the colour-correction parameters to $f_d = f_I = 1.7$. Formally better fits are possible if these parameters are allowed to freely vary, but the improvement over {\tt kerrbb + cutoffpl} is so strong that we deem this unnecessary. The chief cause of model-data discrepancies is now at photon energies $E \geq 10$ keV, where the spectrum is dominated by power-law emission and not the disc. 

\begin{table}
    \renewcommand{\arraystretch}{2}
    \centering 
    \begin{tabular}{|p{.6cm} | p{1cm} p{1.3cm} p{1.2cm} | p{1cm} p{1.2cm}|}
    \hline
       Epoch  & $\dot M$ & $\delta_{\cal J}$ & $\xi$ & $\chi_{r, {\tt fk}}^2$ & $\Delta \chi_{{\tt fk} - {\tt kbb}}^2$  \\
    \hline 
       Nu25 & $1.15^{+0.02}_{-0.02}$ &  $0.043^{+0.001}_{-0.001}$ & $1.91^{+0.05}_{-0.05}$ & ${888.8\over858}$ & $-629.8$ \\ 
       Nu27 & $1.05^{+0.01}_{-0.01}$ &  $0.040^{+0.001}_{-0.001}$ & $1.49^{+0.04}_{-0.04}$ & ${982.1\over 858}$ & $-785.4$ \\ 
       Nu29 & $0.86^{+0.02}_{-0.02}$ &  $0.039^{+0.002}_{-0.002}$ & $1.70^{+0.06}_{-0.06}$ & ${352.3\over360}$ & $-820.7$ \\ 
       Nu31 & $0.79^{+0.01}_{-0.01}$ &  $0.036^{+0.001}_{-0.001}$ & $1.85^{+0.03}_{-0.03}$ & ${518.6\over461}$ & $-816.0$ \\ 
       \hline 
    \end{tabular}
    \caption{Best fitting parameters of the \fk model to each MAXI J1820 epoch, the  fit statistic of this model and the improvement in the fit statistic compared to a {\tt kerrbb + cutoffpl} model. Parameters are in the same units as discussed in the main body of the text. See \citealt{Fabian20} for more details of the individual epochs. Parameter confidence intervals are $90\%$ intervals. The luminosity Eddington ratio is roughly 10$\%$ of the accretion rate parameter for the chosen black hole spin.   }
    \label{epochs}
\end{table}

The fitted parameters are in the same units as discussed in the main body of the text (see also Table \ref{fkt}). See \cite{Fabian20} for more details of the individual epochs. Parameter confidence intervals are $90\%$ intervals. The luminosity Eddington ratio is roughly 10$\%$ of the accretion rate parameter for the chosen black hole spin{, and varies by roughly 50\% over the four observations. The fraction of the total $0.05-50$ keV luminosity sourced from within the ISCO for the four epochs listed was $4.5\%, 4.3\%, 4.1\%$ and $3.8\%$ respectively. The decreasing {bolometric} contribution from within the ISCO traces the slight fall in ISCO stress parameter $\delta_{\cal J}$.  } 

{While the fraction of the total luminosity observed between $0.05-50$ keV fell as the total soft-state disc flux decreased in amplitude, in the $E = 6-10$ keV band the fraction of the total observed flux which was sourced from within the ISCO actually rose, from $\sim 40\%$ to $\sim 60\%$. This highlights the sensitive balance played between all spectral components in determining the relative importance of the plunging region. At this stage it is not yet clear if there are any particular luminosity states in which plunging region emission is negligible.  }

\begin{figure}
    \centering
    \includegraphics[width=\linewidth]{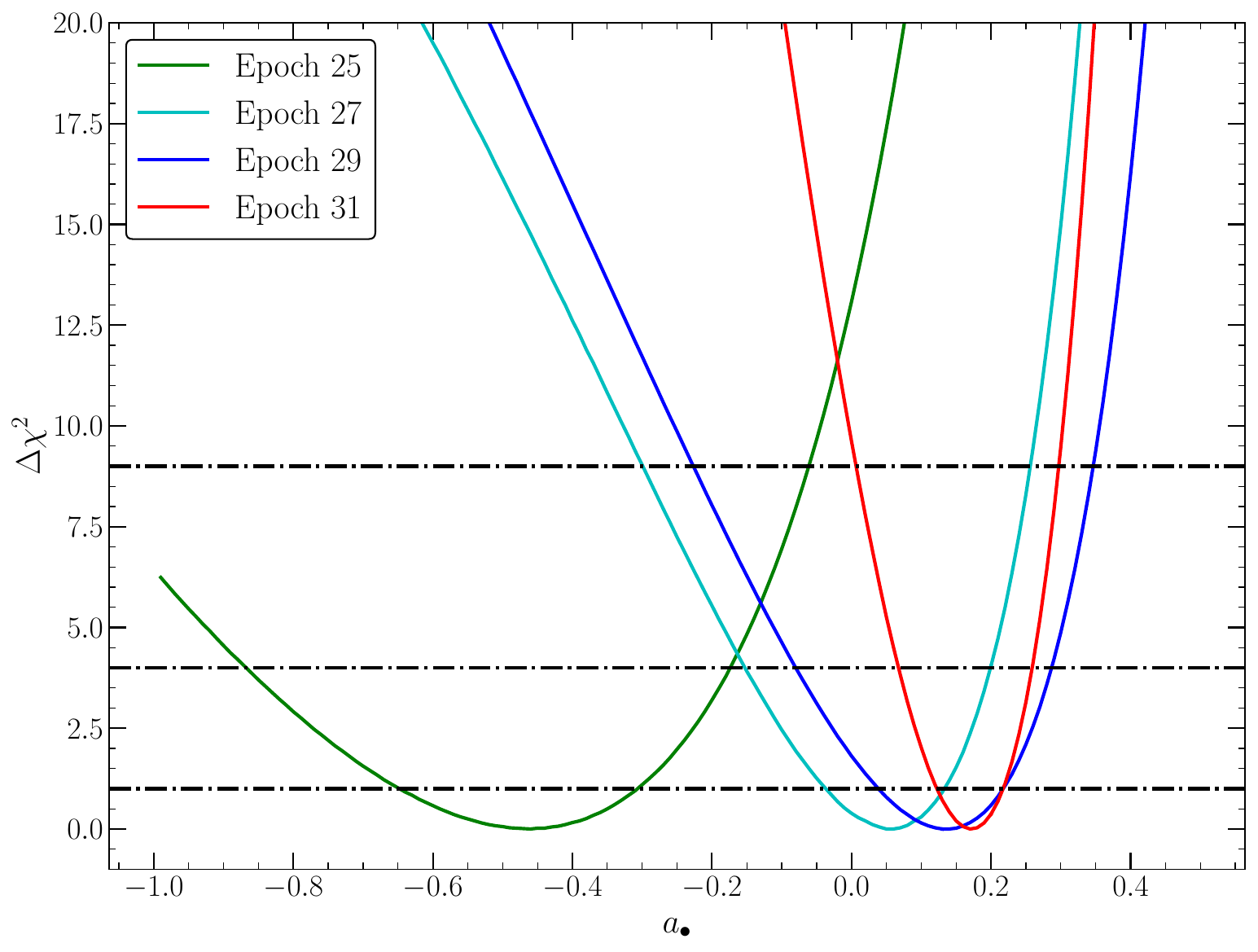}
    \caption{Spin constraints at fixed inclination ($i = 35^\circ$) for each of the different epochs of MAXI J1820 data. Horizontal dot-dashed curves correspond to $1, 2$ and $3 \sigma$ constraints (in order of increasing $\Delta \chi^2$). Epoch 25 is inconsistent with the other epochs at $\sim 2\sigma$, but this is likely the result of the more prominent (and complex) power-law component in this epoch.   }
    \label{fig:diff_epoch_spin}
\end{figure}

{In Figure \ref{fig:diff_epoch_spin} we show the spin constraints at fixed inclination ($i = 35^\circ$) for each of the different epochs of MAXI J1820 data. Horizontal dot-dashed curves correspond to $1, 2$ and $3 \sigma$ constraints for each epoch (in order of increasing $\Delta \chi^2$). Epoch 25 is inconsistent with the other epochs at $\sim 2\sigma$, but this is likely the result of the more prominent {and complex coronal} component in this epoch. It has been suggested \citep[e.g.,][]{Fabian20} that there remains iron emission features in epoch 25, {which will be poorly modelled with a pure power-law, and} which is likely driving the fit statistic in this epoch. As the power-law component fades in relative importance (epochs 25$\to$31), the uncertainty on the spin constraints drops, and settles at the \cite{Buisson19} value measured in the hard state for this inclination.  To verify that epoch 25 is an outlier in all models {(and therefore that it is the coronal complexity driving the fit)}, we fit the model {\tt kerrbb + bbody + cutoffpl} to all 4 epochs and performed an identical test (i.e., we fixed the inclination, and determine {\tt kerrbb} spin posteriors). Epoch 25 is an outlier in this analysis as well, at the same significance level. As in our original {\tt fullkerr} analysis the other 3 epochs showed consistency in their fitted spin parameters,   although the spins required were systematically larger (around $a_\bullet \simeq 0.6$), likely linked to the lack of near-ISCO emission in {\tt kerrbb} models with a vanishing ISCO stress.   }

\begin{figure}
    \centering
    \includegraphics[width=\linewidth]{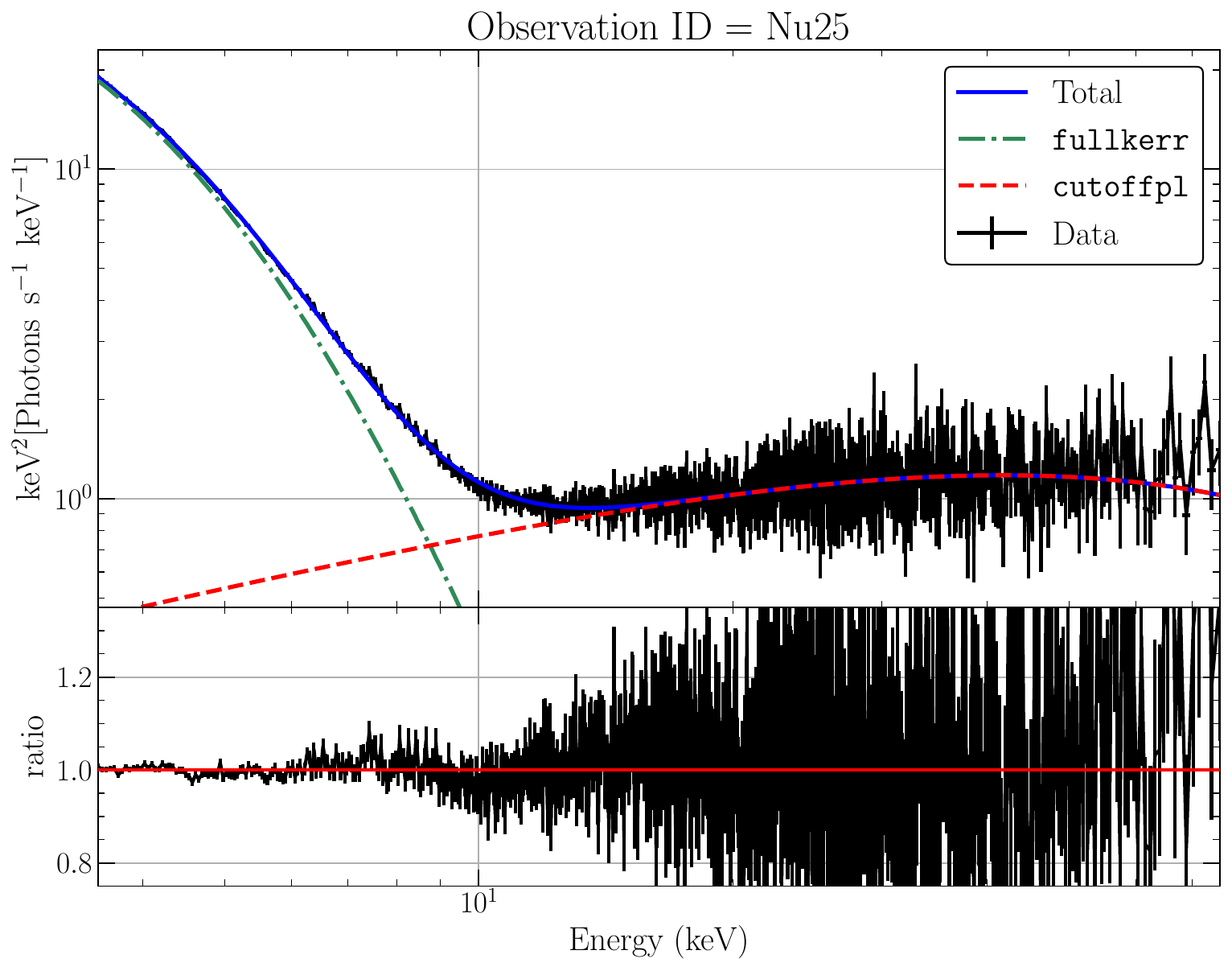}
    \includegraphics[width=\linewidth]{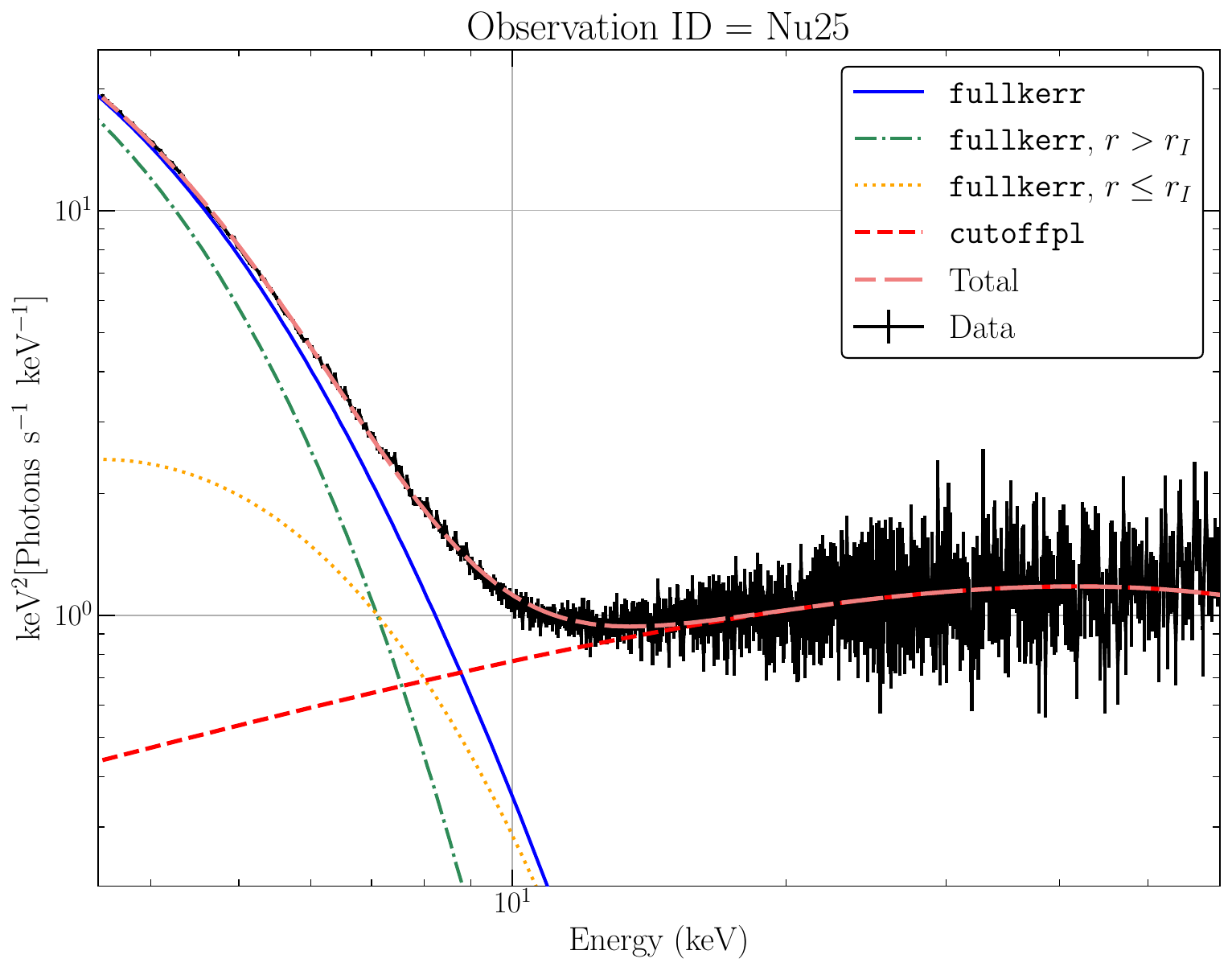}
    \includegraphics[width=\linewidth]{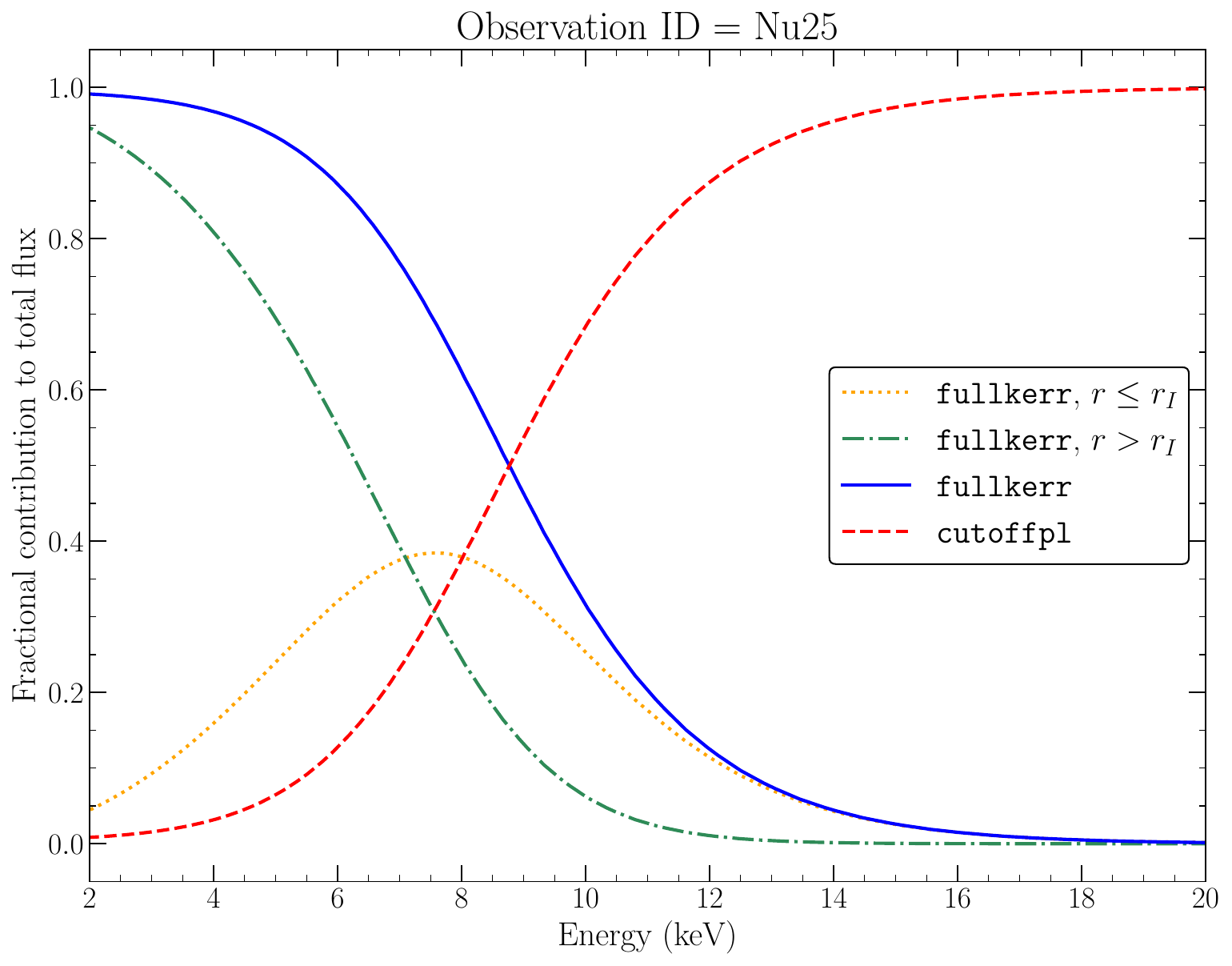}
    \caption{Similar fits to Figures \ref{fig:maxij1820}, \ref{fig:components}, \ref{fig:frac_cont} but for epoch Nu25 \citep[see][for more details]{Fabian20}. See text for fixed and fitted parameter values.  }
    \label{fig:25}
\end{figure}

\begin{figure}
    \centering
    \includegraphics[width=\linewidth]{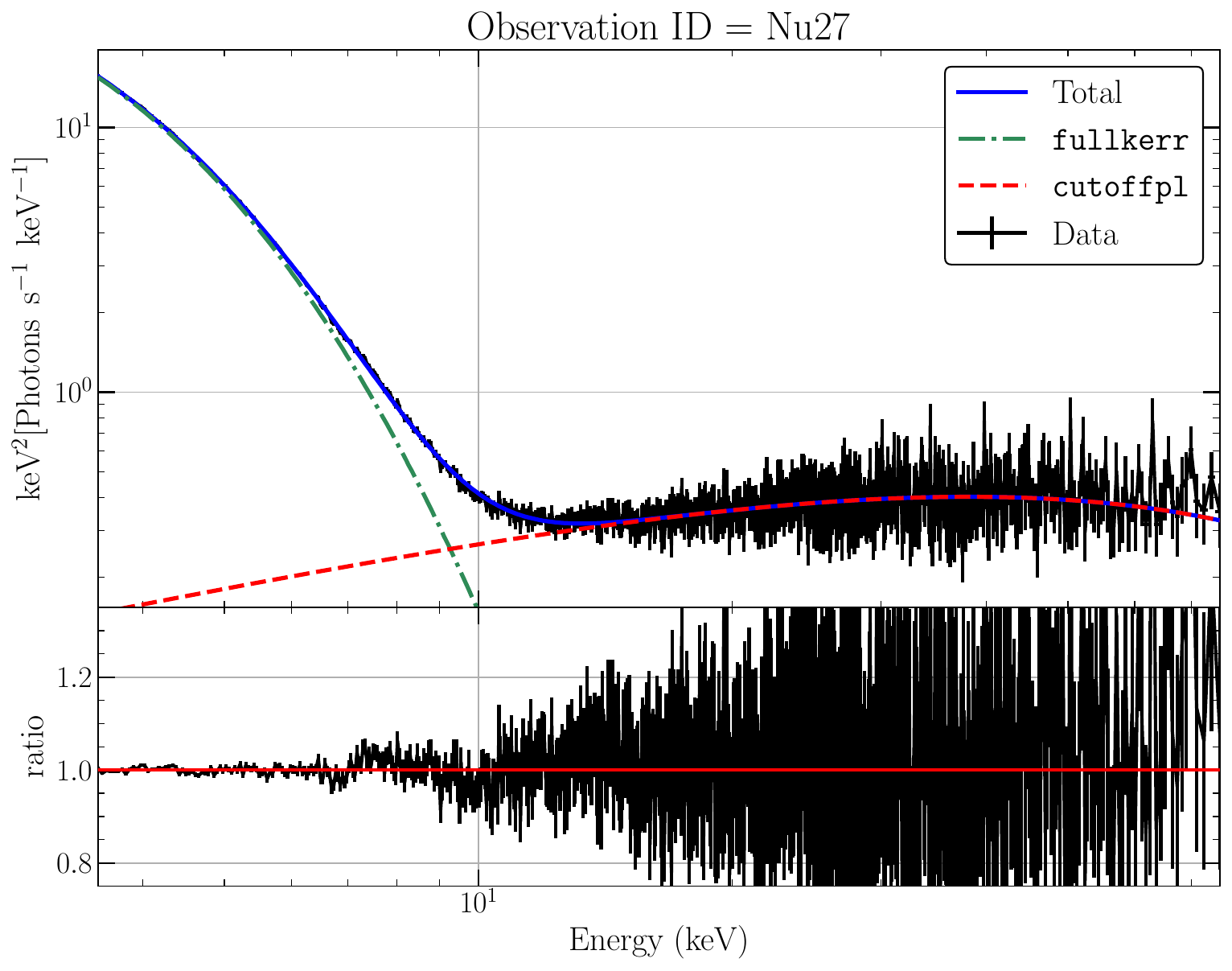}
    \includegraphics[width=\linewidth]{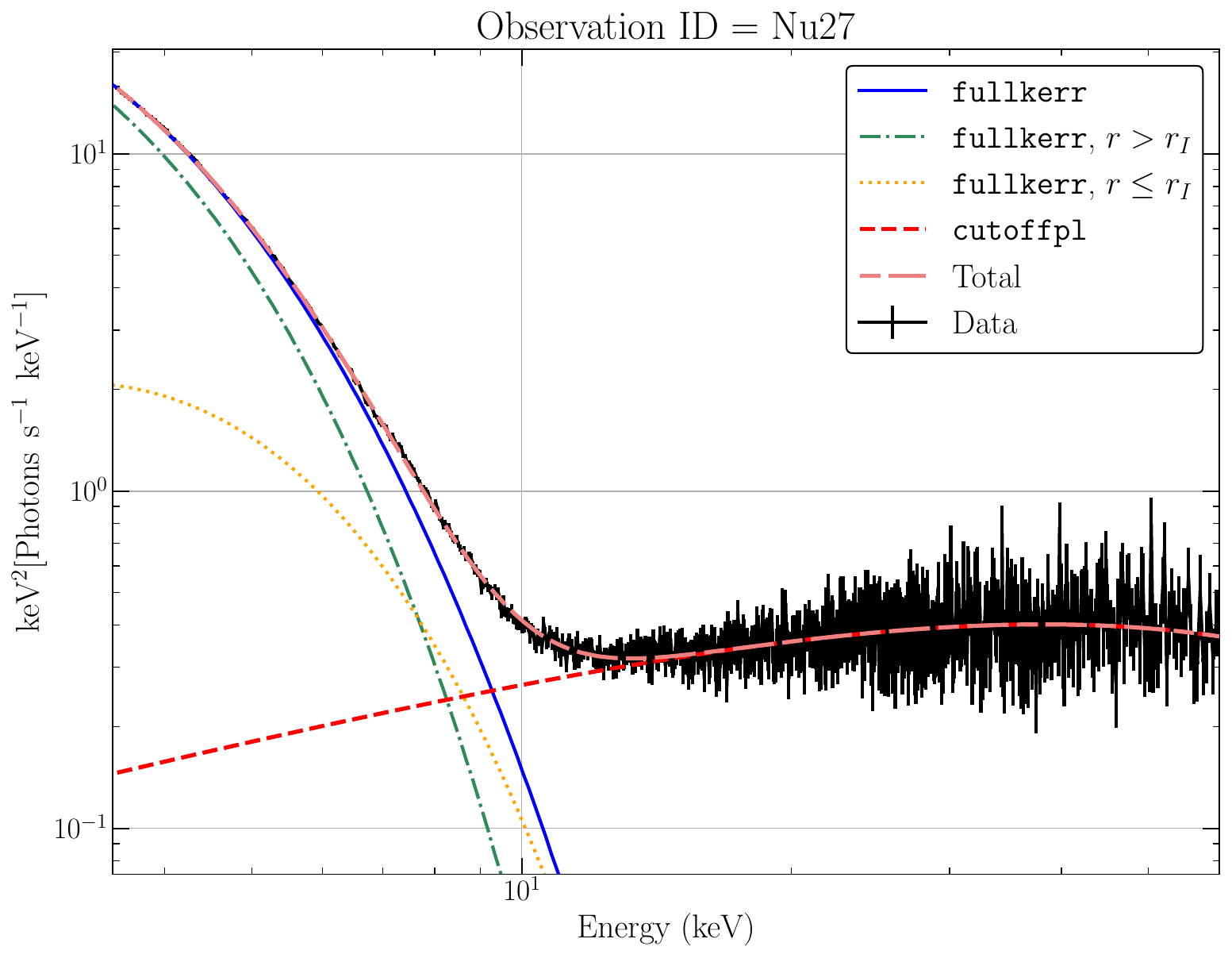}
    \includegraphics[width=\linewidth]{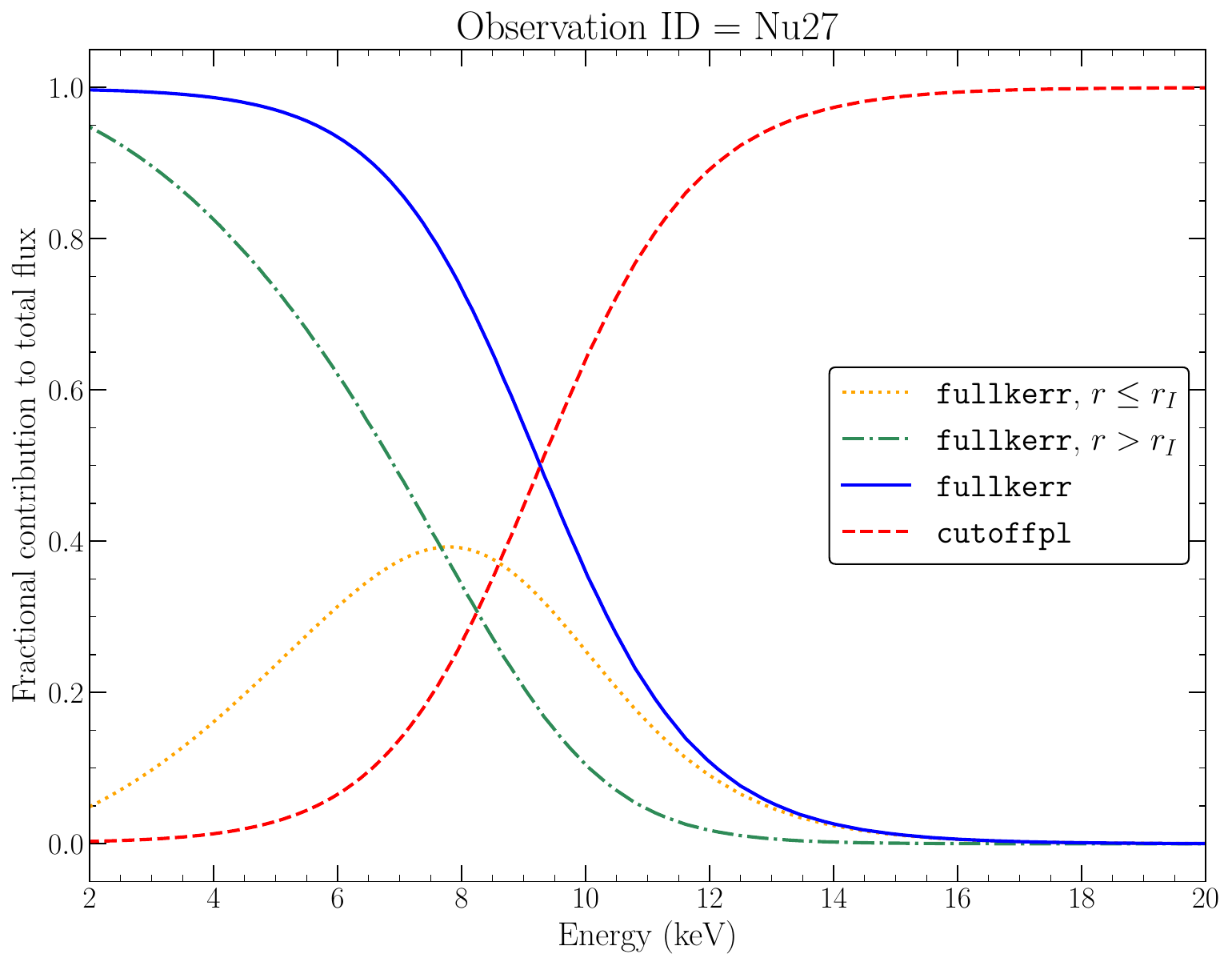}
    \caption{Similar fits to Figures \ref{fig:maxij1820}, \ref{fig:components}, \ref{fig:frac_cont} but for epoch Nu27 \citep[see][for more details]{Fabian20}. See text for fixed and fitted parameter values. }
    \label{fig:27}
\end{figure}

\begin{figure}
    \centering
    \includegraphics[width=\linewidth]{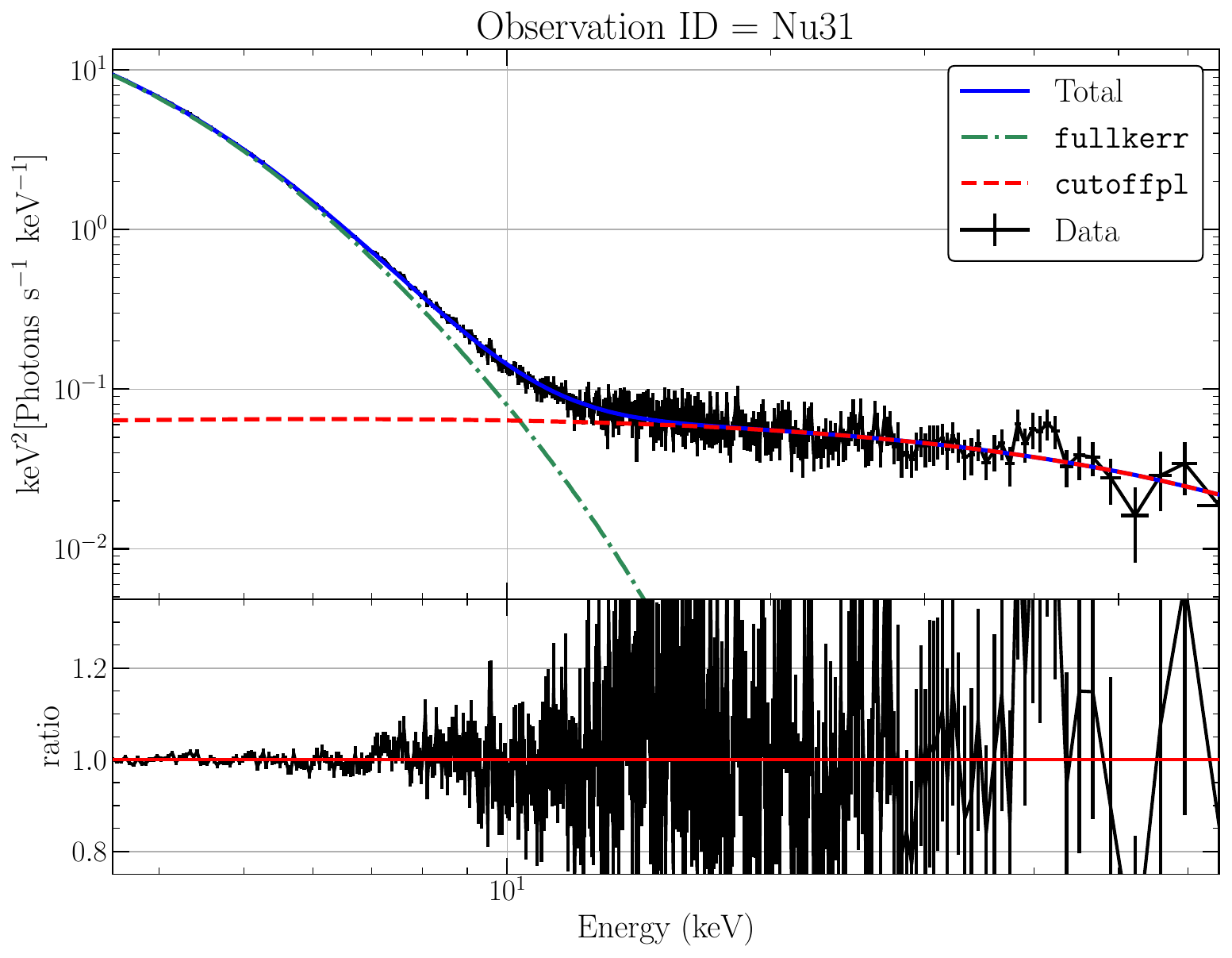}
    \includegraphics[width=\linewidth]{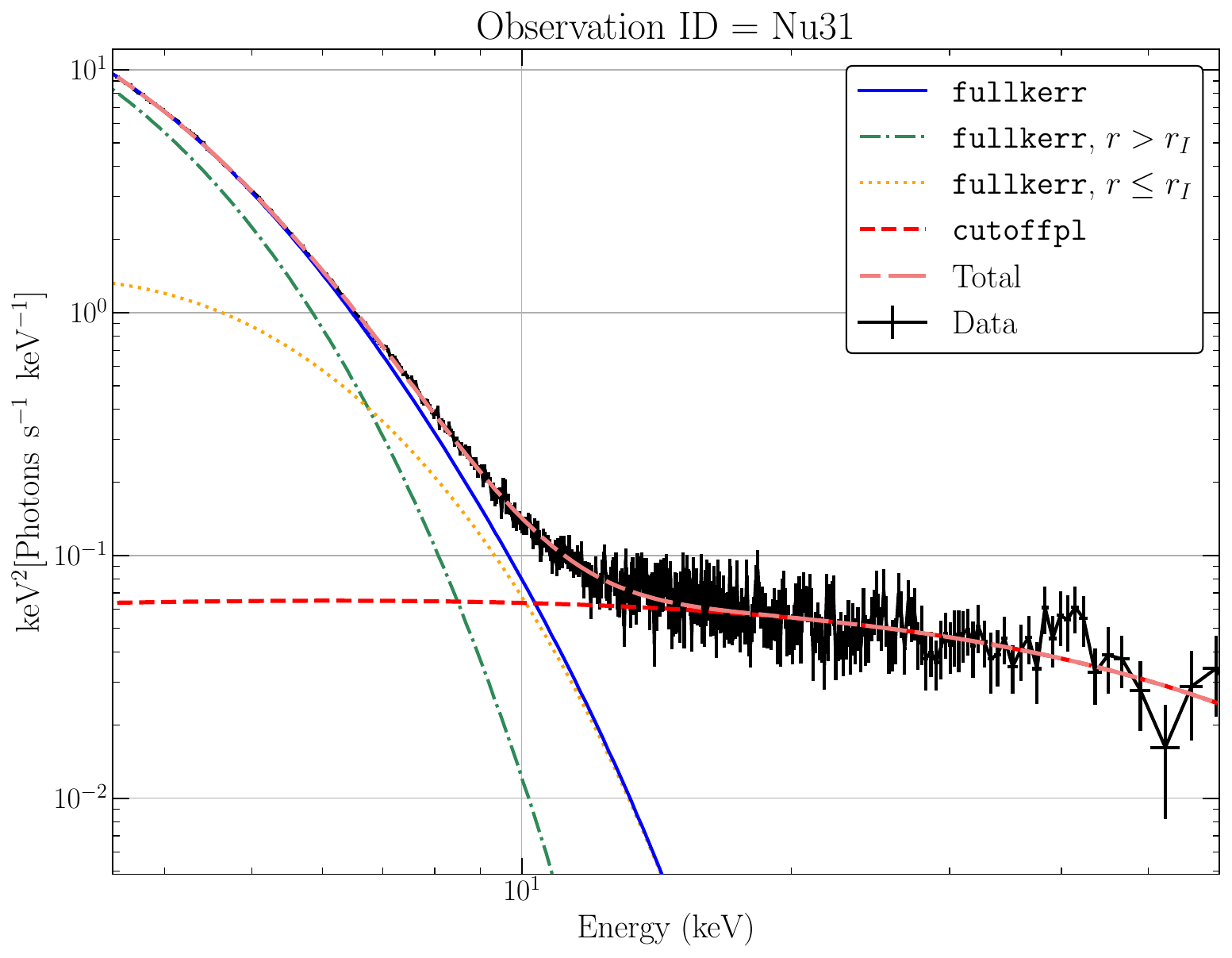}
    \includegraphics[width=\linewidth]{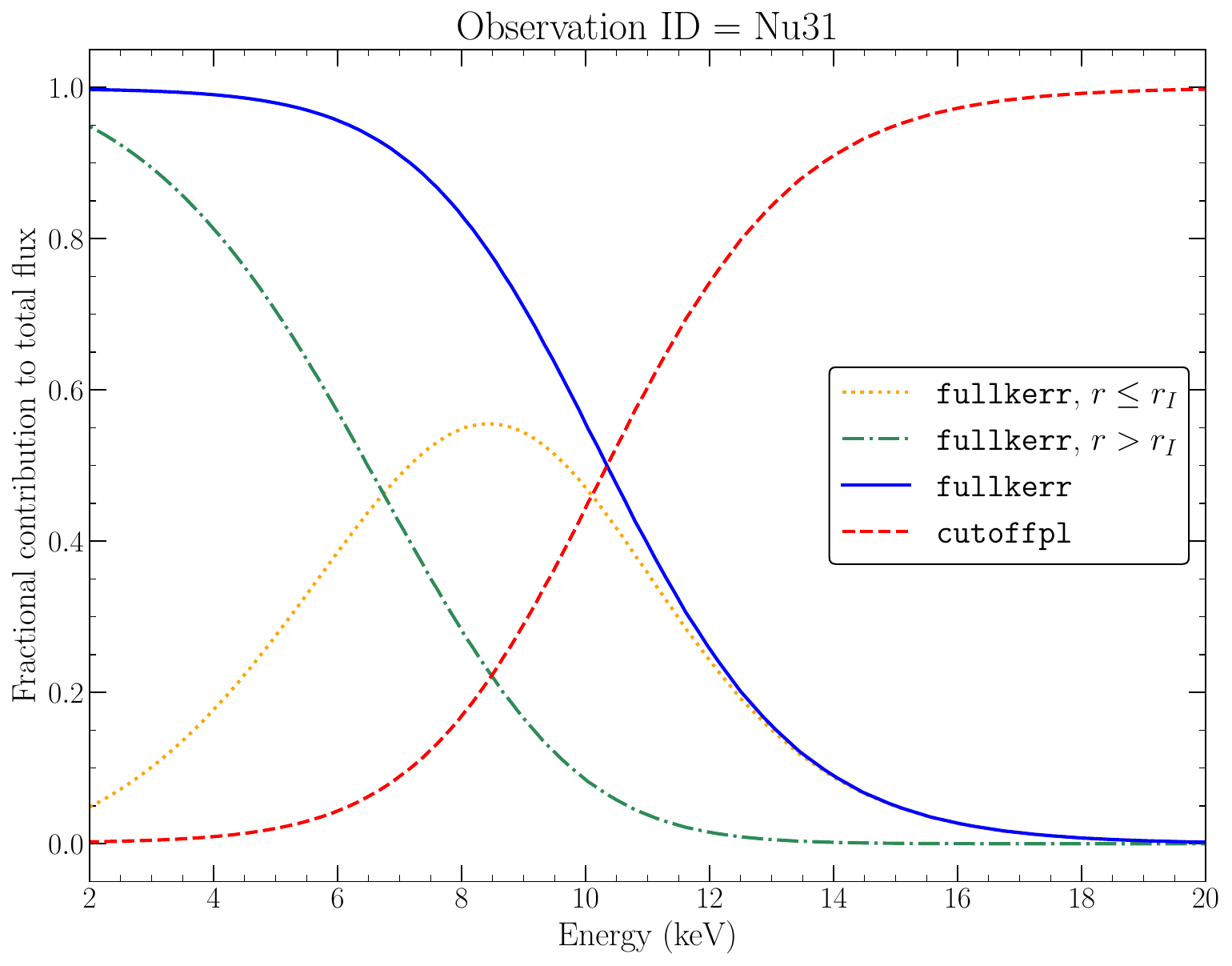}
    \caption{Similar fits to Figures \ref{fig:maxij1820}, \ref{fig:components}, \ref{fig:frac_cont} but for epoch Nu31 \citep[see][for more details]{Fabian20}. See text for fixed and fitted parameter values. }
    \label{fig:31}
\end{figure}

In Figures \ref{fig:25}, \ref{fig:27} and \ref{fig:31} we repeat the analysis presented in Figures \ref{fig:maxij1820}, \ref{fig:components}, \ref{fig:frac_cont} but for epochs Nu25, Nu27 and Nu31 respectively. It is clear to see that MAXI J1820 is well described by a black hole disc, with a substantial intra-ISCO emission component, over a range of epochs.  Interestingly,  the ISCO stress found for each epoch is relatively stable at $\delta_{\cal J} \simeq 0.04$. {As we have discussed in previous sections, the ISCO stress is ultimately sourced from magnetic turbulence within the ISCO, and {\it a priori} it is not obvious that it should remain relatively constant with time. Further detailed modelling of X-ray binaries is encouraged, so that the properties of $\delta_{\cal J}$ can be probed on the population level.  }


\label{lastpage}
\end{document}